\documentclass[english,11pt]{article}

\usepackage{amsmath}
\usepackage{graphicx}
\usepackage{grffile}
\usepackage{latexsym}
\usepackage{color}
\usepackage{amssymb}
\usepackage{amsbsy}
\usepackage{hyperref}
\usepackage[linesnumbered,ruled,vlined,commentsnumbered]{algorithm2e}
\usepackage{authblk}
\usepackage{geometry}
\usepackage{epstopdf}
\usepackage{cite}
\usepackage{booktabs}
\usepackage{array,multirow}
\newtheorem{remark}{Remark}
\usepackage{latexsym}
\usepackage[utf8]{inputenc}
\usepackage[english]{babel}
\usepackage{epstopdf}
\usepackage{hyperref}
\usepackage{xcolor}
\usepackage{amsfonts,amsmath,mathtools,bbm,cool}
\newcommand{\fer}[1]{(\ref{#1})}
\usepackage{tikz}
\usepackage{amsmath,amssymb}
\usepackage{graphicx,subcaption}
\usetikzlibrary{shapes,arrows}
\usetikzlibrary{shapes,arrows,decorations.markings}
\usetikzlibrary{decorations.pathmorphing}

\tikzstyle{block_long} = [rectangle, draw, fill=blue!20,
    text width=10.0em, text centered, rounded corners, minimum height=3em]
\tikzstyle{block_medium} = [rectangle, draw, fill=blue!20,
    text width=6.0em, text centered, rounded corners, minimum height=3em]
\tikzstyle{block_short} = [rectangle, draw, fill=blue!20,
    text width=3.0em, text centered, rounded corners, minimum height=3em]
\tikzstyle{line} = [thick, draw, dashed,  -stealth']

\tikzstyle{block_long} = [rectangle, draw, fill=blue!20,
    text width=13.0em, text centered, rounded corners, minimum height=3.5em]
\tikzstyle{block_medium} = [rectangle, draw, fill=blue!20,
    text width=7.0em, text centered, rounded corners, minimum height=3.5em]
\tikzstyle{block_short} = [rectangle, draw, fill=blue!20,
    text width=5.0em, text centered, rounded corners, minimum height=3.5em]
\tikzstyle{line} = [thick, draw, dashed,  -stealth']
\tikzstyle{chapter} = [rectangle, draw, fill=blue!10,
text width=6.0em, text centered, rounded corners, minimum height=0.5em]

\def\R{\mathbb R}

\def\f{{\bf f}}

\def\cc{w}

\def\w{w}

\def\bbeta{\bar\beta}

\def\cA{\mathcal A}

\def\ve{\varepsilon}

\def\argmin{{\rm arg}\!\min}
\def\be{\begin{equation}}
\def\ee{\end{equation}}
\def\bea{\begin{eqnarray}}
\def\eea{\end{eqnarray}}
\def\beas{\begin{eqnarray*}}
\def\eeas{\end{eqnarray*}}

\font\dc=cmbxti10 %slanted boldface for titles of "definition"

\begin{document}
\title{New Trends on the Systems Approach to Modeling SARS-CoV-2 Pandemics in a Globally Connected Planet}

\author[$1$]{Giulia Bertaglia}
\author[$2$]{Andrea Bondesan}
\author[$3$]{Diletta Burini}
\author[$4$]{Raluca Eftimie}
\author[$5,6$]{Lorenzo Pareschi}
\author[$7,8$]{Giuseppe Toscani}

\affil[$1$]{\it Department of Environmental and Prevention Sciences, University of Ferrara, Italy}
\affil[$2$]{\it Department of Mathematical, Physical and Computer Sciences, University of Parma, Italy}
\affil[$3$]{\it Liceo Statale Classico e Musicale ``A. Mariotti'', Perugia, Italy}
\affil[$4$]{\it Laboratoire de Math\'ematiques de Besançon, CNRS-UMR 6623, Universit\'e de Franche-Comt\'e, France}
\affil[$5$]{\it Maxwell Institute \& Department of Mathematics, Heriot-Watt University, Edinburgh, UK}
\affil[$6$]{\it Department of Mathematics and Computer Science, University of Ferrara, Italy}
\affil[$7$]{\it Department of Mathematics, University of Pavia, Italy}
\affil[$8$]{\it IMATI CNR, Pavia, Italy}

\maketitle

\begin{abstract}
This paper presents a critical analysis of the literature  and perspective research  ideas for modeling the epidemics caused by the SARS-CoV-2 virus. It goes beyond deterministic population dynamics to consider several key complexity features of the system under consideration. In particular, the multiscale features of the dynamics from contagion to the subsequent dynamics of competition between the immune system and the proliferating virus. Other topics addressed in this work include  the propagation of epidemics in a territory, taking into account local transportation networks, the heterogeneity of the population, and the study of social and economic problems in populations involved in the spread of epidemics. The overall content aims to show how new mathematical tools can be developed to address the above topics and how mathematical models and simulations can contribute to the decision making of crisis managers.
\end{abstract}


\tableofcontents

%%%%%%%%%%%%%%%%%%%%%%%%%%%%%%%%%%%%%%%%%%%%
\section{Motivations and plan of the paper}\label{Sec:1}
%%%%%%%%%%%%%%%%%%%%%%%%%%%%%%%%%%%%%%%%%%%%

The  activation of scientific research to contrast the epidemics  generated by \textit{SARS--CoV--2} virus, responsible for the initial \textit{COVID-19} outbreak and the subsequent pandemic, was immediate involving not only virologists and immunologists, within the general framework of biology, but also scientists from mathematical and physical sciences, as well as scientists from social sciences and economics. Various achievements, especially the invention and production of new vaccines, have had a mitigating effect on the impact of the pandemic, but not an eradication effect. For this reason, the research activity in this area still maintains its initial intensity. In the field of biology, intensive research is aimed at understanding how molecular dynamics leads to the evolution of the biological properties of the virus over time, up to mutations, followed by pseudo-Darwinian selection, and how these aspects affect the in-host dynamics in the respiratory tract of individuals.

In fact, the pandemic had an enormous impact on the world's societies by changing our way of thinking, our way of living, our economy and, ultimately, our plans to support our well-being. The pandemic has invaded our planet so that all nations have had to face this epochal change and re-organize the structures devoted to medical care.

The mathematical sciences have contributed not only to the dialog between different scientific environments but also to the dialog between the scientific communities and the political community, as well as to the development of mathematical models capable of describing the dynamics of pandemics in time and space, focusing both on the evolution of pathology within infected individuals and on the spread of pandemics over space. These efforts have led to the development of new mathematical tools towards new multiscale theories of epidemics that go beyond the approach of deterministic population dynamics. These multiscale theories do not focus only on multiscale biological and epidemiological processes, but also on understanding various interactions in a complex interconnected socio-economical world. This requires going beyond classical diffusion theories, as propagation should also consider different transport networks that are somehow related to social activities.
In fact, the whole human system has a variety of specific features that need to be considered in the modeling and simulation of general epidemic spread: from the initial contagion to the in-host viral dynamics in the context of immune activation and anti-viral immune responses, up to the spatial spread of the viral infections by individual movement, which includes also different transportation means (e.g., public transport).

Our paper provides a review and critical analysis of the literature in the field. Through this review we develop a quest towards a new global and multiscale approach to modeling and simulation of the system under consideration. An answer to the following key questions guides such a quest.

\vskip.2cm \noindent \textbf{Key Question 1:} How do in-host dynamics evolve after infection?
\textit{The answer should consider the multiple scales involved in the dynamics, from the molecular (genetic) scale, which refers to sequential mutations of the virus or the molecular phenotype of various cells, to the macroscopic scale of heterogeneous populations, which refers to contagion inside the tissues. In particular, the in-host dynamics involve the microscopic scale of viral particles and host cells, where viral particles replicate, as well as immune cells and molecules involved in innate and adaptive anti-viral immune responses. A class of models that can naturally describe and link these different scales is represented by the kinetic models.}

\vskip.2cm \noindent \textbf{Key Question 2:}   How mathematics can contribute to understanding and modeling the social and economical impact of the epidemic?
\textit{The answer would include the close relationships between the epidemic spreading and the variation of sociality, clarifying at best how social aspects impact over the spreading and at the same time how epidemic spreading impacts over social activities. From the social side, it is evident that various behaviors are closely related to the pandemic spreading, in particular the reaction to lock-down and vaccination policies, which can be measured in terms of the personal opinion. In contrast to the modeling involved in the previous question, here the mathematical modeling does not take into consideration the biological mechanisms responsible for spreading, but the statistical consequences at the level of communities of the elementary rules of single agents, a picture which is typical of the kinetic modeling of physical systems. In particular, a reasonable kinetic modeling of the social consequences of a wrong behavior could help develop suitable controls to improve this behavior.}

\vskip.2cm \noindent \textbf{Key Question 3:} How do epidemics spread spatially across the territory?
\textit{The answer shall include considerations of the distribution of population density in the area and the mobility patterns of commuters for their daily activities. Clearly, information such as the degree of social interactions of individuals, policies, and any restrictions implemented at the governmental level to tackle the spread of the virus also play a crucial role in the modeling of spatial propagation of infectious diseases, intrinsically linking the discussion of this section with that of the previous question. The mathematical modeling, in this case, still starts from a kinetic approach and projects to a macroscopic one in urban areas with a high population density. Discussing the availability and the use of data is also essential to provide realistic predictive scenarios.}

\vskip.2cm

Indeed, we believe that a new vision of epidemics should be developed, based on a multiscale view of the overall dynamics and taking into account the interactions with the external environment in which the epidemics develop. Such an external environment is also a living (rather than inert) matter that is modified by the pandemics and can modify the pandemics. Therefore, we consider a multiscale approach to modeling epidemics through COVID-19, which is viewed as a system interacting with a living and social world.  All these topics will be discussed in the next sections.

\vskip.2cm \noindent Section 2 focuses on the concepts briefly outlined in Section 1 and provides the conceptual framework for the above-mentioned multiscale vision of the contribution that mathematics can make to a new theory of pandemics. We refer to the flowchart in Fig.~1, which is designed to describe a systems approach, where the global system is represented as several interacting subsystems (or blocks, as depicted in Fig.1). The dynamics in each block is described in the figure; these blocks are connected with directional or bi-directional arrows that suggest possible interactions between them, which can be further incorporated into the models. 
 The infection dynamics is shown on the middle line of the figure B1 $\to$ B2 $\to$ B4 $\to$ B5 or $\to$ B6. We start with a healthy population (B1), where the contamination is initiated by infected individuals who join the healthy population along short or long distance networks. Then, blocks B2, B3, B4 correspond to the sequential stages of the dynamics corresponding to the competition between the immune system and the proliferating virus. Blocks B5 and B6 represent the outcome of this competition.  Vaccination programs can mitigate the pathology and lead patients to full recovery.

\vskip.2cm \noindent
Section 3 first defines the concepts of contagion and viral load. It then reviews a multiscale approach to modeling viral contagion and subsequent in-host dynamics leading to the aforementioned conclusion, which can be either the full recovery of the infected patient or death due to viral overpowering of the immune system. The in-host dynamics include the immune competition between virus particles and the immune responses (innate and adaptive). The models consider past Darwinian mutations and selection leading to the emergence of variants, and the activation of the immune system by vaccines.  The approach considers the role of infected-recovered individuals and vaccination programs.  The mathematical tools relate to the further development of the kinetic theory of active particles\cite{[BBD21]} and, more generally, to methods inspired by the mathematics of statistical physics \cite{[ARI19]}.

\vskip.2cm \noindent
Section 4 deals with the mathematical modeling of social and economics dynamics related to the pandemics under consideration. The content deals with some recent attempts to build a new generation of mathematical models which take into account the joint action of a pandemic spreading and the social activities. For the sake of clarity, and to fully appreciate the novelties of the modeling approach, this merging will be restricted to the case of very simple and well-known models of compartmental epidemiology. Discussions and applications involving more complex compartmentalizations will be deferred to the references cited in the section.

\vskip.2cm \noindent
Section 5 focuses on modeling the spatial propagation of epidemics due to the movement of people, corresponding not only to collective dynamics in urban areas, but also to transportation devices, thus taking into account the heterogeneity of the territory and its influence on the spread of pandemics. The models will initially be presented using a basic compartmentalization structure over a geographical network and then extended to include the effects of asymptomatic individuals through numerical applications. Challenges related to data uncertainty and the delicate calibration of model parameters will be discussed. Furthermore, space will be dedicated to presenting new data-driven modeling approaches that leverage the promising performance of machine learning techniques.

%%%%%%%%%%%%%%%%%%%%%%%%%%%%%%%%%%%%%%%%%%%%
\section{On the interactions between pandemics and society}\label{Sec:2}
%%%%%%%%%%%%%%%%%%%%%%%%%%%%%%%%%%%%%%%%%%%%

A picture of a systems approach to modeling and simulation of the COVID-19 virus is shown in Fig.~1, to be considered as a development of the perspective ideas proposed in Ref.~\cite{[BBC20]}. The figure shows the multidisciplinary and multiscale vision that guides the content of this section, which is devoted to the conceptual framework within which the presentation of our work is organized. In particular, the systems approach indicates that the overall dynamics depends on the interactions of the various subsystems. These include not only biology and medicine, but also psychology, social dynamics, and economics. In fact, all subsystems are subject to the action of the epidemic, which affects their dynamics and, as a feedback, modifies the course of the epidemic in time and space. Therefore, a global vision is necessary, as already mentioned in Ref.~\cite{[BBC20]}.

\begin{figure}[h]
% Define block styles
\tikzstyle{decision} = [diamond, draw, fill=blue!20,
    text width=6.5em, text badly centered, node distance=4cm, inner sep=0pt]
\tikzstyle{block1} = [rectangle, draw, fill=blue!20,
    text width=6.5em, text centered, rounded corners, minimum height=4em]
    \tikzstyle{block3} = [rectangle, draw, fill=green!20,
    text width=6.5em, text centered, rounded corners, minimum height=4em]
    \tikzstyle{block2} = [rectangle, draw, fill=red!60,
    text width=6.5em, text centered, rounded corners, minimum height=4em]
     \tikzstyle{block4} = [rectangle, draw, fill=black!60,
    text width=6.5em, text centered, rounded corners, minimum height=4em]
    \tikzstyle{block6} = [rectangle, draw, fill=yellow!60,
    text width=6.5em, text centered, rounded corners, minimum height=4em]
     \tikzstyle{block8} = [rectangle, draw, fill=violet!60,
    text width=6.5em, text centered, rounded corners, minimum height=4em]
    \tikzstyle{block9} = [rectangle, draw, fill=white!60,
    text width=6.5em, text centered, rounded corners, minimum height=4em]
    \tikzstyle{block10} = [rectangle, draw, fill=white!60,
    text width=6.5em, text centered, rounded corners, minimum height=4em]
\tikzstyle{line} = [draw, -latex']
\tikzstyle{cloud} = [draw, ellipse,fill=red!20, node distance=4cm,
    minimum height=2em]
\begin{center}\scalebox{0.80}{
\begin{tikzpicture}[node distance = 3.0cm, auto]
    % Place nodes
     \node [block3] (Vax) {$BV$:\\ Vaccination \\ plans};
     \node [block1, below of=Vax] (Hea) {$B1$: Healthy \\ population};
    \node [block2, below of=Hea] (Inf) {$B2$: Infected, within host};
    \node [block2, below of=Inf] (Hos) {$B3$: Hospital \\ recovery};
    \node [block8, below of=Hos] (I-Care) {$B4$: Intensive\\ care};
    \node [block4, left of=Hos] (Dea) {$B5$: Dead};
     \node [block3, right of=Hos] (Rec) {$B6$: Recovered};
    \node [block6, right of=Hea] (Short) {$B7$: Short range networks};
    \node [block6, left of=Hea] (Long) {$B8$: Long range networks};
       \node [block9, right of=Vax] (Social) {$B9$: Social\\ dynamics};
    \node [block10, left of=Vax] (Economics) {$B10$:\\ Evolutionary\\ economics};
        % Draw edges
     \draw [line width=.5mm,  red, ->] (Hea) -- node  {} (Inf);
     \draw[line width=.5mm,  blue, ->] (Vax) -- node  {} (Hea);
     \draw [line width=.5mm,  red, ->] (Inf) -- node  {} (Hos);
     \draw [line width=.5mm,  red, ->] (Hos) -- node  {} (I-Care);
     \draw[line width=.5mm,  blue, ->] (Inf) -- node  {} (Rec);
     \draw [line width=.5mm,  red, ->] (Inf) -- node  {} (Dea);
     \draw[line width=.5mm,  blue, ->] (Hos) -- node  {} (Rec);
     \draw [line width=.5mm,  red, ->](Hos) -- node  {} (Dea);
     \draw[line width=.5mm,  blue, ->] (I-Care) -- node  {} (Rec);
     \draw [line width=.5mm,  red, ->] (I-Care) -- node  {} (Dea);
     \draw [line width=.5mm, ->] (Hea) -- node  {} (Short);
     \draw [line width=.5mm, ->] (Short) -- node  {} (Hea);
     \draw [line width=.5mm, ->] (Hea) -- node  {} (Long);
    \draw [line width=.5mm, ->] (Long) -- node  {} (Hea);
    \draw [line width=.5mm,  red, ->] (Long) -- node  {} (Inf);
    \draw [line width=.5mm,  red, ->] (Short) -- node  {} (Inf);
    \draw[line width=.5mm, ->] (Hea) -- node  {} (Social);
    \draw[line width=.5mm, ->] (Hea) -- node  {} (Economics);
    \draw[line width=.5mm, ->] (Economics) -- node  {} (Hea);
    \draw[line width=.5mm, ->] (Social) -- node  {} (Hea);
    \draw[line width=.5mm, ->](Social) -- node  {} (Vax);
    \draw[line width=.5mm, ->] (Economics) -- node  {} (Vax);
\end{tikzpicture}}
\end{center}
\caption{The systems approach to a global dynamics}
\end{figure}
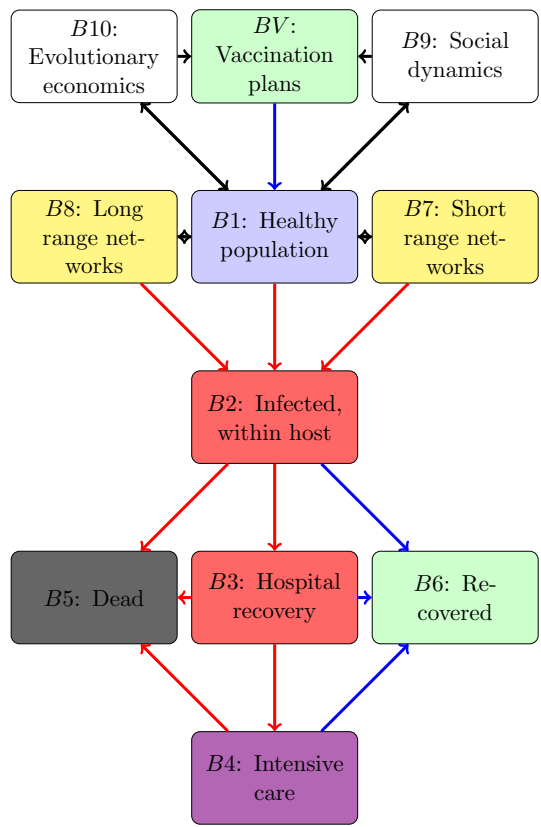

In detail, the central part of the figure, i.e. blocks B1,\ldots, B6, refers to the between-host dynamics considered in Sections 3, 4.  The B2 block includes implicitly a lower-scale dynamics described by the virus particles that can proliferate within the upper respiratory tract of the infected individuals, and the epithelial cells within this respiratory tract. This viral proliferation is countered by the immune system, first by innate immunity and then by the gradual activation of adaptive immunity. Further actions of other subsystems act on the above dynamics, as detailed below:
\begin{enumerate}
\vskip.2cm \item \textbf{Contagion}, which is caused not only by contacts between healthy and infected individuals within the zone where the epidemics are observed, see $B2 \to B1$, but also by interactions caused by entry and exit movements through short and long distance networks, see B7 and B8.

\vskip.2cm \item \textbf{Vaccination programs}, see BV, which mitigate the spread of epidemics and the emergence of variants of the virus, and also affect the intensity with which epidemics develop in the population.

\vskip.2cm \item \textbf{Social Dynamics and Evolutionary Economics}, see B9 and B10, which can have various types of actions on the population, first by acting on the vaccination programs, but also on the way of life of individuals, acting on their way of life, which includes the self-defense ability of the system, as well as the planning of medical actions.

\end{enumerate}

An important feature of the dynamics to be considered is the movement of fully recovered individuals back to healthy individuals due to loss of immunity to the virus. Therefore, they can be infected again, although their ability to contrast the virus is improved by the fact that the immune system keeps a memory through the past learning of the presence of the virus in the host.

The dynamics indicated in the blocks  B1 to B6 is subject to the actions indicated in blocks B7 and B8, which focus on contamination problems and the diffusion of epidemics. This topic is treated in section 4 as part of the overall content that includes the propagation of epidemics in the territory referring to communication networks that identify the main directions of short and long distance connections. This specific topic has been treated in the seminal articles~\cite{Bert3,[BP21],Bert,[BDP21]}. Another topic worth mentioning is contagion dynamics in closed and open environments, for example in crowd dynamics \cite{[ABKT23],[BGQR22],[KQ20]}. The modeling should also include the study of multiphase flows carrying very small droplets \cite{[Seminara]}.

Additional subsystems that play a role in the overall dynamics are represented by blocks B9 and B10 concerning social behavior and economics. These not only affect B1, but can also have an impact on medical interventions, such as vaccination programs. In fact, a vision of the overall medical and social state of the entire population is also taken into account. This issue is discussed in Section 5.

Figure 1 does not show the stone guest of the whole system, i.e. the political government of the nation and aggregation of nations, which acts on each block according to strategies related, at least in principle, to the data emerging from pandemics and society. These data also depend on the strategy of the government and on the reaction of the population \cite{BRBU,[BTZ],ZZ}.

The reasoning proposed in this section and the content of the following ones indicate that the contribution of mathematics should be based on a global systems approach and on the multidisciplinary vision as suggested in Figure 1. Applied mathematicians can contribute by providing tools that represent the different scenarios resulting from actions to mitigate pandemics, including the situation of no action. The selection of possible simulations should be proposed through a dialog with crisis managers, who should also take care to provide the available data needed to derive mathematical models and then validate them.

This complex dialog is not developed in a physical system where the rules of interaction are fixed in time in accordance with the material behavior of such a system. Rather, the interactions take place within a \textit{virtual world} where the rules of interaction are modified by human behavior and thinking, see~Ref. \cite{[Simon]} for the concept of virtual world. We trust that dialog, once made effective, can contribute to the well-being of human societies.

%%%%%%%%%%%%%%%%%%%%%%%%%%%%%%%%%%%%%%%%%%%%
\section{A multiscale-systems approach for in-host dynamics}\label{Sec:3}
%%%%%%%%%%%%%%%%%%%%%%%%%%%%%%%%%%%%%%%%%%%%

The outbreak of the SARS-CoV-2 pandemic posed various challenges to the scientific community, initially in the fields of biology and medicine, generally related to health care and hospitalization management. The interaction between scientists and crisis managers also involved mathematicians, mainly experts in the field of mathematical epidemiology, who were asked to work on models and simulations to provide scenarios of the dynamics of the pandemic under the action of different strategies, such as distancing and blocking actions, considered to mitigate the intensity of the pandemic.

The very first answer was given by technical developments of deterministic SIR (Susceptible-Infected-Recovered) and SEIR (Susceptible-Exposed-Infected-Recovered) models, see Ref.~\cite{[HET00]}. Some interesting results appeared in the literature, for example, see Refs.~\cite{[BFM21],[GBM20]}, somehow related to local situations at regional or national levels, see Refs.~\cite{[BP21],[DGMW22],[FFR21],[GDM21],[LMSW20],[ZBDD21]}. This approach has also developed the study of control problems \cite{[MWM21]}, transport in the territory \cite{[ADKV21],[BDP21],[DM22]}, co-morbidity with other pathologies. On the other hand, scientists quickly understood that the complexity of the system required going beyond stand-alone research, while the search for interdisciplinary research interactions should somehow be encouraged, consistent with the description provided in Section 2. The mathematical sciences have followed this activity, as can be seen in Ref.~\cite{[BBC20]} and in the contents of the special issues~\cite{[BBC21],[Toscani]}.

Therefore, a new frontier research activity has been promoted, linking mathematics with virology, immunology, medicine, see Refs.~\cite{[ARV20],[BDT17],[CFM20C],[CYR20],[TLS21]} for studies that might potentially contribute to mathematical developments. Mathematics has interacted also with related fields such as psychology and economics, see Refs.~\cite{[BD20],DPTZ,[DFV20],[FONT21],[HMF21]}. This frontier looks ahead to a new vision of mathematical epidemiology, where multiscale issues and population heterogeneity are deeply considered. This new trend will be considered in this section with reference to the in-host \& between-host dynamics shown in the central part of Fig.~1, specifically blocks B1, $\ldots$, B6.

First, we critically analyze the state of the art, and then we focus on research perspectives, followed by hints on how to address them. Such critical review  mainly refers to the conceptual framework proposed in Ref.~\cite{[BBC20]}, and developed in Refs.~\cite{[BBO22A],[BBO22B]}, devoted to the dynamics of variants and on the role of vaccination programs.  In addition, we also consider the recent contribution to the modeling of the in-host dynamics of the activation of the immune system to contrast the viral proliferation \cite{[BNER23]}.

The following subsections are devoted to the above topics. We begin with a brief report on the peculiarities of virus properties related to contagion problems and their ability to evade immune defense. An important issue to be considered in the modeling approach is the emergence of variants and their interaction with primary viruses.  In detail, the following topics are discussed in the following subsections:

\vskip.2cm \noindent (i) Description of the biological properties to be considered in the modeling approach from contagion to in-host dynamics. This topic considers a multiscale view of biological dynamics and heterogeneous human behavior.

\vskip.2cm \noindent (ii) Review of the literature on modeling the in-host dynamics of the primary virus, including the emergence of variants and their competition with the primary virus.

\vskip.2cm \noindent (iii) Modeling the dynamics of viral particle replication and the subsequent activation of the immune system.

\vskip.2cm \noindent (iv) Study of the in-host dynamics  specifically related to the level required  for hospital care. The study specifically considers  the in-host dynamics where the decay over time of activated immunity due to vaccination and/or a previous infection once completed with the recovery of the patient.

\vskip.2cm \noindent (v) Critical analysis for further development to include some additional studies on key features of the dynamics.

\vskip.2cm

Then, the review of the existing state of the art, followed by a critical analysis of what has been done and what should be done, leads to the definition of some research perspectives, which are presented in the last subsection. Hints on how to approach these research goals are given to the interested reader.

%%%%%%%%%%%%%%%%%%%%%%%%%%%%%%%%%%%%%%%%%%%%
\subsection{Multiscale view of in-host anti-viral dynamics and between-host epidemics}\label{subsec:3.1}
%%%%%%%%%%%%%%%%%%%%%%%%%%%%%%%%%%%%%%%%%%%%
Over the past two decades, the emerging field of immuno-epidemiology emphasized the importance of integrating knowledge from immunology, parasitology, genetics, epidemiology, ecology, as well as mathematical modeling and statistics \cite{[Hellriegel]}, to create new perspectives and new understanding on the transmission of infectious pathogens across animal/human heterogeneous populations. Therefore, mathematical models that investigate the spread of infectious diseases should consider how inter-individual differences in immune responses (at within-host level) impact epidemiological patterns of infection observed in heterogeneous populations (at between-host level) \cite{[GilchristSasaki],[Martcheva]}. Usually these aspects occur on different time and spatial scales, that need to be considered by the mathematical models.

\begin{quote}
{\dc Why is this multiscale feature important?}
\end{quote}

Indeed, mathematical models should consider that the infection generates a viral load that evolves in time, progressing or regressing through competition with the immune system. Then,  the contagion dynamics for each individual depends on the viral load of each individual in addition to protective devices and fluid mechanical properties, see Refs.~\cite{[BBC20],[SBKT21],[Seminara]}. Then, the diffusion dynamics should take into account the viral load heterogeneously distributed in the territory, as well as the heterogeneous characteristics of the territory, including short and long distance transportation networks.

Therefore, the study of epidemics requires a multiscale approach, where the macroscopic scale corresponds to individuals that may be infected or uninfected, while the microscopic scale corresponds to in-host entities within infected individuals, while  the link between the two scales is played by the dynamics of contagion, which depends on several parameters, including the number of viral particles infecting the individual (the viral load).

The specific features to be considered in the modeling approach are described in the following points, which are preliminary to the modeling. We do not naively claim that the description is exhaustive, as it is limited to the specific assumptions that will actually be included in the modeling approach. In this description we borrow some concepts presented in Ref.~\cite{[BK24]} and refer to the in-host dynamics shown in Fig.~2.
\begin{figure}
% Define block styles
\tikzstyle{decision} = [diamond, draw, fill=blue!20,
    text width=6.0em, text badly centered, node distance=3cm, inner sep=0pt]
\tikzstyle{block1} = [rectangle, draw, fill=blue!20,
    text width=6.0em, text centered, rounded corners, minimum height=4em]
    \tikzstyle{block3} = [rectangle, draw, fill=green!20,
    text width=6.0em, text centered, rounded corners, minimum height=4em]
    \tikzstyle{block2} = [rectangle, draw, fill=red!60,
    text width=6.0em, text centered, rounded corners, minimum height=4em]
     \tikzstyle{block4} = [rectangle, draw, fill=black!60,
    text width=6.5em, text centered, rounded corners, minimum height=4em]
    \tikzstyle{block6} = [rectangle, draw, fill=yellow!60,
    text width=6.5em, text centered, rounded corners, minimum height=4em]
     \tikzstyle{block8} = [rectangle, draw, fill=violet!60,
    text width=6.5em, text centered, rounded corners, minimum height=4em]
    \tikzstyle{block9} = [rectangle, draw, fill=white!60,
    text width=6.5em, text centered, rounded corners, minimum height=4em]
    \tikzstyle{block10} = [rectangle, draw, fill=white!60,
    text width=6.5em, text centered, rounded corners, minimum height=4em]
\tikzstyle{line} = [draw, -latex']
\tikzstyle{cloud} = [draw, ellipse,fill=red!20, node distance=3cm,
    minimum height=2em]
\begin{center}\scalebox{0.80}{
\begin{tikzpicture}[node distance = 3.5cm, auto]
    % Place nodes
    \node [block2] (Inf) {$B2$: Infected, within host};
    \node [block1, left of=Inf] (Hea) {$B1$: Healthy \\ population};
     %   \node [block3, left of=Hea] (Vax) {$BV$:\\ Vaccination \\ plans};
    \node [block2, right of=Inf] (Hos) {$B3$: Hospital \\ recovery};
    \node [block3, above of=Hos] (Rec) {$B6$: Recovered};
    \node [block4, below of=Hos] (Dea) {$B5$: Dead};
    \node [block8, right of=Hos] (I-Care) {$B4$: Intensive\\ care};
        % Draw edges
     \draw [line width=.5mm,  red, ->]  (Hea) -- node  {} (Inf);
   % \path [line] (Vax) -- node  {} (Hea);
     \draw [line width=.5mm,  red, ->]  (Inf) -- node  {} (Hos);
     \draw [line width=.5mm,  red, ->]  (Hos) -- node  {} (I-Care);
    \draw [line width=.5mm, ->]  (Inf) -- node  {} (Rec);
     \draw [line width=.5mm,  red, ->]  (Inf) -- node  {} (Dea);
    \draw [line width=.5mm, ->]  (Hos) -- node  {} (Rec);
     \draw [line width=.5mm,  red, ->]  (Hos) -- node  {} (Dea);
    \draw [line width=.5mm, ->]  (I-Care) -- node  {} (Rec);
     \draw [line width=.5mm,  red, ->]  (I-Care) -- node  {} (Dea);
     \draw [line width=.5mm, ->]  (Rec) -- node  {} (Hea);
   \end{tikzpicture}}
\end{center}
\caption{Flow chart of the viral infection and its impact on the between-host (epidemiological-level) dynamics.}
\end{figure}
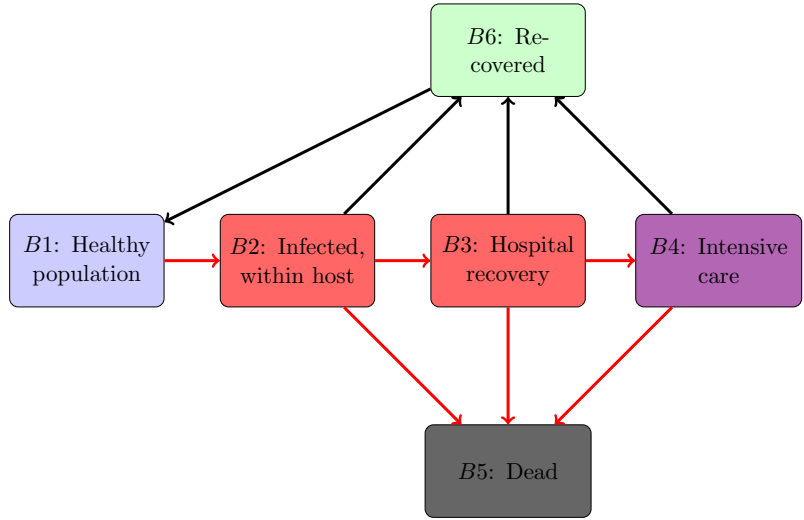

\begin{itemize}

\vskip.2cm  \item \textbf{General framework:}  The in-host dynamics should be modeled at the micro scale, within the upper respiratory tract, and should focus on the competition between the virus and the immune system. The infection may progress (or regress)  and may end with a \textit{full recovery} or with the \textit{death} of the patient.

\vskip.2cm \item \textbf{Viral load:}  The number of viral particles detected in  the upper respiratory tract defines the \textit{viral load}. The development of the pathology corresponds to the dynamics of the \textit{viral load} starting from the  so-called \textit{initial viral load}.

\vskip.2cm \item \textbf{Social distance:} This term covers various factors, mainly the physical distance between individuals and technical defenses such as masks. In general, such distance is related to the awareness of contagion.

\vskip.2cm \item \textbf{Contagion:} The probability of contagion depends on several factors, such as the interaction between healthy and infected individuals;  the \textit{social distance} between them; the characteristics of the area where the infection occurs (temperature, humidity...); and, most importantly, the number of copies of the virus  (viral load) that can reach  a susceptible person.

\vskip.2cm \item \textbf{Proliferation:}  The virus interacts with the human epithelial cells through spike proteins. Then, it introduces a fragment of DNA into them,  where thousands of new viral copies are produced. These copies then invade additional epithelial cells.

\vskip.2cm \item \textbf{Affinity:} The speed with which the virus detects cells that can be infected is a viral property called \textit{affinity}. The viral affinity is specific to each variant of the virus.

\vskip.2cm \item \textbf{Dynamics of variants:} During the rapid production of new viruses  many mistakes are made in the copying (transcription) of the viral RNA. Most of them (mutations) result in the production of less effective virus particles. In a few cases, new variants with more invasive properties may emerge.  Darwinian competition among virus particles may favor their spread.

\end{itemize}

In addition to the above descriptions, the approach should also consider some specific features of the immune defense against infections. The immune system is an emblematic example of how complex  specific subsystems of living organisms can be. The anti-viral immune response is formed of an early non-specific phase that involves the innate immunity, and a later antigen-specific phase that involves adaptive immunity.
\begin{itemize}
\item \textbf{Innate immunity:} This is the first line of defense against external pathogens (including viruses), and is formed of physical and chemical barriers, cells, cytokines and antiviral proteins that inhibit or eliminate infection with no or little pathogen specificity and with no or little generation of a long-lasting protective memory \cite{[Weber2021_AntiviralInnateImmunity]}. The innate immunity comes into action in the first 5-7 days following the infection. Once the pathogen is recognized (mainly by dendritic cells), innate immune cells such as macrophages, neutrophils and NK cells try to eliminate the pathogen or to keep the infection under control until the activation of the adaptive immunity. A typical feature of anti-viral innate immunity is the production of (type-I and type-III) interferons by various cells in the organism, which can have both direct and indirect antiviral effects \cite{[Weber2021_AntiviralInnateImmunity]}. These key cytokines are also involved in the development of adaptive immunity.

\vskip.2cm \item \textbf{Adaptive immunity:} The early-stage non-specific innate immunity activate the specific adaptive immune responses, which then take between 6-10 days to develop. These adaptive responses are carried out through \cite{[Alberts2002]}: (i) B cells that are activated and secrete antibodies (immunoglobulins), which bind to and inactivate the viral antigen that stimulated their secretion; (ii) T cells that kill virus-infected cells (as these cells react against foreign antigens that are presented to them on the surface of host cells). Adaptive immunity is characterized by the formation of B cell and T cell immune memory, which ensures a faster activation of the adaptive immune response the next time the same pathogen is encountered. Mutations in the pathogen's proteins can lead to the appearance of new variants. In the context of SARS-CoV-2 it seems that memory adaptive responses can still recognize and act against some viral variants.
\end{itemize}

During the SARS-CoV-2 pandemics, it has become clear that the severe and fatal infection cases were characterized by a dysfunction of the anti-viral immune responses (i.e., impaired or delayed innate and adaptive immune responses). Therefore, to understand the heterogeneity of viral infections at the level of the whole population, we need to understand first anti-viral immune responses at the level of each individual in the population.

In this study we do not naively claim that we can provide an exhaustive description of the interactions between the virus and the anti-viral immune responses, but we simply focus on those aspects of the innate and adaptive immune dynamics that are useful to derive a mathematical description  of the key outputs of the virus-immune competition.

A detailed description of these specific aspects is given in the following subsections. Here, we only point out that the parameters and variables of the system are heterogeneously distributed in the cell populations and also take into account  the past pathological history of the individual.

%%%%%%%%%%%%%%%%%%%%%%%%%%%%%%%%%%%%%%%%%%%%
\subsection{On the multiscale modeling of the in-host/between-host dynamics and pathology}\label{subsec:3.2}
%%%%%%%%%%%%%%%%%%%%%%%%%%%%%%%%%%%%%%%%%%%%

The current state of the art in multiscale approaches covers, at least in part, the various characteristics of the complex dynamics under consideration. We refer to the first approach proposed in Ref.~\cite{[BBC20]} and the subsequent developments in Ref.~\cite{[BBO22A],[BBO22B]}
to provide a brief overview of the results obtained, and then we show further developments of these results.

We consider, mainly with reference to~\cite{[BBO22A]}, the dynamics of interactions between the primary virus and a variant. We then discuss how the model can be simplified or further developed to include the role of vaccination programs. The dynamics we consider are visualized in Figure 3, which shows the dynamics of the virus towards increasing its proliferative level and that of the immune system contrasting this action and returning cell populations to their homeostatic (healthy) states. Red arrows  indicate the action of the virus, while black arrows indicate that of the immune system.

%%%%%%%%%%% Figure 3
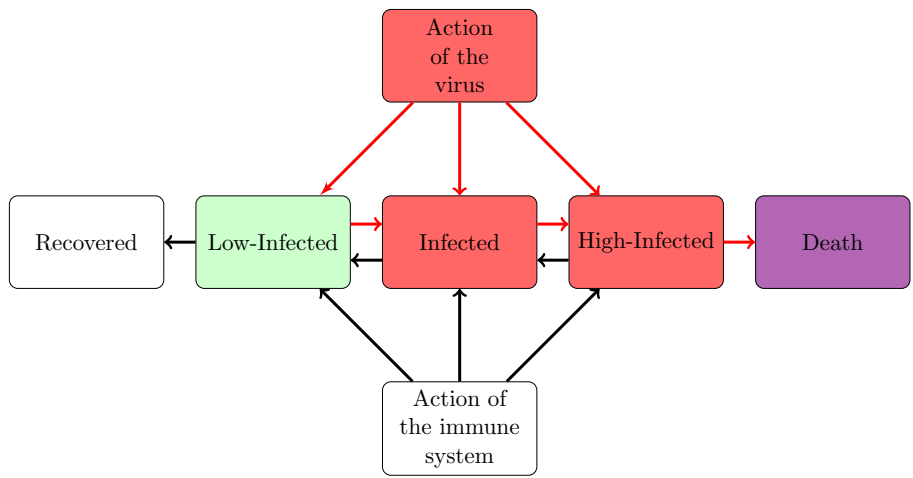
\begin{figure}[!h]
% Define block styles
\tikzstyle{decision} = [diamond, draw, fill=blue!20,
    text width=4.5em, text badly centered, node distance=3cm, inner sep=0pt]
\tikzstyle{block1} = [rectangle, draw, fill=blue!20,
    text width=6em, text centered, rounded corners, minimum height=4em]
    \tikzstyle{block3} = [rectangle, draw, fill=green!20,
    text width=6em, text centered, rounded corners, minimum height=4em]
    \tikzstyle{block2} = [rectangle, draw, fill=red!60,
    text width=6em, text centered, rounded corners, minimum height=4em]
     \tikzstyle{block4} = [rectangle, draw, fill=violet!60,
    text width=6em, text centered, rounded corners, minimum height=4em]
    \tikzstyle{block6} = [rectangle, draw, fill=white!60,
  % text width=7em, text centered, rounded corners, minimum height=4em]
   %\tikzstyle{block7} = [rectangle, draw, fill=grey!60,
    text width=6em, text centered, rounded corners, minimum height=4em]
   % text width=7em, text centered, rounded corners, minimum height=4em]
    % \tikzstyle{block8} = [rectangle, draw, fill=violet!60,
\tikzstyle{line} = [draw, -latex']
\tikzstyle{cloud} = [draw, ellipse,fill=red!20, node distance=3cm,
    minimum height=2em]
    \begin{center}\scalebox{0.80}{
\begin{tikzpicture}[node distance = 3.1cm, auto]
    % Place nodes
    \node [block2] (Inf) {Infected};
    % \node [block8, above of=Inf] (Virus) {Virus};
     \node [block2, above of=Inf] (Virus) {Action \\of the\\ virus};
    \node [block3, left of=Inf] (Inf-Low) {Low-Infected };
    \node [block6, left of=Inf-Low] (Rec) {Recovered  };
    \node [block2, right of=Inf] (Inf-High) {High-Infected};
     \node [block4, right of=Inf-High] (Death) {Death};
   \node [block6, below of=Inf] (Immune) {Action of the immune system};
   % \node [block6, below of=Inf-High] (Imm*j+1) {$f^{j+1}_{3,4,5}$:  Action immune $\gamma$};
   % \path [line] (Inf*j+1) -- node  {} (Inf*j);
     \draw [line width=.5mm, ->] ([yshift=-.3cm] Inf.west) -- ([yshift=-.3cm] Inf-Low.east);
     \draw [line width=.5mm,  red, ->] {}([yshift=.3cm] Inf.east) -- ([yshift=.3cm] Inf-High.west);
     \draw [line width=.5mm, ->] (Inf-Low) -- node  {} (Rec);
     \draw [line width=.5mm,  red, ->]  ([yshift=.3cm] Inf-Low.east) -- ([yshift=.3cm] Inf.west);
     \draw [line width=.5mm, ->] ([yshift=-.3cm] Inf-High.west) -- ([yshift=-.3cm] Inf.east);
    \draw [line width=.5mm,  red, ->] (Inf-High) -- node {} (Death);
    \draw [line width=.5mm, ->]  (Immune) -- node  {} (Inf);
    \draw [line width=.5mm, ->]  (Immune) -- node  {} (Inf-High);
    \draw [line width=.5mm, ->]  (Immune) -- node  {} (Inf-Low);
     \draw [line width=.5mm,  red, ->] (Virus) -- node  {} (Inf);
     \draw [line width=.5mm,  red, ->] (Virus) -- node  {} (Inf-High);
     \draw [line width=.5mm,  red, ->] [line] (Virus) -- node  {} (Inf-Low);
          \end{tikzpicture}}
       \end{center}
\caption{Immune defense (black arrows to left) contrasts  virus action (red arrows to right). The block on the middle line refers to individuals, while the upper and lower block to the cell-virus scale.}
\end{figure}

The modeling approach in the papers cited above, consistent with the description in Section 3.1, considers the following:

\vskip.1cm \noindent \textbf{1.}  The approach is multiscale, where the macroscale corresponds to individuals that may be infected or uninfected, while the microscale corresponds to in-host entities within infected individuals.

\vskip.1cm \noindent \textbf{2.} The probability of contagion depends on the level of infection, i.e. the \textit{viral load}, as well as the \textit{social distance} between individuals.

\vskip.1cm \noindent \textbf{3.} Competition between the \textit{proliferating virus} and the \textit{immune system} describes how the level of infection can progress (or regress) due to the prevalence (or lack of prevalence) of the virus over the immune defense.

\vskip.1cm \noindent \textbf{4.}  Mutations and selection leading to the emergence of new variants of the virus are modeled by a post-Darwinian dynamics. The probability of mutations increases with increasing viral charge and frequency of interactions.

\vskip.1cm \noindent \textbf{5.} Vaccination programs and past infection increase the ability to defend, but this property is not a permanent state as it decays over time as already observed in~\cite{[BBO22B]}.
\vskip.1cm
The multiscale epidemiological-level model presented below is a technical variant of the one in Ref. \cite{[BBO22A]}. Specifically, five FSs denoted by the subscripts $i=1, \ldots, 5$, which are carriers of a pathological state, include an additional microstate corresponding to the level of pathology denoted by the superscript $j$. In accordance with Fig.~3, we have chosen five steps, i.e. $j=1,\ldots, 5$.

The approach is developed by using the following parameters:

\vskip.1cm \noindent  $\alpha = \alpha (t) \in [0,1]$ defines the level of social distance. It includes the  ``locking'' action $0 < \alpha_\ell < 1$ and the de-locking action   $\alpha_d$ with $\alpha_\ell < \alpha_d < 1$. The limit values $\alpha = 0$ and $\alpha = 1$ correspond to full isolation and absence to social distance, respectively.

\vskip.1cm \noindent  $\gamma$ is the  activated within host immunity.

\vskip.1cm \noindent  $\kappa_j$, with $j = 1, \ldots, 5$,  defines the levels of pathology which  technically refers to the level of proliferative activity of the virus.

\vskip.1cm \noindent  $\lambda > 0$ models the increase of proliferative activity of a variant of the the primary virus: $\kappa_j(\lambda) = \kappa_j(1 + \lambda)$.
\vskip.1cm
The representation of the system is given by the following probabilities, which are given by the ratio between each $f_i^j = f_i^j(t)$ divided by the maximum number of allowed cells in each functional subsystem.

\vskip.1cm \noindent $i=1, \hskip.2cm j=1$: \textit{Healthy} individuals with state $f_1^1(t; \kappa_1)$.

\vskip.1cm \noindent $i=2,  \hskip.2cm j=2,\ldots, 5$:  \textit{Level of infection  by the primary virus} $f_2^j(t; \kappa_j,  \gamma)$.

\vskip.1cm \noindent $i=3,  \hskip.2cm j=2,\ldots, 5$:  \textit{Level of infection  by a  variant} $f_3^j (t, \kappa_j (1 + \lambda),  \gamma)$.

\vskip.1cm \noindent $i=4$: \textit{Recovered individuals}  $f_4=f_4(t)$, who succeed  to go back, from  past-infected state $f^2$, to the state $j=1$.

\vskip.1cm \noindent $i=5$: \textit{Dead individuals} $f_5=f_5(t)$ who reach the state  $j=5$ moving from past-infected state $f^4$.

The mathematical model translates the above assumptions, visualized in the flowchart of Figure 3, into a differential system. A detailed description of these assumptions follows:

\begin{itemize}

\vskip.2cm \item The virus multiplies by feeding on epithelial cells and increases the level of pathology from $j$- to $(j+1)$-level depending on $\kappa_j$ and, on the level of $\lambda$, on the new variant. The immune system reduces the virus from $j$- to $(j-1)$-level depending on the parameter $\gamma$.

\vskip.2cm \item Infection by variants refers to $f_1$ and by infected individuals to $f_4$ starting from a small initial state of infected by variant $\ve_v$. The increase of the proliferative ability of the variant with respect to the primary virus is modeled by the factor $(1 + \lambda)$.

\vskip.2cm \item The tendency to recover considers the inflow, from healthy individuals, of individuals from $2$-FS and $3$-FS, with state $j=2$, into $1$-FS, corresponding to $j=1$, due to the action of the immune system.

\vskip.2cm \item The dynamics related to the need for hospitalization are modeled by the variables $f_{2,3}^j$ with $2 < j < m -1$.

\vskip.2cm \item The dynamics of the trend towards death is caused by the influx of $2$-FS and $3$-FS, in $j=m-1$, into $5$-FS, corresponding to $j=m$, due to the effect of viral proliferation.

\end{itemize}

A scaling useful towards qualitative simulations is as follows:
\begin{equation}
\kappa_j = j \dot \kappa, \hskip2cm  \beta = \frac{\gamma}{\kappa}, \hskip2cm t = \kappa \cdot t_r,
\end{equation}
where $t_r$ is the real time.

The, technical calculations lead to the following  mathematical model:
\be\label{variants}
\begin{split}
\frac{\partial f_1(t)}{\partial t}&=-\alpha(t) \displaystyle{\sum_{j=2}^{j=m}}\, j\, f_1(t)\left(f_3^j(t)+(1+\lambda)\,f_4^j\right),\\
\frac{\partial f_2^j(t)}{\partial t} &= \alpha(t) \, \displaystyle{\sum_{s=2}^{j=m}} \, s \, f_1(t) \, f_2^s(t) + \left((j-1) \,  f_2^{j-1}(t) - j\, f_2^j(t)\right)\\
&+  \beta  \, \left(f_2^{j+1}(t)  -  f_2^{j}(t)\right), \\
\frac{\partial f_3^j(t)}{\partial t} &=   \alpha  (t) \, \displaystyle{\sum_{s=2}^{s=m}} \, s \, (1 + \lambda)\, f_1(t) \, f_3^s(t)  \\
&+  (1 + \lambda)\left((j-1) \,  f_3^{j-1}(t) - j\,  f_3^j(t)\right) + \beta  \, \left(f_3^{j+1}(t)  -  f_3^{j}(t)\right),\\
\frac{\partial f_4(t)}{\partial t} &= \beta \left(f_2^2(t) + f_3^2(t)\right), \\
\frac{\partial f_5(t)}{\partial t} &= 4 \, f_2^{m-1}(t) + 4(1 + \lambda)\,f_4^{m-1}(t).
\end{split}
\ee

Some limits and extensions of this model are defined in the following remarks.

\begin{remark}
The specialization of the level of pathology depends on the choice of $m$.  For example, for $m=5$, $j=2, 3, 4$ can correspond to home care, hospital care, and advanced care, respectively. The case $m=6$ may also include asymptomatic individuals.
\end{remark}

\begin{remark}
Heterogeneous behavior of healthy individuals, such as that associated with different immune defenses, can be modeled by dividing $f_1$ into groups with different defenses, such as those associated with age. A temporary high level of defense can be acquired either by vaccination programs or by past infection. 
\end{remark}

\begin{remark}
Vaccination programs can be modeled by adding the population of vaccinated individuals who acquire n improved defense ability analogous to that recovered after infection.
\end{remark}

The model includes a small number of parameters related both to the characteristics of the biological system and to the actions of the crisis managers. In fact, the goal of the approach is mainly to describe the scenarios that can occur for variable values of the parameters. Indeed, the mathematical tools developed in Refs.~\cite{[BBO22A],[BBO22B]} lead to a useful description of the scenarios that can appear during the pandemics. Some of these scenarios provide useful clues to the design of policies that might mitigate the damage of pandemics. For example, the model shows that vaccination programs should be accompanied by social distancing measures. In particular:
\begin{enumerate}
\item Increasing values of $\alpha_d$ lead to a second wave with high values of the density of infected individuals, thus reducing the benefit of the vaccination program.

\vskip.2cm \item The variant already becomes prevalent during the first wave, while during the second wave it completely dominates the primary virus. This behavior is enhanced by the $\lambda$ aggressiveness of the variant.  It is also reinforced by $\alpha_d$.

\vskip.2cm \item Vaccination programs reduce the number of infected individuals. The effect occurs during the lock-down period and increases after the lock=down. Increasing the activation of the vaccine will decrease the number of infected individuals.
\end{enumerate}

The simulations available in Refs.~\cite{[BBO22A],[BBO22B]} provide useful qualitative information, but do not yet cover all case studies. Therefore, the search for an exhaustive description of the dynamics that can occur in pandemics would be a useful perspective for simulations. The modeling approach reviewed in these papers provides a useful description of the different dynamics that can occur during pandemics. On the other hand, the description of the immune defense is limited to a parameter that does not describe the gradual activation of adaptive dynamics, which is an important feature of the immune response \cite{[COOP10],[MF18]}; see also Ref.~\cite{[VBEA20]}. This specific topic will be discussed in the next subsection.

%%%%%%%%%%%%%%%%%%%%%%%%%%%%%%%%%%%%%%%%%%%%
\subsection{In-host dynamics with virus proliferation and immune system activation}\label{subsec:3.3}
%%%%%%%%%%%%%%%%%%%%%%%%%%%%%%%%%%%%%%%%%%%%

To address the limitations of the model \eqref{variants}, in the following we present a multiscale mathematical model for anti-viral immune responses at in-host level. To this end, we follow the approach in Ref.~\cite{[BNER23]} and consider the following functional subsystems (FS) that could depend of a microscale variable $u\in [0,1]$: (1) a population of virus particles described by variable $g_{1}(t,u)$, (2) a population of infected epithelial cells described by $g_{2}(t,u)$, (3) a generic anti-viral immune cell population $g_{3}(t,u)$. For simplicity, we assume that the virus particles and infected cells are homogeneous populations, not characterized by a particular activity status (and thus we use $g_{1,2}(t)= \int_{0}^{1}g_{1,2}(t,u)du$), while the immune cells depend also on a microscale variable $u\in[0,1]$ that describes their activation status. The time-evolution of these variables as a result of proliferative/destructive interactions and conservative interactions (for $g_{3}$ only) that could lead to changes in the activity status, is given by the following equations:
\be\label{in-host-model}
\begin{split}
\displaystyle \frac{\partial g_1(t)}{\partial t}&= \bar{\gamma} \bar{N_{c}} g_{2}(t)- \bar{\mu} g_{1}(t) \int_{0}^{1}K(u) g_{3}(t,u)du \\
\displaystyle  \frac{\partial g_2(t)}{\partial t} &= \bar{\alpha} g_{1}(t) \left(\bar{N_{H}}-g_{2}(t) \right) - \bar{\gamma} g_{2}(t)-\bar{\mu} g_{2}(t) \int_{0}^{1}K(u) g_{3}(t,u)du, \\
 \displaystyle  \frac{\partial g_3(t,u)}{\partial t} &=  \bar{\kappa}(u) g_{3}(t,u)g_{1}(t)-\bar{\lambda} g_{3}(t,u) + \bar{\beta } g_{1}(t)\int_{0}^{1}\mathcal{A}(u_{*} \rightarrow u) g_{3}(t,u_{*})du_{*}.
   \end{split}
\ee

These equations incorporate the following assumptions:
\begin{itemize}
\item Viral particles ($g_{1}$) are produced (at a rate $\bar{\gamma }$) by infected epithelial cells ($g_{2}$) that burst out and release an average of $N_{c}$ viral particles per cell. These viral particles are eliminated at a rate $\bar{\mu}$ by immune cells. The kernel $K(u)$ gives the anti-viral effect of immune cells, which depends on their activity status. In Ref.~\cite{[BNER23]} it was assumed that $K(u)=(1+0.1 \:u)$, with $\nu>0$ (to describe the action of both innate immunity for $u=0$ and adaptive immunity for $u>0$).

\vskip.2cm \item Infected immune cell population ($g_{2}$) in the upper respiratory tract is growing at a rate $\bar{\alpha}$ until it reaches the critical size $\bar{N_{H}}$. Beyond this size the virus .... The cells die at a rate $\bar{\gamma}$, and are eliminated by the immune cells at a rate $\bar{\mu}$.

\vskip.2cm \item Immune cells ($g_{3}$) are proliferating at a rate $\bar{\kappa}$ which could depend on their activation state $u$. In Ref.~\cite{[BNER23]} this rate was chosen  to be $\bar{\kappa}(u)=10^{-9}(1+u)ml/(day \times RNAs)$. The immune cells die at a rate $\bar{\lambda}$ and become activated in the presence of the virus at a rate $\bar{\beta}$.
\end{itemize}

\begin{remark}
In the above in-host equations~\eqref{in-host-model} the parameters have an upper bar, to distinguish them more easily from the parameters appearing in the between-host model \eqref{variants}.
\end{remark}

\begin{figure}[!t]
\centering
\includegraphics[width=5.2in]{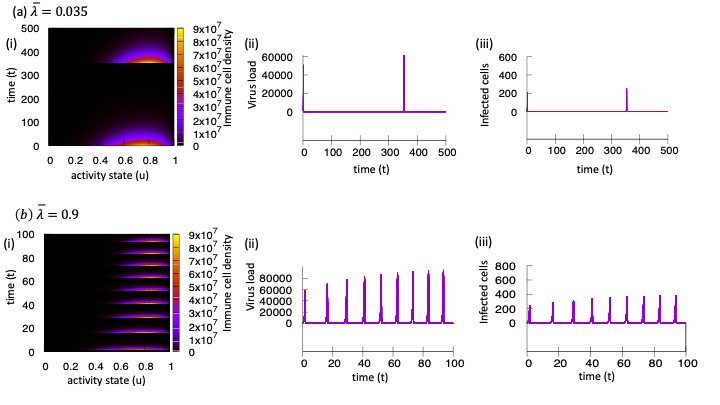}
\caption{Re-infection dynamics of model \eqref{in-host-model}, as we increase the death rate $\bar{\lambda}$ of immune cells (and so we reduce their persistence). Sub-panels (i) show the immune response (with different levels of activation $u$); sub-panels (ii) show the virus load; finally sub-panels (iii) show the infected cells. Parameters values are: (a) $\bar{\lambda}=0.035$; (b) $\bar{\lambda}=0.085$; (c) $\bar{\lambda}=0.9$. The rest of parameters are\protect\cite{[BNER23]}:
$\bar{N_{H}}=10^{5}$, $\bar{N_{c}}=10^3$, $\bar{\alpha}=1.44\times 10^{-6}$, $\bar{\gamma}=0.69315$, $\bar{\mu}=10^{-6}$, $\bar{\lambda}=0.01$, $\bar{\beta}=0.01$, $\nu=0.1$. }
\label{Fig_InHost1}
\end{figure}

\begin{figure}[!t]
\centering
\includegraphics[width=5.4in]{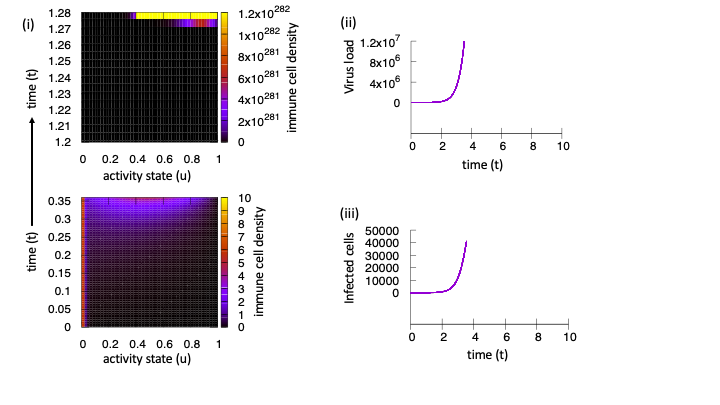}
\caption{Blow-up dynamics of model \eqref{in-host-model}, when the immune cells do not detect and kill the virus particles or the infected immune cells (i.e., $\bar{\mu}=0$). Sub-panels (i) show the immune response for time $t \in [0,0.35]$ (top sub-panel) and $t\in [1.2, 1.28]$ (bottom sub-panel). Sub-panel (ii) shows the virus load. Finally sub-panel (iii) shows the infected cells. Parameters values are\protect\cite{[BNER23]}:
$\bar{\mu}=0,$ $\bar{N_{H}}=10^{5}$, $\bar{N_{c}}=10^3$, $\bar{\alpha}=1.44\times 10^{-6}$, $\bar{\gamma}=0.69315$, $\bar{\mu}=10^{-6}$, $\bar{\lambda}=0.01$, $\bar{\beta}=0.01$, $\nu=0.1$. }
\label{Fig_InHost2}
\end{figure}

The multiscale model \eqref{in-host-model} was shown to exhibit different behaviors: from temporary viral elimination and re-infection (see Fig.~\ref{Fig_InHost1}), to blow-up in the immune response (see Fig.~\ref{Fig_InHost2}). \\
%Below we discuss in more detail these two behaviors:
In Figure~\ref{Fig_InHost1} we see that the time of viral re-infection (as well as the density of virus particles and density of infected cells) depends on the persistence of immune response:  lower  immune cell death rates (e.g., $\bar{\lambda}=0.035$ in panels (a)) are associated with longer anti-viral immune protection, while higher immune cell death rates (e.g., $\bar{\lambda}=0.9$) are associated with very fast re-infections. In addition, it seems that during re-infections the immune responses are characterized by higher activation states, i.e., at each re-infection the immune cell population is more activated against the viral pathogen. Obviously, this behavior does not take into consideration the mutation in SARS-CoV-2 (not incorporated into the model), which leads the virus to escape immunity \cite{[Harvey21]}. We note that, for model simplicity, the immune cells considered in \eqref{in-host-model} do not differentiate between effector, effector-memory and memory cells, and thus the death of immune cells means also the loss in anti-viral immune protection.\\
%%%%%% Figure
%%%%%%
In Figure~\ref{Fig_InHost2} we see that when the immune cells cannot eliminate the virus particles (i.e., $\bar{\mu}=0$), the over-activation of these cells can lead to immune blow-up. Note here that the blow-up occurs first in the immune response, and not in the number of virus particles or the infected cells (which are at high levels, but much lower than the level of immune response).
This behavior corresponds to the cytokine storm that has been observed in some COVID-19 infections (when there is an acute overproduction and uncontrolled release of pro-inflammatory cytokines) \cite{[Montazersaheb]}.
%%%%%% Figure
%%%%%%

%%%%%%%%%%%%%%%%%%%%%%%%%%%%%%%%%%%%%%%%%%%%
\subsection{Additional study of the in-host dynamics and immune competition}\label{subsec:3.4}
%%%%%%%%%%%%%%%%%%%%%%%%%%%%%%%%%%%%%%%%%%%%

This subsection reports on the main concepts proposed in Ref.~\cite{[BK24]} to derive models of the in-host immune competition that take into account the decay over time of activated immunity due to vaccination and/or a previous infection once completed with the recovery of the patient.

This model is specifically related to the level required for hospital care. Consider a system of four functional subsystems, $FS_i$, with subscripts $i=1,\ldots, 5$, where $i=1$ and $i=5$ correspond to healthy and dead individuals, respectively. Further, $FS_2$ and $FS_3$ denote individuals infected by the primary virus and variants, while $FS_4$ and $FS_5$ correspond to immune activation with respect to $FS_2$ and $FS_3$, respectively. Each FS contains $m$ internal states, denoted by $j=1,\ldots, m$, to intermediate states of each functional subsystem.

Discrete states are used, so that the state of each functional subsystem is given by $f_{ij}$, interpreted as the dependent variable of the system. The derivation is obtained by numerically balancing the number of particles in the elementary volume of the state space, i.e. the number growth is equal to the inflow minus the outflow in the above volume. Once the differential structure is obtained, simulations provide a quantitative description of the dynamics provided by the model. The mathematical structure corresponding to the above assumptions is as follows:
\begin{equation}\label{BK}
\partial_t f_{ij} =  \sum_{p,h = 1}^{5} \sum_{q,k = 1}^m  \eta_{pq}^{hk}[\f]\, \cA_{pq}^{hk}[\f](p \to i, q \to j) f_{pq} f_{hk}
 - f_{ij}\sum_{h= 1}^5 \sum_{k = 1}^m \eta_{ij}^{hk}[\f] f_{hk},
\end{equation}
where $\eta_{pq}^{hk}$ models the interaction rate for interactions of $f_{pq}$ with  $f_{hk}$ and $\cA_{pq}^{hk}$, models the transition probability from the state $pq$ to $ij$.

This framework corresponds to a system of $5 \times 4$ ordinary differential equations, which are considered to describe the dynamics of the dependent variables $f_{ij}$. The dynamics is obtained by solving the associated initial value problem, where we consider a population in which a very small fraction of individuals correspond to be infected by a primary virus and by a variant in $FS_2$ and $FS_3$, respectively.

Details of the assumptions that generate the model are given in Ref.~\cite{[BK24]}. In terms of the blocks in Figure 6, the dynamics are as follows:\vspace{-0.2cm}
\begin{itemize}
\item \textit{Contagion};
\item \textit{Vaccination programs} that increase the level of immune defense as the action and decay of the immune level by transition of immune cells to the high level of defense;
\item \textit{Natural decay of the immune defense} when it is not activated by the virus;
\item \textit{Activation of the virus and the immune system} with transition to a more aggressive state, while the immune cells activate from learning the presence of the virus.
\end{itemize}
%%%%%%%%%%% Figure 6
\begin{figure}[!h]
% Define block styles
\tikzstyle{decision} = [diamond, draw, fill=blue!20,
    text width=4.5em, text badly centered, node distance=3cm, inner sep=0pt]
\tikzstyle{block1} = [rectangle, draw, fill=blue!20,
    text width=6em, text centered, rounded corners, minimum height=4em]
    \tikzstyle{block3} = [rectangle, draw, fill=red!20,
    text width=6em, text centered, rounded corners, minimum height=4em]
    \tikzstyle{block2} = [rectangle, draw, fill=red!60,
    text width=6em, text centered, rounded corners, minimum height=4em]
     \tikzstyle{block4} = [rectangle, draw, fill=violet!60,
    text width=5em, text centered, rounded corners, minimum height=4em]
    \tikzstyle{block6} = [rectangle, draw, fill=white!60,
   text width=6em, text centered, rounded corners, minimum height=4em]
   \tikzstyle{block7} = [rectangle, draw, fill=green!60,
    text width=6em, text centered, rounded corners, minimum height=4em]
   \tikzstyle{line} = [draw, -latex']
\tikzstyle{cloud} = [draw, ellipse,fill=red!20, node distance=3cm,
    minimum height=2em]
    \begin{center}\scalebox{0.90}{
\begin{tikzpicture}[node distance = 3.5cm, auto]
    % Place nodes
    \node [block2] (Inf) {Infected\\medical\\care};
         \node [block2, above of=Inf] (Recovered) {Recovered\\from\\infection};
    \node [block3, left of=Inf] (Inf-Low) {Infected\\home care};
    \node [block6, left of=Inf-Low] (Population) {Susceptible\\ hidden\\ symptomatic};
    \node [block2, right of=Inf] (Inf-High) {High level\\intensive care};
    \node [block7, above of=Population] (Vaccination) {Vaccination programs};
    \node [block4, above of=Inf-High] (Death) {Death};
     \draw [line width=.5mm, ->] ([yshift=-.3cm] Inf.west) -- ([yshift=-.3cm] Inf-Low.east);
     \draw [line width=.5mm,  red, ->] {}([yshift=.3cm] Inf.east) -- ([yshift=.3cm] Inf-High.west);
      \draw [line width=.5mm, ->]  ([yshift= -.3cm] Inf-Low.west) -- ([yshift= -.3cm] Population.east);
      \draw [line width=.5mm, red, ->]  ([yshift= .3cm] Population.east) -- ([yshift=.3cm]  Inf-Low.west);
     \draw [line width=.5mm,  red, ->]  ([yshift=.3cm] Inf-Low.east) -- ([yshift=.3cm] Inf.west);
     \draw [line width=.5mm, ->] ([yshift=-.3cm] Inf-High.west) -- ([yshift=-.3cm] Inf.east);
    \draw [line width=.5mm,  red, ->] (Inf-High) -- node {} (Death);
    \draw [line width=.5mm, green,  ->] (Recovered.west) -- node  {} (Population.north);
    \draw [line width=.5mm, green, ->] (Vaccination.south) -- node  {} (Population.north);
    \draw [line width=.5mm, blue, ->] (Inf) -- node  {} (Recovered);
    \draw [line width=.5mm, blue, ->] (Inf-High) -- node  {} (Recovered);
    \draw [line width=.5mm, blue, ->] (Inf-Low) -- node  {} (Recovered);
          \end{tikzpicture}}
       \end{center}
\caption{Contagion, in-host dynamics and medical care. Immune defense (black arrows to left) contrasts  virus action (red arrows to right). The block on the middle line refers to individuals, while the upper and lower block to the cell-virus scale.}
\end{figure}
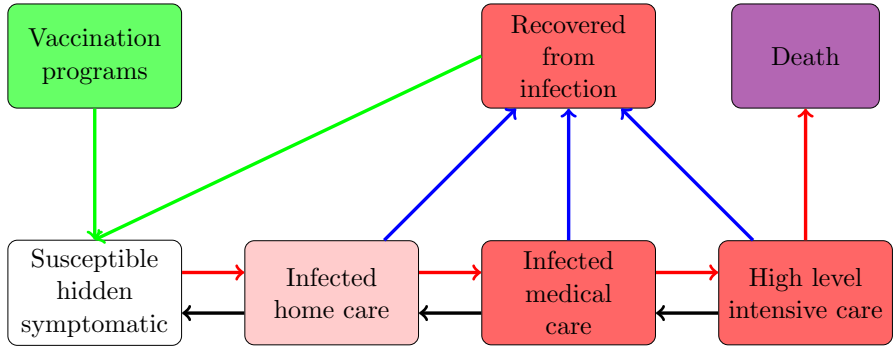

%%%%%%%%%%%%%%%%%%%%%%%%%%%%%%%%%%%%%%%%%%%%
\subsection{Research perspectives about in-host dynamics}\label{subsec:3.5}
%%%%%%%%%%%%%%%%%%%%%%%%%%%%%%%%%%%%%%%%%%%%

The development of methods to model in-host dynamics has been reviewed in this section, starting with Ref.~\cite{[BBC20]}, which initiated a multiscale approach to modeling the dynamics of competition between a proliferating virus and the immune system. Mathematical models refer to the central horizontal line of blocks in Figure 3. These models can provide useful information about the different levels of pathology, which is important for defining a plan for medical treatment. In fact, different medical interventions correspond to different levels of pathology.

The scientific relevance of the study of the in-host dynamics was rapidly identified by \textbf{The Royal Society},  which promoted the organization of a voluntary and interdisciplinary working group, called \textbf{RAMP}, coordinated  by Mark Chaplain of the University of St. Andrews, UK, see Ref.~\cite{[RAMP]}. The specific task of RAMP was the study of the in-host dynamics, while the first achievements promoted by this group was the interdisciplinary paper~\cite{[BBC20]}, where various hints to research activity were given and indeed followed by applied mathematicians.

The strategic intuition of this initiative was confirmed by the real evolution of the general epidemics, which led to the description given in section 2 and suggested the search for research paths substantially new with respect to those of deterministic population dynamics. In terms of research perspectives, the usual strategy is to look at the new achievements in order to understand which are the relevant problems that are still open and which are the clues to tackle them. Accordingly, some key problems are presented below, selected also according to the scientific knowledge of the authors.

\vskip.2cm \noindent $\bullet$ \textit{The in-host space dynamics.} This topic is the natural evolution of the research achievements reviewed in this section. One possible approach is the heuristic modeling developed in Ref.~\cite{[BPTW19]} under the assumption that the transport dynamics is induced by a reaction-diffusion system. In fact, the micro-macro derivation method proposed in Ref.~\cite{[BC22]} shows that the dynamics can be translated into macroscopic equations. On the other hand, it is worth understanding if the problem can be studied by modeling the spatial dynamics of cells and virus particles at the microscopic scale by developing the approach proposed in Ref.~\cite{[CDKS23]}.

\vskip.2cm \noindent $\bullet$   \textit{Further development of the multiscale approach.} A very challenging problem is to extend the multiscale vision by understanding the interaction between the molecular scale and the cellular and viral scale. In particular, understanding how genetic instabilities can generate variants of the initial virus. Studies of these dynamics have been initiated in Refs.~\cite{[ARV20],[CFM20C],[CYR20],[MOSR20]} to understand the properties at the molecular scale, followed by the study of the complex dynamics at the molecular scale in Refs.~\cite{[BDT17],[RBDS18],[TLS21]}. Arguably, further developments could help to address the above challenging problem.

\vskip.2cm \noindent $\bullet$   \textit{Contagion dynamics.} Looking at the higher scale to study contagion between individuals, it soon becomes apparent that the in-host dynamics lead to the calculation of the viral load in the respiratory tract. Then the probability of transmission increases as the viral load increases. Thus, both physical distance and viral load affect the probability of transmission. Another key issue is the spatial spread of infection. As studied in Ref.~\cite{[BP21]}, this dynamics depends on the collective dynamics over transportation networks. The overall study is a challenging problem that may lead to a new vision of epidemics.

\vskip.2cm \noindent $\bullet$  \textit{Vaccination programs.} A kinetic theory approach has been developed in Ref.~\cite{[BBO22B]}, where  the last section indicates the problems that are still left open. In particular, the fact that the action of the vaccination decays with time and thus loses its efficacy in  a period of about twelve months. This behavior is somewhat analogous to that of individuals who recover from an infection. In fact, the immune systems is activated in these individuals, but the ability to learn the presence of the virus decays over time.

\vskip.2cm \noindent $\bullet$ \textit{On the dialog between mathematics and crisis managers.} The interest in deriving mathematical models is motivated by biology, specifically by the interest in describing by equations the complex dynamics of the system under consideration, but also by the need to contribute to the problems that ongoing epidemics pose to society. For example, the description provided by models cannot be limited to the number of susceptible, infected, recovered and dead. In fact, an important information is the level of pathology related to the availability of hospitals. Other interactions between society and epidemics concern the psychological impact \cite{[FONT21]}, and the impact on productivity ~\cite{[BD20],[HMF21]}. These types of interactions require further improvement of the modeling of interactions \cite{[BE24]}, as the world where interaction rules governing opinion dynamics develop evolves over time \cite{[Simon]}.

%%%%%%%%%%%%%%%%%%%%%%%%%%%%%%%%%%%%%%%%%%%%
\section{Social aspects of pandemic spreading}\label{Sec:4}
%%%%%%%%%%%%%%%%%%%%%%%%%%%%%%%%%%%%%%%%%%%%

As briefly discussed in the Introduction, a pandemic is a social phenomenon. Its growth is heavily influenced by different social features, like the intensity of individuals' relationships or their personal opinion about the public health measures to be taken, and it has itself a strong impact on society, favoring for example the development of new forms of work and business activities. In particular, the importance of the social aspects has been outlined by the sudden pandemic caused by the SARS-CoV-2 virus, which lead the central governments to adopt major public restrictions, like social distancing and lock-down policies. Such containment measures have been aimed at reducing the epidemic peak, in order to guarantee the sustainability of the national health care systems during periods of sharp increases in the number of infected individuals. During the SARS-CoV-2 pandemic it was further observed that, as cases of infection intensified, collective adherence to these so-called non-pharmaceutical interventions (NPIs) was particularly crucial to ensure public health in the absence of an effective pharmacological treatment.\cite{APZ,BC22,Bert3,Vig,[ZBDD21]}. The success of the measures relied heavily upon the individuals' compliance with them, which in turn depended on their personal opinions about the necessity of adopting such social restrictions \cite{BRBU,DC,Tc,Tsao}. Improving individuals' response to health protections thus becomes a central issue in devising measures that could effectively lead to virtuous changes in the daily social habits \cite{DFD}.

Once it is established that the spread of an epidemic is intimately linked with the social aspects, one immediately realizes that the time scales at which the two phenomena evolve are different. In fact, the time taken by the individuals to react against the disease by increasing their own personal protection is much faster than the typical time of the epidemic spreading. Therefore, the study of the SARS-CoV-2 pandemic has directed research toward a multiscale framework that takes into account the interaction between different time and spatial scales, ranging from the scale of the virus, to that of individuals, and even beyond, to the scale of a population's collective behavior \cite{[BBC20]}. An interdisciplinary point of view that has been able to develop rapidly, thanks to the contributions of epidemiologists, immunologists and economists in combination with those of mathematicians, statisticians and computer scientists.

In what follows, we will present in details some recent attempts to build a new generation of mathematical models which take into account the joint action of a pandemic spreading and the social activities. For the sake of clarity, and to fully appreciate the novelties of the modeling approach, we will consider this merging in the case of very simple and well-known models of compartmental epidemiology.

\subsection{Classical epidemiological models}
\label{classic_SIR}
%%%%%%%%%%%%%%%%%
\begin{figure}
\centering
    {
    \includegraphics[scale = 0.4]{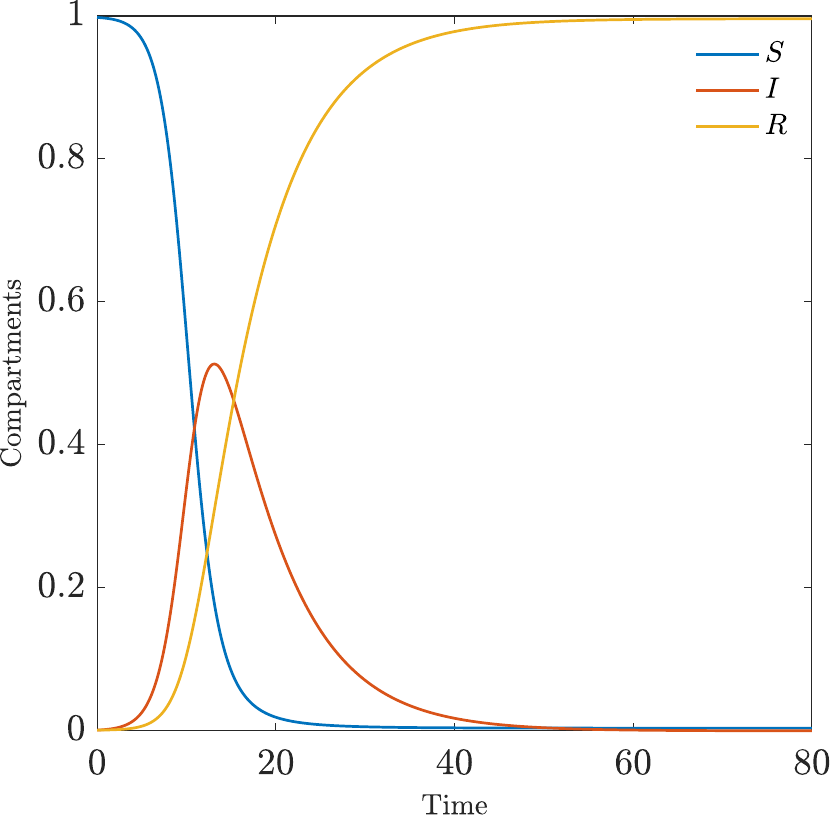} \hspace{0.5cm}
    \includegraphics[scale = 0.4]{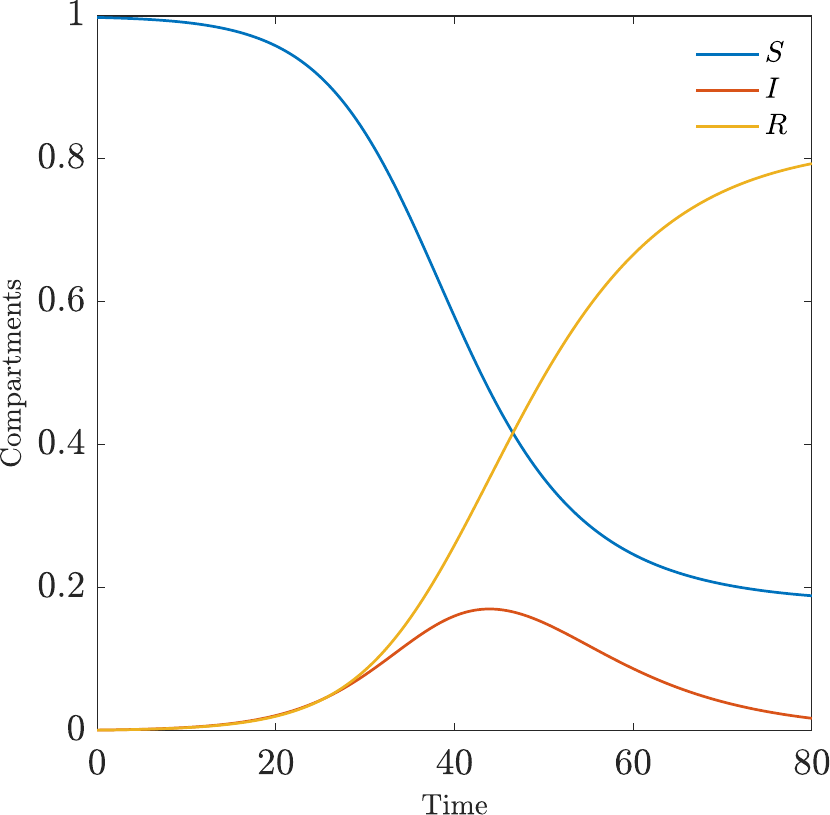}\\
    }
\caption{The left figure refers to the evolution of the SIR pattern \eqref{SIR-b} in the absence of social distancing, where we have chosen a transmission rate $\beta_T = 0.8$. The right figure, obtained by taking a lower $\beta_T = 0.3$, relates instead to an evolution in the presence of social distancing, and shows the flattening of the contagion curve. In both cases, the parameter $\gamma_I$ is fixed to $1/7$.
}\label{fig:test1_2}
\end{figure}
%%%%%%%%%%%%%%%%%

At the time when the SARS-CoV-2 pandemic occurred, mathematical models for understanding the spread of an epidemic were a well-established scientific field of applied mathematics, whose first significant results date back to almost a century ago \cite{[HET00]}. A distinguished contribution is the pioneering work of Kermack and McKendrick on compartmental epidemiology \cite{KMK}, in which a still widely used model was introduced, assuming that the initial population is divided into susceptible (S), who may contract the disease, infected (I), who have already contracted it and may transmit it, and recovered (R), who are either cured from the disease (and immune) or deceased. The model is given by a system of ordinary differential equations prescribing the evolution over time $t > 0$ of the percentages $S(t)$, $I(t)$ and $R(t)$ of the population that belong to the three compartments mentioned above. It reads
  \be\label{SIR-b}
    \begin{split}
\frac{dS(t)}{dt}  &= -\beta_T S(t)I(t),
\\
\frac{dI(t)}{dt}  &= \beta_T S(t)I(t)  - \gamma_I I(t),
\\
\frac{dR(t)}{dt}  &= \gamma_I I(t),
\end{split}
    \ee
and is often referred to as the SIR model, using the acronym for the categories just introduced.

The assumption made by Kermack and McKendrick is that of ``homogeneous mixing''. This corresponds to assuming that each individual has the same probability of infecting any other individual in the population, and that the increase in the infected is directly proportional to the product between the percentage of infected and susceptible, with proportionality constant $\beta_T$, translating a transmission rate. Furthermore, it is assumed that each individual has the same recovery rate $\gamma_I$, allowing to switch from infected to recovered. These parameters determine the evolution of the model's compartments, and it is shown that the dynamics of the infectious class depends on the ratio $R_0 = \beta_T/\gamma_I$, the so-called basic reproduction number \cite{[HET00]}. Figure \ref{fig:test1_2} shows an example of the solution provided by this model in the absence and in the presence of social distancing, characterized by a reduction in the value of $\beta_T$.

Although important from a modeling perspective, the hypothesis of ``homogeneous mixing'' turns out to be unrealistic, since it does not take into account some essential aspects of the infection spreading, like the intensity and nature of someone's contacts with family members, co-workers, schoolmates, a group of friends, or even their spatial movements. In addition, it does not consider that age and any medical conditions can fundamentally alter the course of the infection and the likelihood of recovery.

For these reasons, starting from the classical SIR model, more articulated ones have been proposed to more realistically represent the population by dividing it into additional compartments. In particular, several models have introduced features that are related with the mobility or the age of the subjects, along with additional sociological factors determining the interactions between individuals that can promote the transmission of the infection, see for example Refs. \cite{Buonomo,Giordano}. The dynamical evolution of these supplementary social characteristics, however, has rarely been considered (except for the birth and death rates in the context of age-dependent models\cite{[HET00]}) and the resulting models are always described by systems of ordinary differential equations.

On the other hand, modeling approaches that make use of systems of partial differential equations, inspired by classical kinetic theory, have been recently introduced and studied in the social sciences and have shown the ability to accurately describe complex social phenomena, like the formation of opinions among individuals, the creation of wealth distributions, or the emergence of contacts in social networks. See, for example, the monograph \cite{PT} and the volume \cite{NPT}. Similarly, models based on partial differential equations have proved to be essential in modeling the dynamics of the movement of individuals at different scales \cite{Piccoli}.

In addition, a problem that cannot be disregarded in the case of the SARS-CoV-2 pandemic is the uncertainty in the official data regarding the number of infected individuals. Specifically, the heterogeneity of procedures used to test for disease positivity, the delays in recording and reporting the results, and the high proportion of asymptomatic patients make the construction of predictive scenarios to be highly affected by randomness \cite{JP}.

In the following part, we will further discuss the relevance of the aforementioned social dynamics within an epidemic and some recent developments about their mathematical modeling by partial differential equations. Although, for simplicity in the exposition, the models will be introduced in the context of classical SIR-type compartmental dynamics, an extension to more complex epidemic systems could be done in a similar way. In particular, this summary only partially represents the enormous amount of research that has been done recently in this area, and we refer to the bibliography and to the specific literature mentioned in it for more details.

\subsection{Epidemics and social interactions}
\label{contact}

The integration of the classical SIR model (or more detailed compartmental models) with the different social aspects mentioned above is a problem that has been addressed by the research community, for the most part, through a methodology that is proper to statistical mechanics, and more particularly to kinetic theory \cite{Pulvirenti}. To better elucidate this approach, it is useful to describe in details how the evolution of the pandemic described by the classical SIR model can be studied more realistically by introducing into the spreading mechanism a dependence of the distribution on some social characteristics, like the number of contacts, the age, the opinion, or the wealth of individuals.

The model can be constructed from Kermack and McKendrick's classical SIR by making the number of susceptible, infected and recovered individuals depend on the social characteristic of interest, denoted by the variable $w \in \mathcal{I}$ that belongs to any interval $\mathcal{I} \subseteq \R$. Then, the quantities $f_J(w,t)$, with $J \in \{S,I,R \}$, represent the statistical distributions of the social characteristic in the three different compartments at some time $t > 0$.

The total distribution of the social characteristic $f(\cc,t)$ is given by
 \[
  f(\cc,t) =  f_S(\cc,t) + f_I(\cc,t) + f_R(\cc,t), \qquad \int_{\mathcal{I}} f(\cc,t) d\cc = 1.
    \]
In particular, the percentages $S(t)$, $I(t)$ and $R(t)$ of the population in the three compartments, measured at time $t >0$, are obtained from the statistical distributions $(f_J(w,t))_{J \in \{ S, I, R \} }$ by integration with respect to the variable $\cc$.

The evolution of the distributions $f_J(\cc,t)$, $J \in \{S,I,R\}$, is finally obtained by combining the SIR-type epidemic process with the dynamics of the social characteristic, according to the system
    \be
    \begin{split}
\frac{\partial f_S(\cc,t)}{\partial t} &= -K_T(f_S,f_I)(\cc,t) +  \frac 1\tau Q_S(f_S)(\cc,t),
\\
\frac{\partial f_I(\cc,t)}{\partial t} &= K_T(f_S,f_I)(\cc,t)  - \gamma_I(\cc) f_I(\cc,t) + \frac 1\tau Q_I(f_I)(\cc,t),
\\
\frac{\partial f_R(\cc,t)}{\partial t} &= \gamma_I(\cc) f_I(\cc,t) + \frac 1\tau Q_R(f_R)(\cc,t).
\end{split}
\label{eq:kin-main}
    \ee
where the kinetic-like operators $Q_J(f_J)$, $J\in\{S,I,R\}$, define the changes in the social characteristic due to the microscopic interactions between individuals. Notice moreover that in this setting the recovery rate of the infected $\gamma_I(\cc)>0$ may be dependent on the social characteristic itself, while the transmission of the infection is now governed by the function
\be
 K_T(f_S,f_I)(\cc, t) = f_S(\cc,t) \int_{\mathcal{I}} \beta_T(\cc,\cc_*)f_I(\cc_*,t) \,d\cc_*,
 \label{eq:K-def}
 \ee
 where $\beta_T(\cc,\cc_*) \geq 0$ describes the spread of the infection between an infected and a susceptible with social characteristics $\cc$ and $\cc_*$, respectively.
Note that in the classical SIR model, the function $\beta_T$ is assumed to be constant. The choice in \eqref{eq:K-def} also implies a different definition of the basic reproduction number (see Ref. \cite{[HET00]}). Due to the conservation of the total number of individuals in the operators $Q_J(f_J)$, $J\in \{ S,I,R\}$, one defines a time-dependent effective reproduction number
\be
R_0(t)=\frac{\int_{\mathcal{I} \times \mathcal{I}}\beta_T(\cc,\cc_*) f_S(\cc,t) f_I(\cc_*,t) \,d\cc_*\,d\cc}{\int_{\mathcal{I}}\gamma_I(\cc) f_I(\cc,t)\,d\cc},
\label{eq:Rt}
\ee
providing the change in the number of infected individuals at time $t > 0$.

The scaling parameter $\tau >0 $ defines the time scale for the relaxation of the statistical distribution for the social characteristic toward the corresponding equilibrium densities $f^{\mathrm{eq}}_J$, solutions of the functional relations $Q_J(f^{\mathrm{eq}}_J) = 0$, $J\in\{S,I,R\}$.
In what follows, we will briefly describe three particularly significant cases in which this modeling approach has been used: the case in which the social characteristic is represented by the number of daily contacts between individuals, the case in which it stands for the opinion and, finally, the case in which it corresponds to the wealth (or more generally the economic status) of the individuals.

%%%%%%%%%%%%%%%%%
\begin{figure}
\centering
 {
    \includegraphics[scale = 0.35]{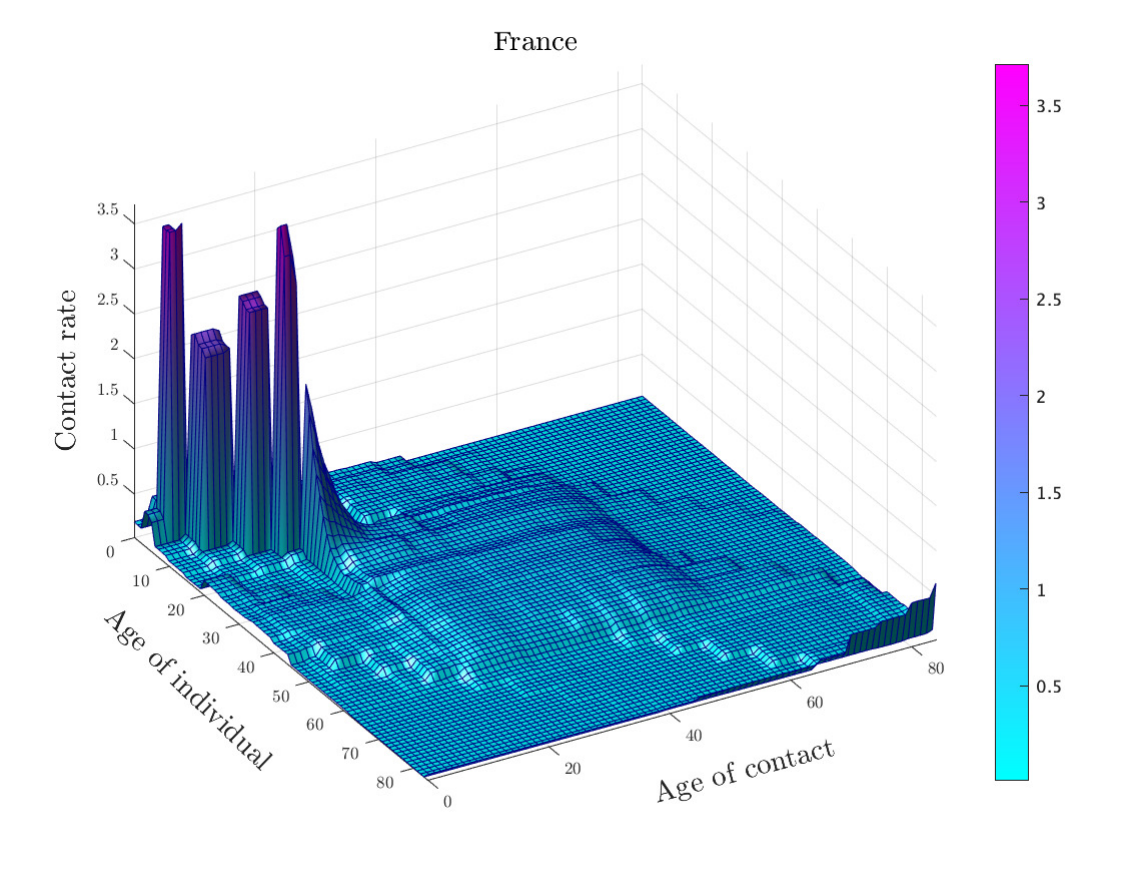} \hfill \includegraphics[scale = 0.35]{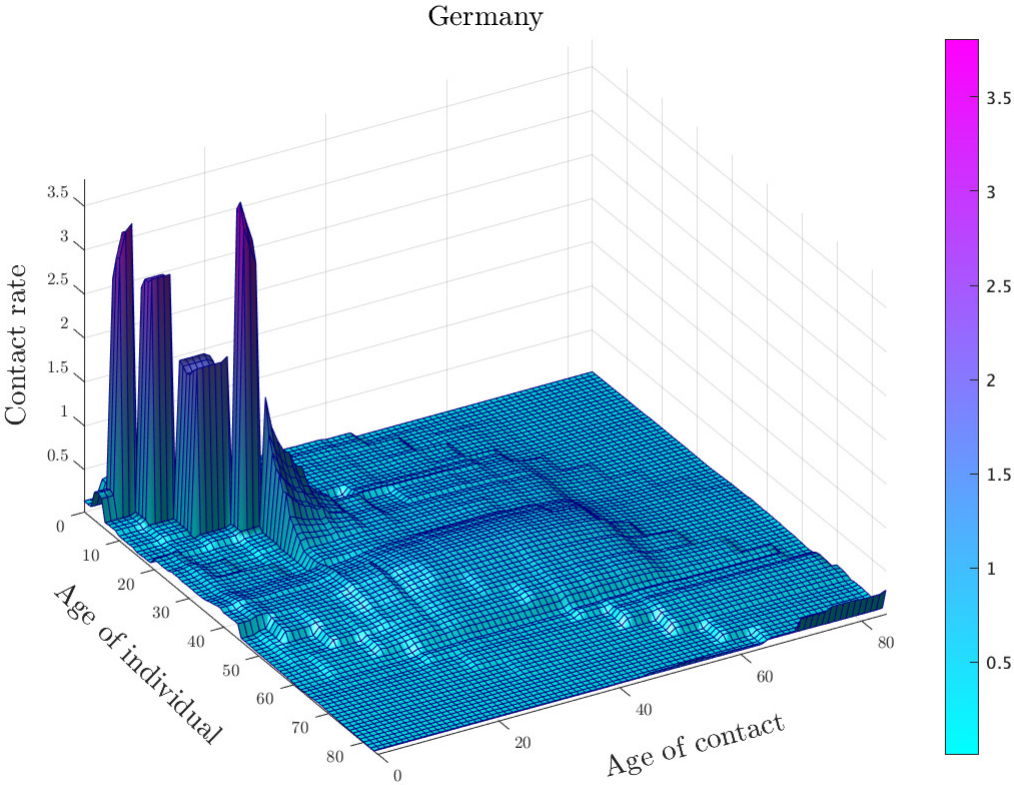} \\[4mm]
    \includegraphics[scale = 0.35]{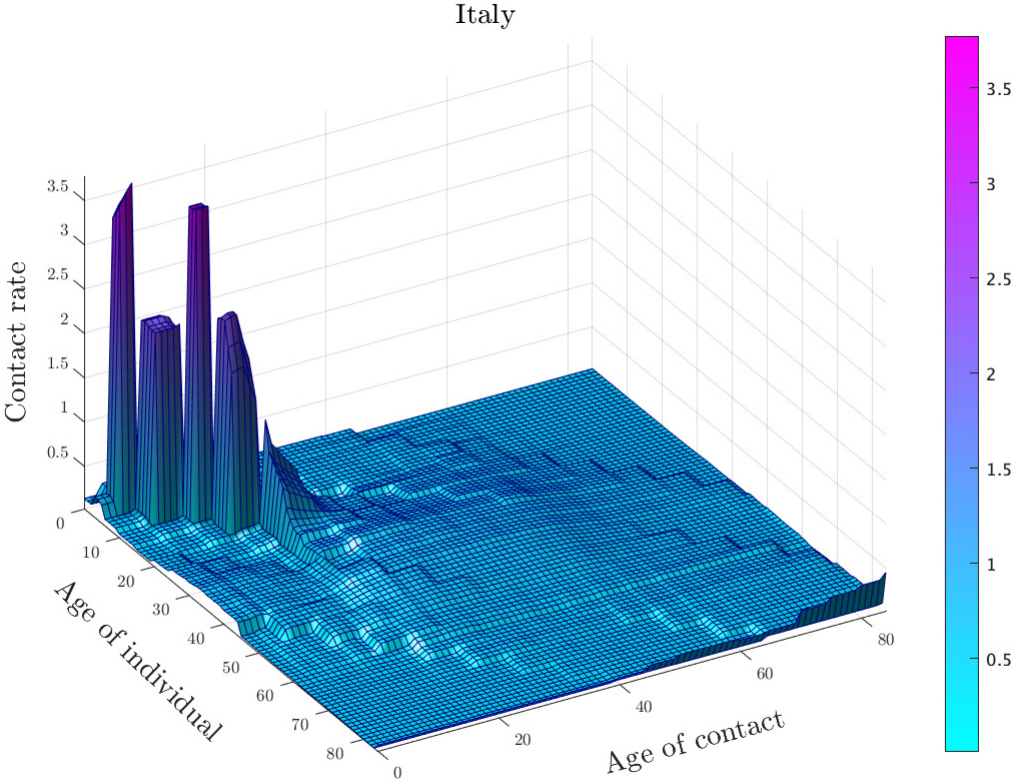} \hfill \includegraphics[scale = 0.35]{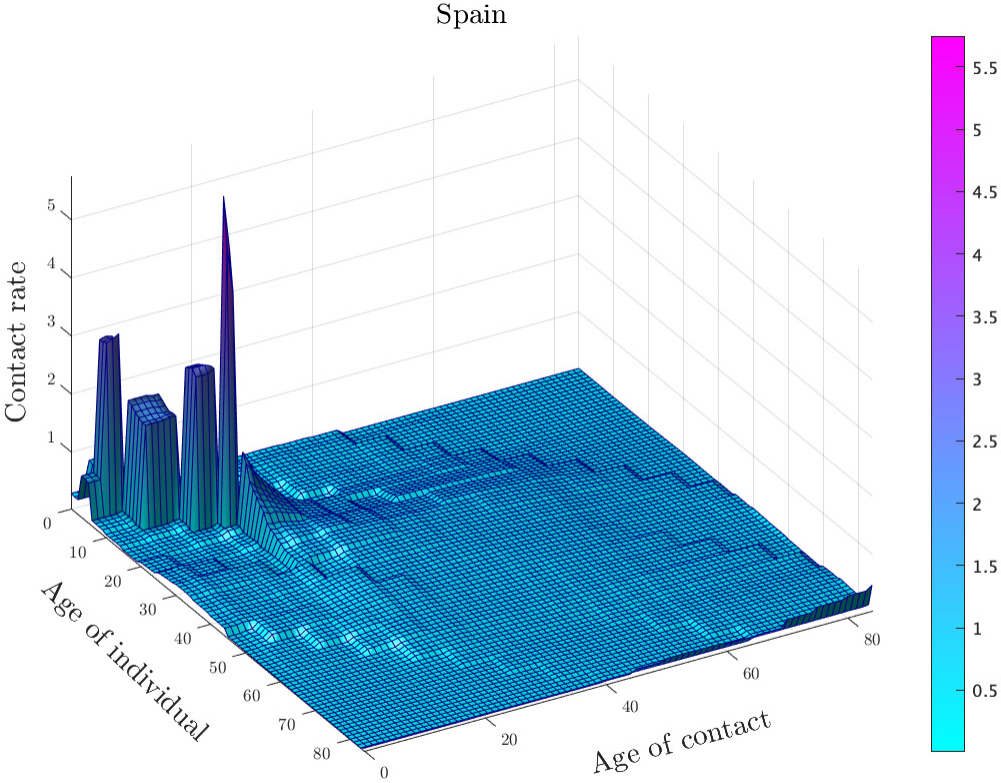} \\[4mm]
     \includegraphics[scale = 0.35]{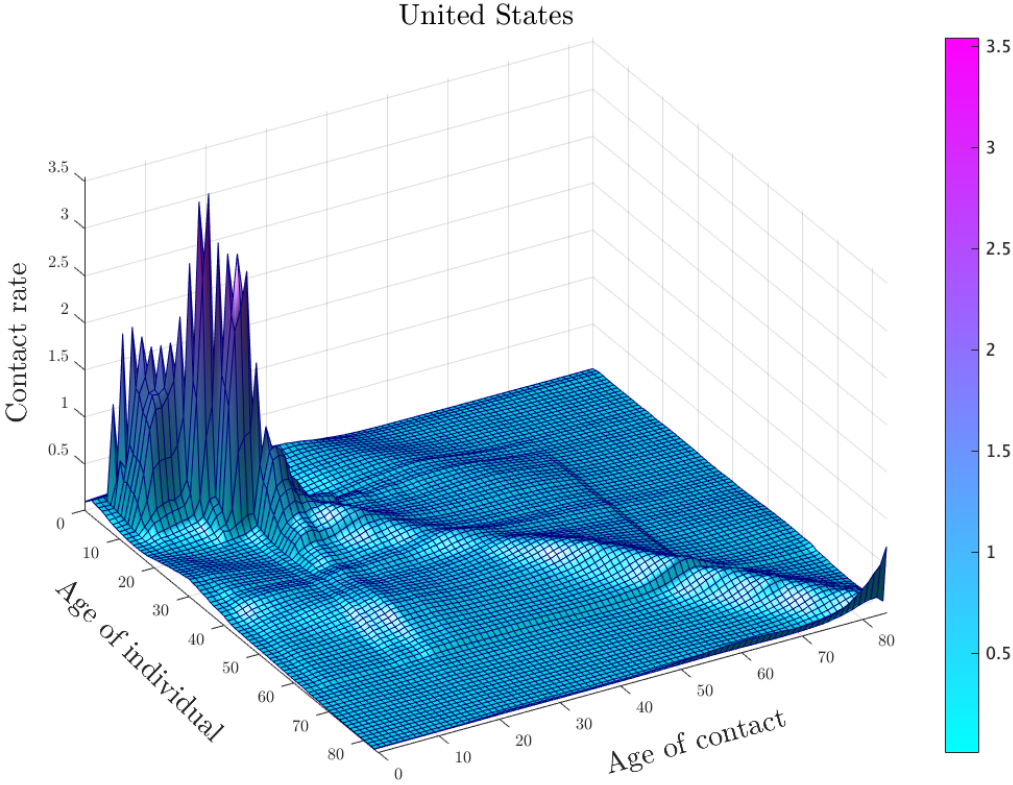} \hfill \includegraphics[scale = 0.35]{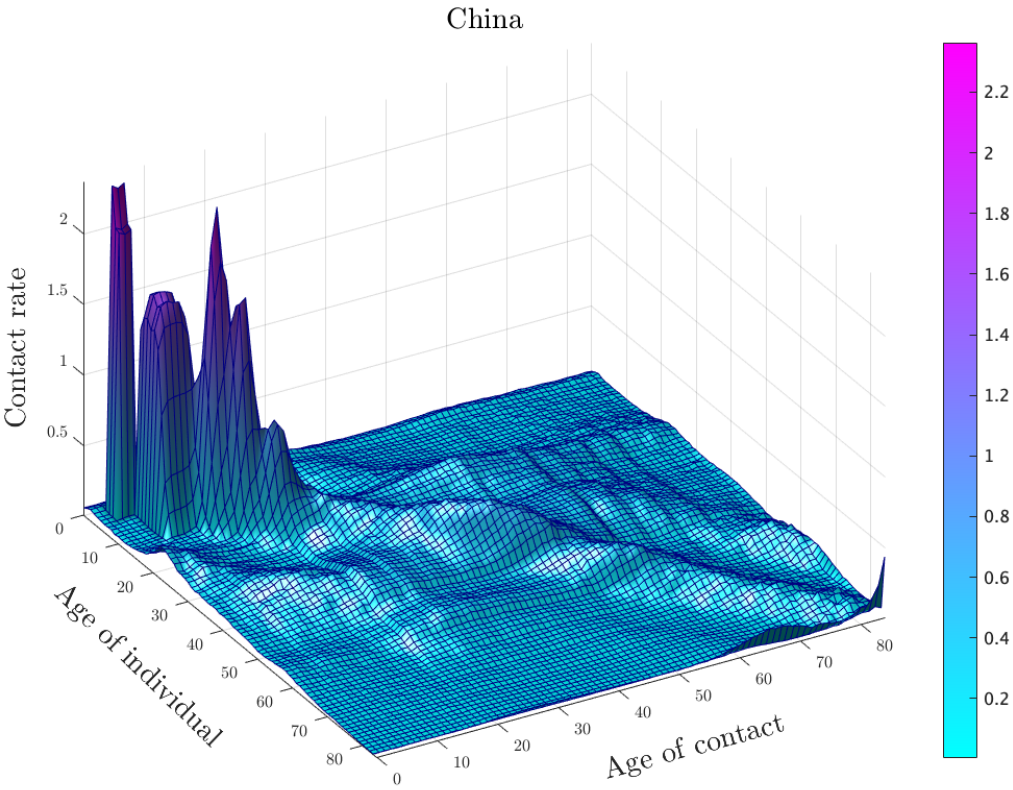} \\[4mm]
    \includegraphics[scale = 0.35]{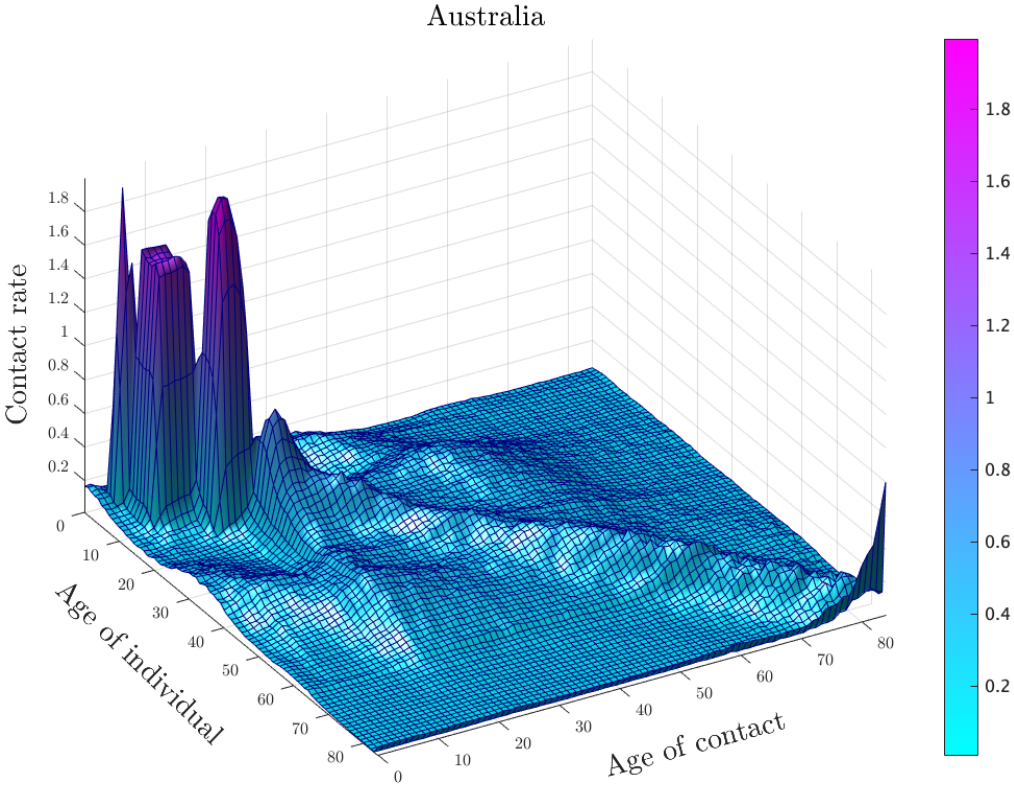} \hfill \includegraphics[scale = 0.35]{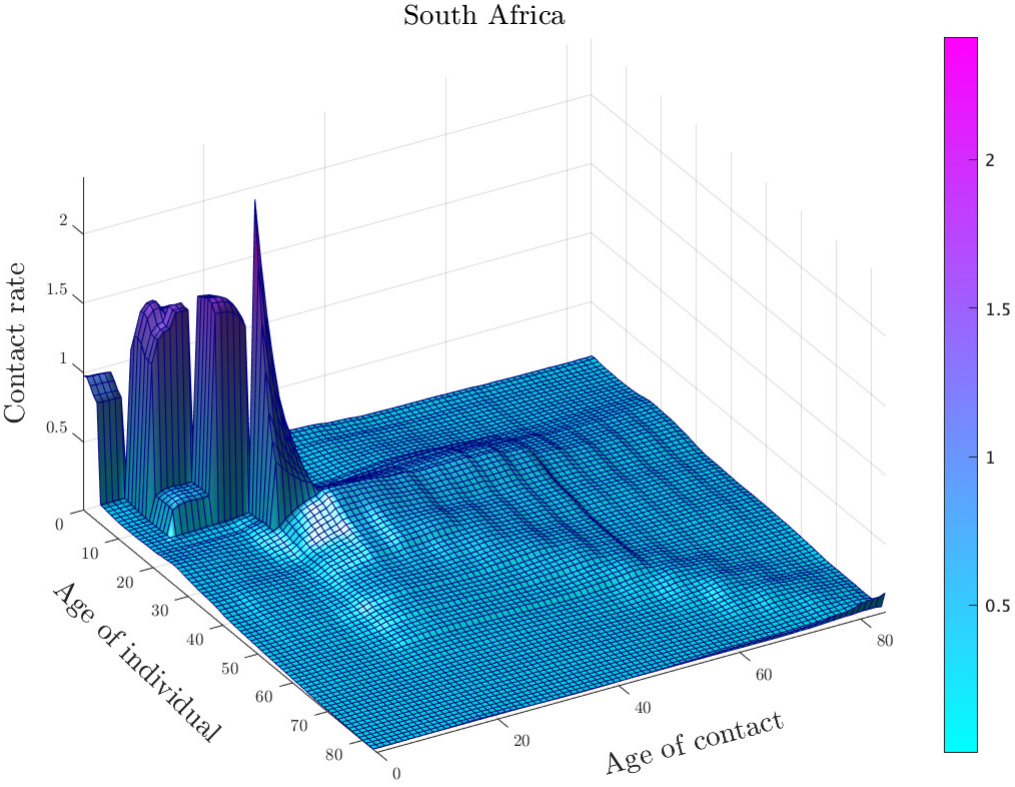}
}
\caption{Intensity of social contacts (weighted over household, community, school and business contacts) in different countries, depending on the age of the individuals involved in the interaction. One may observe very similar patterns characterizing the contact matrices.}\label{fig:sociali_1}
\end{figure}
%%%%%%%%%%%%%%%%%

\subsection{The role of social contacts} \label{sec:s-sir}

The number of social contacts between individuals clearly plays an important role in the spread of a disease. From a formal point of view, family, school, or work contacts can be characterized by appropriate interaction matrices, according to the age of the individuals involved in the social activity under consideration \cite{Colombo}. The corresponding contact patterns appear to be very similar for European and non-European countries \cite{Plos,PREM,[MLP]}, and can thus be assumed as universal.
Using the data collected in Ref. \cite{[MLP]} via censuses and surveys throughout 35 countries, one can infer the statistical distributions of population's daily social contacts stratified by age. We show in Figure \ref{fig:sociali_1} a collection of the resulting contact matrices, for some representative countries in different areas of the world. Such matrices of contacts were used in Refs. \cite{APZ,APZ2} in the case of age-dependent SIR-type models, to specifically study the problem of introducing pharmaceutical control measures as a function of the available data, along with the effect of uncertainty in these data.

Taking this into account, a statistical description was developed in Ref. \cite{S-SIR}, based on an uneven ``mixing'' in the contagions between infected and susceptible individuals in which, unlike the classical models, the probability of contagion was taken to be proportional to the average contacts between individuals in the population, while the increase in the infected was directly proportional to the product between the averages of daily contacts for infected and susceptible individuals. In such a model, the kinetic variable $w$ naturally ranges in the interval $\mathcal{I} = \R_+$ and, for example, the amount
 \[
 \int_3^{10} f_I(\cc,t) d\cc,
 \]
translates the percentage of infected having between $3$ and $10$ daily contacts.
Furthermore, in \eqref{eq:K-def} a function of the form $\beta_T(\cc,\cc_*)= \bar{\beta}_T \cc \cc_*$ was considered,
where the infection transmission rate was chosen to be proportional (with proportionality constant $\bar{\beta}_T$) to the number of contacts $\cc_*$ of the infected and $\cc$ of the susceptible individuals. The function $\gamma_I(w)=\bar{\gamma}_I$ was instead assumed to be independent of the number of contacts.

Without going into technical details, the operators $Q_J$, $J\in \{S,I,R\}$, describe the daily variations in the distributions of contacts $f_J(\cc,t)$, $J\in \{S,I,R\}$. Formally, by the methods of classical kinetic theory, they can be written as linear Boltzmann-type integral operators and can be subsequently approximated by Fokker--Planck-type operators of the form (see Refs. \cite{FPTT17,PT})
\begin{equation}
 \begin{split}
 Q_J(f_J)(\cc,t) =
 \frac{\partial}{\partial w}\left[\frac{\lambda}{2}\left(\frac w{ m_J(t)} -1\right)f_{J}(\cc,t) +\frac{\sigma}{2} \frac{\partial}{\partial w} (w f_J(\cc,t))\right],
\end{split}
\label{eq:fok1}
\end{equation}
where
\[
m_J(t)=\frac1{J(t)}\int_{\R_+} w f_J(w,t) \,dw,\quad J\in \{S,I,R\},
\]
denotes the average number of contacts in the $J$-th compartment.  Here, the positive constant $\lambda$ characterizes the rate of alignment to the average number of contacts, while the quantity $\sigma > 0$ defines the stochastic variance of this process via a spread that is proportional to the number of contacts.
In the case of the Fokker--Planck equations \eqref{eq:fok1}, it is possible to show that the equilibrium distributions $f^{\mathrm{eq}}_J$, solutions of the equations $Q_J(f^{\mathrm{eq}}_J)=0$, $J\in \{S,I,R\}$, are of Gamma-type, in agreement with what has been found experimentally (see for example Ref. \cite{Plos}).
From the knowledge of the equilibrium states, and owing to the rapid timescale (small values of $\tau$) at which people adapt their behavior to the risky situation, a simplified description of the evolution can be obtained by analogy with the derivation of fluid dynamical models from the kinetic theory of gases, described by the Boltzmann equation and by the resulting hierarchy of equations for the moments of the particle distribution \cite{Cer}.

In particular, the presence of a contact function with a varying rate of contagion does not allow one to obtain directly a closed system of moments, by integrating in the variable $\cc$ the kinetic equations of the distributions $f_J(\cc,t)$, $J\in \{S,I,R\}$, since the zeroth-order moments vary in dependence of the first-order ones. The closure can be recovered only if one knows the explicit form of the stationary distribution of contacts. For a Gamma-type densities, in fact, the knowledge of the zeroth- and first-order moments is sufficient to fully characterize the distribution. By looking at a regime of $\tau \ll 1$, the model \eqref{eq:kin-main} can thus be described in terms of a closed system of six ordinary differential equations, giving the evolution of the percentages $J(t)$ and of the mean numbers of contacts $m_J(t)$, with $J\in \{S,I,R\}$. Such a system takes the form
 \be
    \begin{split}
\frac{dS(t)}{dt}  &= -\bbeta_T \, m_S(t) m_I(t) S(t)I(t), \\[2mm]
\frac{dI(t)}{dt}  &= \bbeta_T \, m_S(t) m_I(t) S(t)I(t)  - \bar{\gamma}_I I(t), \\[2mm]
\frac{dR(t)}{dt}  &= \bar{\gamma}_I I(t), \\[2mm]
\frac{d m_S(t)}{d t} &= - \frac{\bbeta_T}{\nu} m_{S}(t)^2 m_I(t) I(t), \\[2mm]
\frac{d m_I(t)}{d t} &=   \bbeta_T m_{S}(t) m_I(t)  \left( \frac{1+\nu}{\nu} m_S(t) - m_I(t) \right) S(t), \\[2mm]
\frac{d m_R(t)}{d t} &=  \bar{\gamma}_I \frac{I(t)}{R(t)}\left( m_I(t) - m_R(t)\right).
\end{split}
\label{eq:ssir}
\ee
 Unlike the classical SIR model, in \eqref{eq:ssir} the probability of infection is considered to be proportional to the average contacts of individuals in the population, and the increase in infected is directly proportional to the product between the averages of daily contacts of infected and susceptible individuals.
In addition to the parameters $\bbeta_T$ and $\bar{\gamma}_I$, which are analogous to the corresponding ones in the classical SIR system, the social SIR model sees the presence of the additional parameter $\nu=\lambda/\sigma$, characterizing the heterogeneity of the population \cite{BBT}. In particular, as the variance $\sigma$ tends to zero (absence of heterogeneity), the parameter $\nu\to+\infty$ and the contacts' mean of the susceptible individuals remains constant. In that case, the system reduces to the classical SIR model, with $\beta_T$ replaced by $\bbeta_T m_S(0)m_I(0)$.

%%%%%%%%%%%%%%%%%
\begin{figure}
\centering
{\includegraphics[scale = 0.4]{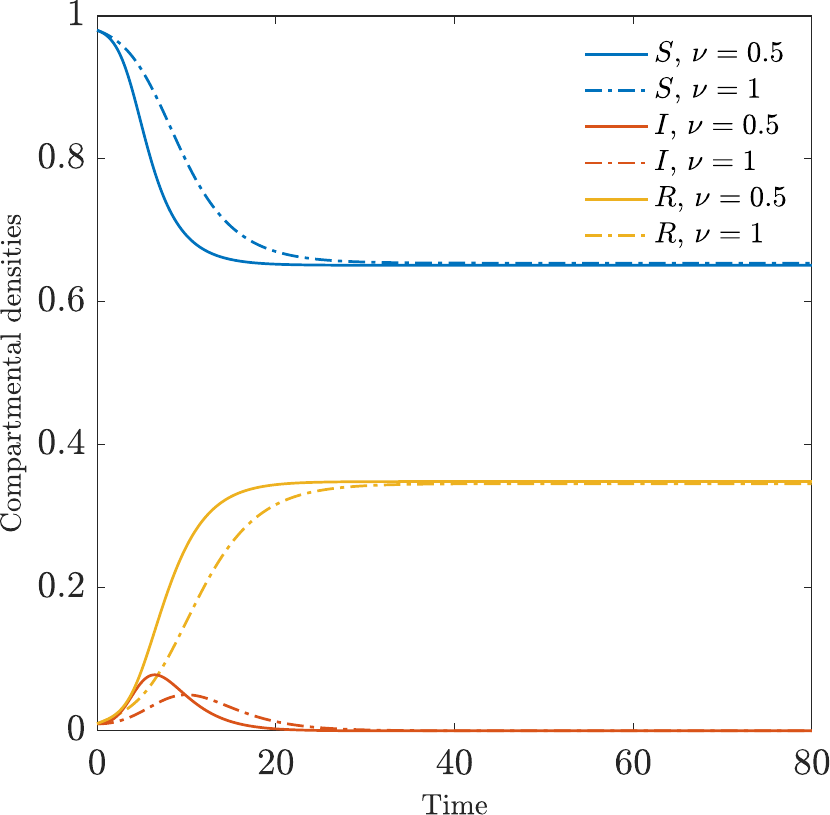} \hspace{0.5cm}
\includegraphics[scale = 0.4]{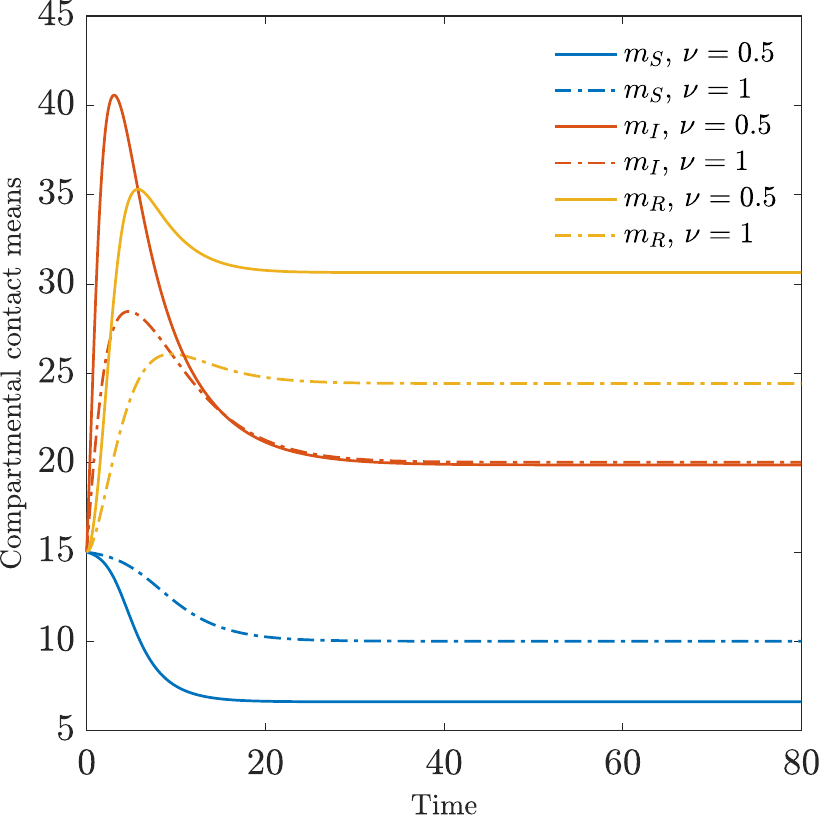}}
\caption{
On the left, evolution of the percentages of susceptible, infected, and recovered in the system \eqref{eq:ssir} for two different values of the heterogeneity parameter $\nu = 0.5, 1$. On the right, the corresponding evolution of the compartmental contact averages. Here, we have chosen the epidemiological parameters $\bbeta_T = 0.2$ and $\bar{\gamma}_I = 1/2$.
}
\label{fig:system}
\end{figure}
%%%%%%%%%%%%%%%%%

In Figure \ref{fig:system} we have plotted the evolution for the percentages of susceptible, infected and recovered (left), as well as for their mean contact numbers (right), starting from 15 daily contacts on average and for two different values of the parameter $\nu = 0.5, 1$. The right figure shows a rapid growth in the number of contacts for the infected, immediately after the start of the epidemic, suggesting that the people with the most likelihood to pass into the infected compartment are those who statistically have a high number of contacts. In addition, we may observe a natural decrease in the number of contacts among susceptible individuals, which is very pronounced in the case of high heterogeneity. In the figure on the left, it can be also be seen that in the presence of higher heterogeneity in the number of contacts, the peak of the infected is brought forward in time. For concrete applications of the model \eqref{eq:ssir} in real clinical settings, we refer to Refs. \cite{ATS,[ZBDD21]}.

\subsection{Consensus formation and pandemic spreading}

We shall detail here how to adapt the strategy that has been concisely presented in Section \ref{sec:s-sir}, to the case in which the social variable of interest, in connection with the pandemic spreading, is represented by the personal opinion of individuals \cite{ZZ}.  Each agent is endowed with an opinion variable $w \in \mathcal{I}$ which varies continuously in the interval $\mathcal{I} = [-1,1]$, where $-1$ and $1$ denote the two opposite beliefs regarding the protective behavior. In particular, the value $w = -1$ is associated with agents that do not believe in the necessity of protections (like wearing masks or reducing daily contacts), whereas agents having opinion $w = 1$ are in complete agreement with the protective measures in place. It will be also assumed that individuals characterized by a high protective behavior are less likely to get the infection.

Following the notations of Section \ref{sec:s-sir}, we denote with $f_J(w,t)$ the distribution of opinions at time $t > 0$ of the agents in the compartment $J \in \{S,I,R\}$. The kinetic realization of system \eqref{SIR-b} for the coupled evolution of opinions and disease is given once again by the system \eqref{eq:kin-main} \cite{ZZ},
% \begin{equation}
%\label{eq:kinetic_opinion}
%\begin{split}
%\frac{\partial f_S(w,t)}{\partial t} &= -K_T(f_S,f_I)(w,t) + \dfrac{1}{\tau} Q^\opinion_{S}(f_S)(w,t), \\
%\frac{\partial f_I(w,t)}{\partial t} &= K_T(f_S,f_I)(w,t)  - \gamma_I f_I(w,t) + \dfrac{1}{\tau} Q_{I}(f_I)(w,t), \\
%\frac{\partial f_R(w,t)}{\partial t} &= \gamma_I f_I(w,t) + \dfrac{1}{\tau}Q_{R}(f_R)(w,t),
%\end{split}
%\end{equation}
where now the $Q_J(f_J)$, $J\in \{S,I,R\}$, denote the kinetic operators quantifying the variation of the agents' opinions in each compartment. These are Fokker--Planck-like operators, associated with kinetic equations of the form
 \be\label{op-FP}
 \frac{\partial f_J(w,t)}{\partial t} = Q_J(f_J)(w,t)= \frac{\sigma}{2}\frac{\partial^2 }{\partial w^2}\left((1-w^2)
 f_J(w,t)\right) +\lambda \frac{\partial }{\partial w}\left((w -\mu)f_J(w,t)\right),
 \ee
for any given $J \in \{ S, I, R \}$. The parameter $\mu$ is such that $-1 < \mu < 1$, while $\sigma$ and $\lambda$ are  positive constants characterizing the intensities of respectively the diffusion and the compromise, which are commonly considered to be the two main components of the change in opinion. Suitable boundary conditions at the extremal points $w = \pm 1$ then guarantee conservation of mass for the solutions \cite{PT}. \\
The  steady states of equations \fer{op-FP}, reached exponentially fast in time for a wide range of the three parameters $\mu$, $\sigma$ and $\lambda$ \cite{FPTT17}, are solutions of the first-order differential equations
\[
 %\label{ch6:FPstaz2}
\frac{\nu}{2}\frac{\partial}{\partial w}\left((1-w^2)f^{\mathrm{eq}}_J(w)\right) + (w -\mu) f^{\mathrm{eq}}_J(w)= 0,
 \]
where $\nu =\sigma/\lambda$. In case a mass density equal to unity is chosen, such equilibrium states are explicitly given by the Beta densities
 \be\label{beta}
f^{\mathrm{eq}}_J(w)= C_{\mu,\nu} (1-w)^{-1 + \frac{1-\mu}\nu} (1+w)^{-1 + \frac{1+\mu}\nu},
 \ee
with $C_{\mu,\nu} > 0$ being a normalization constant. Since $-1 <\mu<1$, each $f^{\mathrm{eq}}_J$ is integrable on the interval $(-1,1)$. Note that $f^{\mathrm{eq}}_J$ is also continuous on $(-1,1)$, and as soon as $\nu > 1+|\mu|$, the distribution tends to infinity as $w \to \pm 1$. \\

Moreover, the spread of the infection in system \eqref{eq:kin-main} depends on the presence of the local incidence rate
\begin{equation*}
K_T(f_S,f_I)(w,t) =  f_S(w,t) \int_{-1}^1 \beta_T(w,w_*) f_I(w,t) dw,
\end{equation*}
that drives the transmission of the infection in terms of the opinion variable $w$. In this context, the function $\beta_T(w,w_*)$ is a nonnegative decreasing function measuring the impact of the protective behavior in the interactions between susceptible and infectious agents. A leading example for $\beta_T(w,w_*)$ is provided by
\begin{equation}
\label{eq:kappa_def}
\beta_T(w,w_*) = \dfrac{\bbeta_T}{4^{\alpha}} (1-w)^{\alpha} (1-w_*)^{\alpha},
\end{equation}
where $\bbeta_T > 0$ is the baseline transmission rate characterizing the specific epidemics and $\alpha>0$ is a parameter linked to the efficacy of the protective measures. Thus, the local incidence rate, and consequently the evolution of the disease, is fully dependent on the joint opinion of agents toward the protective behaviors. In particular, we remark that for $\alpha\equiv 0$ the influence of opinion dynamics on the epidemiological model disappears, and in this case the evolution of the mass fractions $J(t)$, $J \in \{ S,I,R\}$, obeys the standard SIR model \ref{SIR-b}.

We moreover notice that in the simple setting $\alpha = 1$, the local incidence rate takes the form
\[
K_T(f_S,f_I)(w,t) = \dfrac{\bbeta_T}{4}(1-w)f_S(w,t)(1-m_I(t))I(t)\ge 0, \qquad I(t)\ge 0,
\]
where $m_I(t)$ represents the mean opinion of infected individuals at time $t > 0$. Note that $K_T(f_S,f_I)(w,t)  \equiv 0$ whenever $m_I(t)\equiv 1$ for some $t > 0$, or in the hypothetical case where all susceptible agents are concentrated on the maximal protective behavior $w = 1$.

The knowledge of the equilibrium state, coupled with the rapid time scale at which people adapt their behavior to the risky situation (small values of $\tau$), eventually allows to obtain from the kinetic model \fer{eq:kin-main} an equivalent of the macroscopic system \fer{eq:ssir}. As in the previous case, the closure is made possible by the nature of the stationary distributions of opinions \eqref{beta}. Indeed, for a Beta-type density, knowledge of the corresponding zeroth- and first-order moments is once again sufficient for its complete determination.

\subsection{The economic impact of a pandemic}

The COVID-19 pandemic has highlighted the close link between economics and health in the context of emergency management. Assessing the impact of an epidemic phenomenon on a country's economy has proven to be one of the key aspects to be taken into account when designing suitable containment strategies. From a mathematical point of view, it is important to emphasize that developing a systematic approach to study the effects on the economies of countries facing a pandemic is objectively a very complex problem and that a mathematical model can only provide rough indications of the possible consequences, based on simplified assumptions about the model parameters.

The basic idea is to trace these phenomena back to the evolution of a country's so-called wealth distribution, which measures how many people belong to increasing income brackets.
In this case, as proposed in Ref. \cite{DPTZ}, assuming that the evolution of the densities obeys to system \fer{SIR-b} and that individuals in the various compartments act differently in the economic process, the general structure of the model has always the form \eqref{eq:kin-main}, where now the status of each individual in the compartment is uniquely determined by their wealth $w \in \mathcal{I} = \f_R^+$ (measured in terms of a reference currency).
Thus $f_S(\w,t)$, $f_I(\w,t)$ and $f_R(\w,t)$ denote the wealth distributions of the susceptible, infected and recovered populations.

The function $\beta_T(w,w_*)$ represents in this case the contact rate between susceptible and infected with riches $w$ and $w_*$, respectively. For example, a contact rate in the form of a decreasing function of $|w-w_*|$ expresses the fact that individuals with different degrees of wealth live in different environments, and their contact is thus limited. The choice of a wealth-dependent recovery rate $\gamma_I=\gamma_I(w)$ can be motivated by considering that greater economic availability might allow access to better hospitals and care, ensuring higher chances of healing.

Similarly to the previous sections, the operators $Q_J(f_J)$, $J\in \{S,I,R\}$, describe the change in wealth across classes due to financial transactions. Unlike the case of social contacts, they are related to each other through the different averages, meaning that the transactions occur differently depending on the epidemic compartment of individuals.
Formally, such terms turn out to be nonlinear Boltzmann integral operators, which can be approximated by Fokker--Planck equations of the form (see Ref. \cite{DPTZ})
\be\label{eq:fok2}
 Q_J (f_J)(w,t) = \frac{\partial}{\partial w}\left[
\left(\lambda_J w-\bar{m}(t)\right)f_J(w,t)+\frac{\sigma(t)}{2} \frac{\partial}{\partial w} (w^2 f_J(w,t)) \right],
\ee
where we have set
\[
\bar{m}(t)=\lambda_S m_S(t)S(t)+\lambda_I m_I(t)I(t)+\lambda_R m_R(t)R(t).
\]
The description of financial transactions, reported here only qualitatively, is based on the choice of two parameters. The first one defines the so-called safeguard threshold characterized by $\lambda_J \in [0,1]$, $J\in \{S,I,R\}$, i.e. the maximum percentage of money that the individual is willing to employ in a transaction, and the second one is associated with the risk inherent to the transaction, characterized by the variance $\sigma(t)$ through a spread that is proportional to the square of the individual's wealth. The time dependence of the variance was postulated by assuming that when there is a significant spread of the epidemic, the variance of the risk tends to increase. This is in agreement, for example, with the reactions of financial markets to the announcements of new numbers of infected people, which was observed in different countries during the COVID-19 pandemic.

The model makes it possible to study the distribution of wealth in the population at the end of a pandemic event, depending on the parameters of risk $\lambda_J$, $J\in\{S,I,R\}$, and $\sigma(t)$, which can be precisely correlated with the presence of the epidemic. In the case of small time scales $\tau \ll 1$, i.e. when the relaxation time in the behavior of individuals with respect to business activities is very rapid compared with the time of the disease's evolution, a macroscopic approximation based on the equilibrium states solving $Q_J(f^{\mathrm{eq}}_J)=0$, $J\in\{S,I,R\}$, can still be adopted. In this case, it turns out that the $f^{\mathrm{eq}}_J$ are inverse Gamma densities, in accordance with what was observed in the late 19th century by the economist Vilfredo Pareto, about the form of income distributions. Specifically, the tails of the wealth distribution are characterized by a polynomial decay measured by the Pareto index, which generally takes values between $1.5$ and $3$, where small values of this index correspond to more pronounced economic inequalities. For an extended discussion of the subject, and for its description in a kinetic framework, we refer to Ref. \cite{PT}.

The economic inequalities described by Pareto-type power laws may be accompanied by additional phenomena, to the detriment of precise segments of the population. One example of inequality is that arising from sudden economic shocks that typically harm the so-called middle class. The number of individuals who form the middle class can be roughly measured in this context as the number of people within the range defined by the poorest 20\% and the richest 20\% of individuals, which we denote by $M_C(t)$. Extreme instability events can result in a bimodal income distribution; in fact, in such cases the population tends to split into two groups, with an impoverishment of the middle class.
%%%%%%%%%%%%%%%%%
\begin{figure}
\centering
\includegraphics[scale = 0.35]{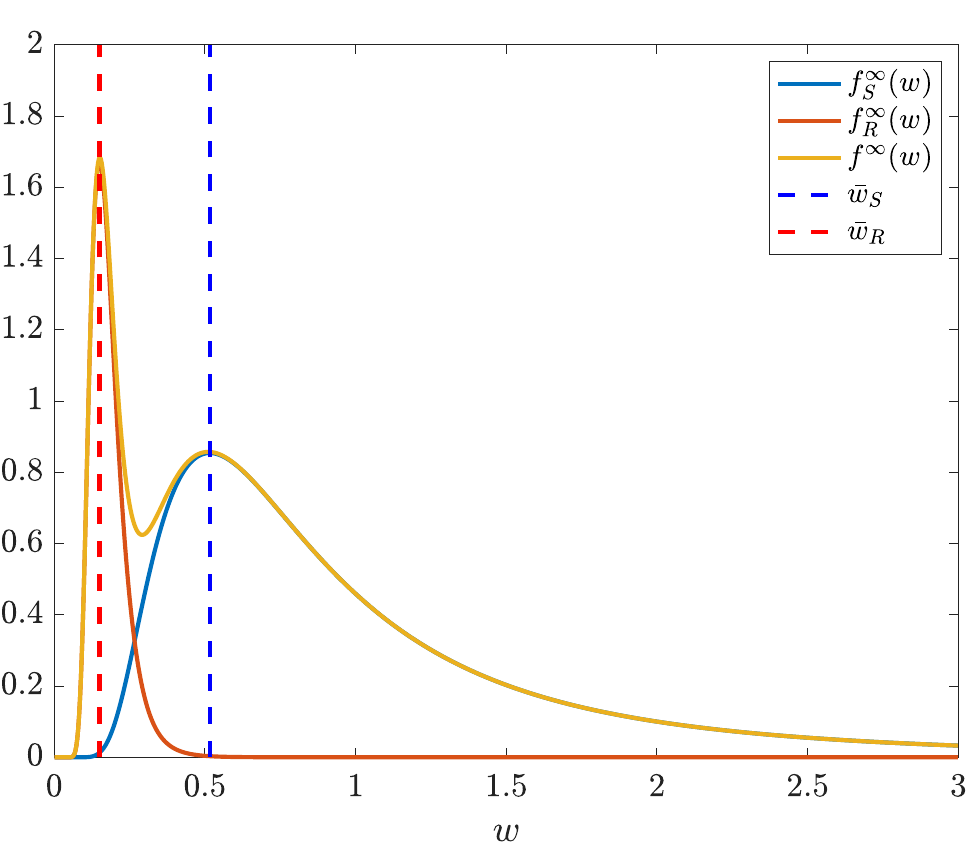}\hskip .5cm
\includegraphics[scale = 0.47]{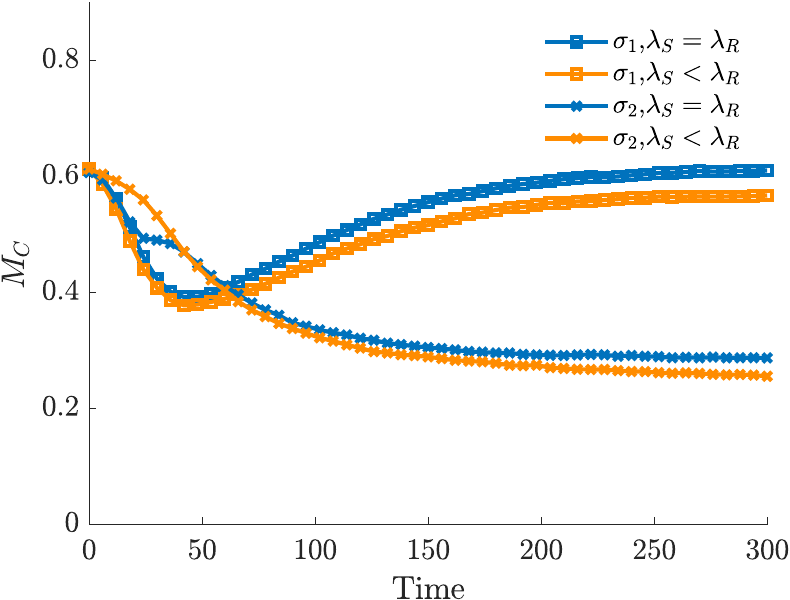}
\caption{On the left, bimodal distribution of wealth in the model of interaction between epidemic and wealth of individuals, and on the right, the phenomenon of the shrinking middle class as the epidemic progresses (see Ref. 32). }
\label{fig:Gini_MC}
\end{figure}
%%%%%%%%%%%%%%%%%

It is interesting to observe how the previous model, for appropriate parameter choices, leads to the formation of bimodal type curves, as shown in Figure \ref{fig:Gini_MC} on the left. In the right figure, we also report the evolution of the mean class $M_C(t)$ corresponding to two different measures of risk
\[
\sigma_1(t)= \sigma_0(1+\alpha I(t)),\qquad \sigma_2(t)=\sigma_0\left(1+\alpha\int_0^t I(s)\,ds\right),
\]
where $\alpha>0$ measures the impact of the infection on the variance of economic transactions.
We can clearly see how emerging inequalities mainly affect the middle class, which is steadily decreasing for $\sigma_2$, where the risk retains memory of the epidemic phenomenon, while it undergoes a transient decrease for $\sigma_1$.

\subsection{Research perspectives about the social aspects of a pandemic}

Aiming to elucidate the basic ideas behind the mesoscopic approach, the previous discussion has been mainly focused on the most simple socio-epidemiological kinetic-based SIR models. However, there exists an increasing literature in this direction, rapidly developing to produce more articulate and detailed descriptions. The changes in the social behavior of individuals and the evolution of an epidemic are in fact strictly connected mechanisms that mix in a very complex way, explaining the need for improved models. The COVID-19 pandemic has been a motor for this development and it is regarded as the primary contemporary example of the interplay between epidemics and social features.
Indeed, the pandemic has been characterized by several waves of infection, whose temporal dynamics can be related not only with the natural mutations of the virus, but  with the central governments' NPIs/PIs (containment measures based on lock-downs and masks' use, vaccination campaigns, etc.). The populations' reaction toward them, accepting or not to adopt protective behaviors or not, or to change or not social attitudes, is nowadays recognized to be an important aspect of the pandemic evolution.
This suggests the construction of new mathematical models able to quantify the possible scenarios consequent to the degree of adherence of the social behavior to the imposed restraint measures.  On this aspect there are two recent attempts, which aim to describe such phenomena from a kinetic point of view, possibly paving the way to future works in this direction.

\vskip.2cm \noindent $\bullet$ \textit{Contagion dynamics driven by utility processes.} To characterize the spreading dynamics, Ref. \cite{[BPT]} introduces a new compartmental Boltzmann-type model by borrowing tools from microeconomics, complementing previous approaches constructed from game theory \cite{[Bohl],[CMM]}. Specifically, the interactions between the agents and the virus are described by a microscopic state vector $(\mathbf{x},\mathbf{v})$, where the individual's socio-physical condition $\mathbf{x} = (x_1,x_2)$ depends on the personal resistance to the disease, measured by the positive quantity $x_1$, and on the degree of sociality, measured by the positive quantity $x_2$, while the viral impact $\mathbf{v} = (v_1,v_2)$ depends on the level of contagiousness $v_1$ and on the severity of the disease $v_2$ (see also Ref. \cite{[DLT],DLT23} for a similar description, taking into account the agents' viral load). The transmission of the infection is then driven by binary exchanges  based on the Cobb--Douglas utility functions, and the Edgeworth box to identify the common trade area where the benefit increases for both agents \cite{Edg}. This approach provides a valuable tool for gaining insights into the dynamics of infectious diseases and their transition from high-risk pandemics to low-risk endemic states, shedding light in particular on the mechanisms that steer these phenomena toward endemicity, via the oscillatory nature observed in successive pandemic waves. It is important to remark that the analysis in Ref. \cite{[BPT]} is complementary to that of Section \ref{subsec:3.2}, since it only relies on the statistical behavior of the possible in-host dynamics, by leaving biological details completely unexplored.

%%%%%%%%%%%%%%%%%
\begin{figure}[th!]
\centering
{\includegraphics[scale = 0.4]{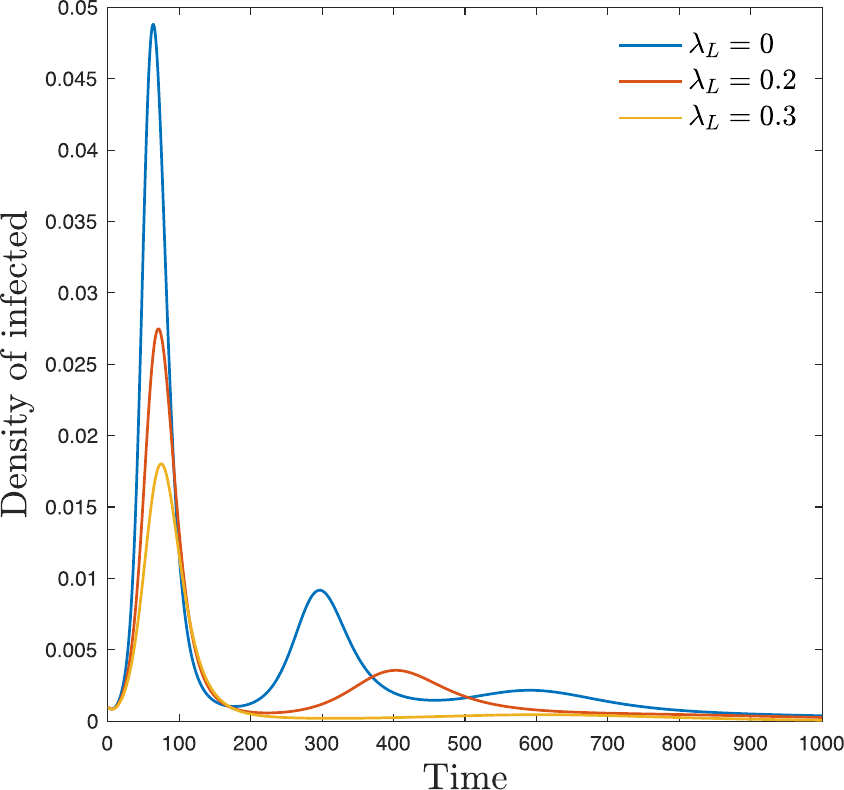} \hspace{0.5cm}
\includegraphics[scale = 0.4]{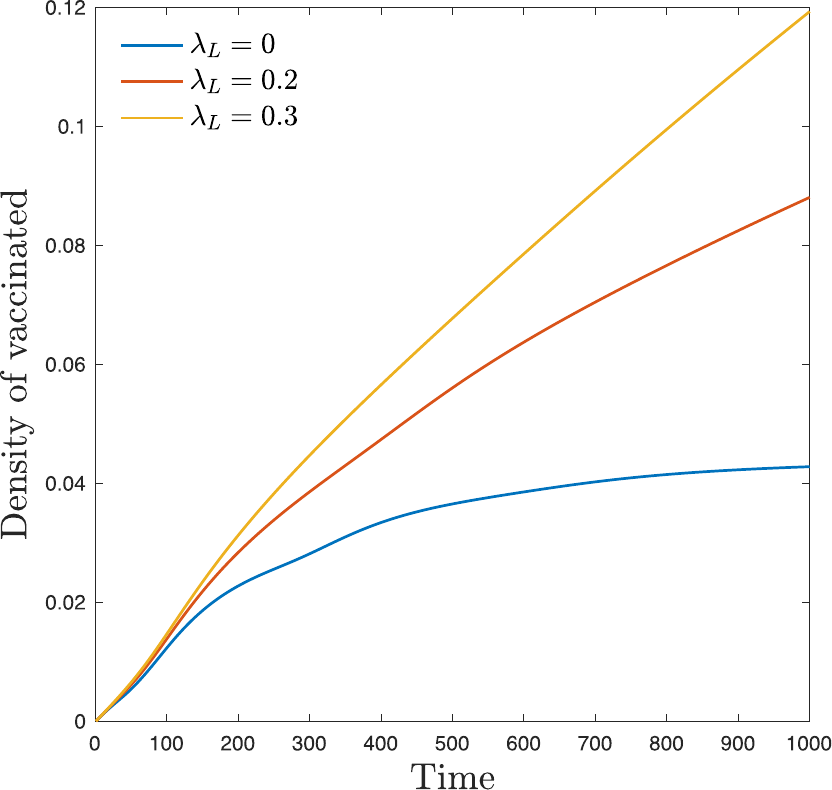}}
\caption{
Evolution of the percentages of infected (left) and vaccinated (right) individuals, as prescribed by the opinion-based kinetic compartmental model proposed in Ref. \protect\cite{[BTZ]}. The dynamics shows creation of waves in the density of infected, due to the spontaneous reaction of individuals toward the containment measures, depending on the current state of the pandemic. We also observe the positive impact of the leaders on this latter: as their influence on the population increases with the parameter $\lambda_L = 0, 0.2, 0.3$, the intensity of the epidemic waves is progressively reduced and the number of vaccinated grows.
}
\label{fig:SIRV-VH}
\end{figure}
%%%%%%%%%%%%%%%%%

\vskip.2cm \noindent $\bullet$ \textit{Modeling vaccine hesitancy.} The second attempt, which can open interesting perspectives about the possible application of mathematical modeling to the social forecasting of health policies, deals with the modeling of vaccine hesitancy, an issue that policy makers need to face every time a massive viral spreading occurs, and particularly observed during the COVID-19 pandemic \cite{DM24,Fplos23,BPTZ22}. Starting from the idea that any modifications in individuals' opinions should be regarded as a spontaneous behavioral change induced by the surrounding environment \cite{[BDOG],[Poletti2009]}, the approach of Ref. \cite{[BTZ]} considers an opinion-based SIRV-type model (with a further subgroup $V$ of vaccinated individuals) similar in structure to system \eqref{eq:kin-main}, where the evolution of the different compartments $J \in \{ S, I, R, V \}$ is described by distribution functions $f_J(w,t)$, $w \in [-1,1]$, interacting between them and with a static density of leaders $f_L(w)$. The balance of their microscopic interactions is accounted for via Fokker--Planck-type operators of the form
\begin{multline*}
Q_J(f_J,f_L)(w,t) = \frac{\sigma}{2} \frac{\partial^2 }{\partial w^2}\left((1-w^2) f_J(w,t)\right) \\[1mm]
 + \frac{\partial}{\partial w}\left( \left( w - \lambda \mu(I(t)) -\lambda_L \int_{-1}^1 f_L(w) dw \right) f_J(w,t)\right),\end{multline*}
where the positive coefficients $\lambda \le 1$ and $\sigma$ measure, respectively, the compromise and the self-thinking propensities of individuals, while the parameter $\lambda_L$, satisfying the constraint $\lambda + \lambda_L = 1$, characterizes the intensity of the leaders' influence on the population. The average reaction of the agents to the current state of the epidemics, encoded by the density of infected individuals $I(t)$, is modeled by the function
\begin{equation*}
\mu(s) = 1 - 2(1 - s)^\alpha, \qquad s \in [0,1],
\end{equation*}
where $\alpha > 0$ translates the degree of risk perception. In particular, for large numbers of infected the population tends to adopt a protective behavior (since in this case $\mu \to 1$) in an effort to reduce the cases, while a reduction in the epidemic spreading pushes the population toward more risky behaviors, as the compartmental means are now driven by $\mu$ to $-1$. The presence of leaders acts as a counterbalance to $\mu(I(t))$ and, depending on the parameter $\lambda_L$, it is able to move the population's average opinion toward more neutral or positive positions. An example of this feature is shown in Figure \ref{fig:SIRV-VH}, where the epidemic waves produced by the model in the evolution of infected can be lowered by suitably increasing the value of $\lambda_L$. A similar positive impact is observed on the number of vaccinated, which grows in correspondence to a stronger role of the leaders.

%%%%%%%%%%%%%%%%%%%%%%%%%%%%%%%%%%%%%%%%%%%%
\section{Spatial propagation of the infection}\label{Sec:5}
%%%%%%%%%%%%%%%%%%%%%%%%%%%%%%%%%%%%%%%%%%%%

The models discussed so far, while invaluable for capturing the temporal dynamics of infectious disease spread and the impact of population heterogeneity, operate under the simplifying assumption of spatial homogeneity within a deterministic framework.
Yet, spatial considerations become paramount when addressing interventions that vary across different regions of a country. Additionally, uncertainties surrounding initial conditions and modeling parameters, particularly within the realm of social sciences, necessitate careful consideration in numerical simulations, especially when informing public health decisions.
%However, the spatial dimension becomes crucial when addressing interventions that vary across different regions of a country. Moreover, uncertainties regarding initial conditions and modeling parameters, especially within the realm of social sciences, require careful consideration when employing numerical simulations, particularly for making public health decisions.

%%As already discussed, the foundation of epidemiological modeling finds its origins in the seminal work of Kermack and McKendrick, who introduced compartmental models driven by systems of ODEs \cite{KMK}. These models,
%while invaluable for capturing the temporal dynamics of infectious disease spread and the impact of the heterogeneity of the population, operate under the simplifying assumption of spatial homogeneity within a deterministic setting.
%Typically, characterizing the average behavior of a population provides an initial reliable representation of the development of an epidemic.
%However, the spatial dimension becomes of paramount relevance when addressing interventions that vary across different regions of a country. Furthermore, uncertainties surrounding initial conditions and modeling parameters, especially within the realm of social sciences, necessitate their careful consideration when employing numerical simulations, particularly when they are used to take public health decisions.

In the following sections, we build on insights from kinetic transport theory to present recent advances achieved in hyperbolic models of epidemic spread. These models offer an innovative approach to investigating epidemic propagation by accounting for the spatial movement and interactions of distinct population groups at different spatio-temporal scales.
%, accounting for heterogeneity of the territory in which the epidemic spread occur.
%We will expand upon the integration of network spatial structures into the discussed kinetic modeling framework, enhancing the ability to capture the intricate dynamics of epidemic spread across interconnected communities.
An essential aspect of this extension involves distinguishing between the behaviors of commuting and non-commuting individuals. This differentiation prevents the emergence of unrealistic scenarios where all individuals move over long distances in the computational domain, enabling a more accurate description of population movements within and between regions.
Subsequently, we will address data uncertainty and calibration of model parameters. To tackle the latter, we will discuss recent approaches based on machine learning techniques and physics-informed neural networks.
%To this aim, we introduce robust methodologies for model calibration and parameter estimation based on neural networks, advancing towards a modeling framework that increasingly relies on data-driven insights.

\subsection{Modeling of pandemic spread within a geographical area}\label{Sec:5.1}
To mathematically describe the space-time evolution of an epidemic as a system of interacting individuals, we can consider two main approaches. The first involves designing a model with two-dimensional spatial dependence to characterize disease spread in a geographical area \cite{Bert3,[BDP21],Vig,Vig1}. This approach has the potential to provide a fine description of the dynamics but  must contend with the lack of detailed data on the territory. Another approach is based on leveraging the characteristics of the territory, where individuals commute along transportation networks. This latter choice allows for modeling to be restricted to one spatial dimension along the main connections between cities \cite{[BP21],Bert}.

In the context of this work, we focus on the latter methodology, and we present the models by sticking within the framework of the compartmental approach. As already done in the previous sections, we account for the simplest compartmental subdivision of the population,  the SIR model \cite{[HET00],KMK}. Nevertheless, the proposed strategy can be straightforwardly extended to more complex compartmentalization, better suited for studying specific real epidemic phenomena, as presented in Refs. \cite{ABBDPTZ21,BertBiFi,[BP21]}.
We then consider densities $f_J^\pm(x,t)$, $x\in \Omega\subset \mathbb{R}$, $t>0$, $J \in \{S,I,R\}$,  of individuals moving in two opposite directions with velocities $\pm \lambda_J$, $J \in \{S,I,R\}$, depending on the compartment.
The number of susceptible, infectious, and recovered individuals at location $x$ and time $t$ are then represented by the densities
\be
\begin{split}
s(x,t) &= f_S^+(x,t)+f_S^-(x,t), \\
i(x,t) &= f_I^+(x,t)+f_I^-(x,t), \\
r(x,t) &=f_R^+(x,t)+f_R^-(x,t),
\end{split}
\label{eq.macro-densities}
\ee
respectively, expressed in relative value with respect to the size of the entire reference population $N = S(t) + I(t) + R(t)$, with
\[
S(t) = \int_\Omega s(x,t)\,dx, \quad I(t) = \int_\Omega i(x,t)\,dx, \quad R(t) = \int_\Omega r(x,t)\,dx.
\]
The spatio-temporal dynamics of the population is defined through the following SIR-type kinetic model with discrete velocities\cite{BertProc,BertBiFi,BertAPNN,Bert}:
\be
\begin{split}
	\frac{\partial f_S^{\pm}}{\partial t} \pm \lambda_S \frac{\partial f_S^{\pm}}{\partial x} &= - F(f_S^{\pm},i) \mp \frac{1}{2\tau_S}\left(f_S^+ - f_S^-\right),\\
	\frac{\partial f_I^{\pm}}{\partial t} \pm \lambda_I \frac{\partial f_I^{\pm}}{\partial x} &=  F(f_S^{\pm},i) -\gamma_I f_I^{\pm} \mp \frac{1}{2\tau_I}\left(f_I^+ - f_I^-\right),\\
	\frac{\partial f_R^{\pm}}{\partial t} \pm \lambda_R \frac{\partial f_R^{\pm}}{\partial x} &= \gamma_I f_I^{\pm} \mp \frac{1}{2\tau_R}\left(f_R^+ - f_R^-\right),									
\end{split}
\label{eq.SIR_kinetic}	
\ee
where the parameter $\gamma_I$ is the recovery rate (see Section \ref{classic_SIR}), while the transmission of the infection is governed by an incidence function $F(\cdot,i)$, assumed to be\cite{Bert}
\begin{equation}
F(g,i)=\beta_T \frac{g i^p}{1+\kappa i}.
\label{eq:incF}
\end{equation}
Notice that the classical bi-linear case corresponds to $p = 1$, $\kappa=0$. Here, the transmission rate $\beta_T$, accounting for both number of contacts and probability of transmission, may vary based on governmental control actions that might be implemented as the disease progresses, such as mandatory wearing of masks, shut-down of specific work/school activities, or full lock-downs \cite{Giordano,Vig1}. Furthermore, in this incidence function, $\kappa$ acts as incidence damping coefficient based on the self-protective behavior of individuals that arises from the awareness of the risk associated with the epidemic \cite{[BP21]}.
In general, even though in our notations we have ignored the spatial dependency for the sake of simplicity, the epidemic parameters $\beta_T$, $\gamma_I$ and $\kappa$ can be heterogeneous in space, depending on the position $x$. Similarly, the relaxation times $\tau_S$, $\tau_I$, and $\tau_R$.

There are two aspects of particular interest worth noting. The first regards the property of conservation of the total mass of the system, which appears evident when summing up all the equations in \eqref{eq.SIR_kinetic} and integrating over the whole domain $\Omega$ under the assumption of no inflow/outflow boundary conditions.
In addition, we can observe that when assuming the same boundary conditions, summing up only the third and fourth equations in \eqref{eq.SIR_kinetic} and integrating in space we obtain that
\[
\frac{\partial}{\partial t} \int_{\Omega} i(x,t)\,dx =  \int_{\Omega} F(s,i)\,dx-\int_{\Omega} \gamma_I(x) i(x,t)\,dx \geq 0
\]
when
\be\label{eq.R0_space}
R_0(t)=\frac{\int_{\Omega} F(s,i)\,dx}{\int_{\Omega} \gamma_I(x) i(x,t)\,dx} \geq 1.
\ee
Thus, we have just obtained the definition of the reproduction number for system \eqref{eq.SIR_kinetic}, which gives information on the space averaged instantaneous variation of the number of infective individuals at time $t>0$ \cite{Bert}. Notice that when no spatial dependency is attributed to variables and parameters and $\kappa= 0$ is fixed, definition \eqref{eq.R0_space} leads to the basic reproduction number of the classical SIR ordinary differential equations (ODEs) model (see Section \ref{classic_SIR} and Refs. \cite{CS78,[HET00],KMK}).

%Leaving aside the source terms that determine the evolution on the epidemic side (i.e., those that depend on the epidemic parameters), it can easily be noticed that system \eqref{eq.SIR_kinetic} finds its roots in the Goldstein--Taylor model \eqref{eq.GT}. In fact, for each population compartment considered, the dynamics of spatio-temporal evolution follows model \eqref{eq.GT}, with the only difference being that the scaling parameter is now identified with the symbol $\tau$ and the velocities $\lambda$ are left generic.

Now, it is possible to rewrite system \eqref{eq.SIR_kinetic} in macroscopic form in terms of the total densities of each compartment, $s$, $i$, and $r$, and the relative fluxes $j_S = \lambda_S \left(f_S^+ - f_S^-\right)$, $j_I = \lambda_I \left(f_I^+ - f_I^-\right)$, and $j_R = \lambda_R \left(f_R^+ - f_R^-\right)$, obtaining
\be
\begin{split}\label{eq.SIR_macro}
	\frac{\partial s}{\partial t} + \frac{\partial j_S}{\partial x} &= -F(s,i) ,\\
	\frac{\partial i}{\partial t} + \frac{\partial j_I}{\partial x} &= F(s,i) -\gamma_I i,\\
	\frac{\partial r}{\partial t} + \frac{\partial j_R}{\partial x} &= \gamma_I i ,\\
	\frac{\partial j_S}{\partial t} + \lambda_S^2 \frac{\partial s}{\partial x} &= -F(j_S,i)  -\frac{j_S}{\tau_S}, \\
	\frac{\partial j_I}{\partial t} + \lambda_I^2 \frac{\partial i}{\partial x} &= \frac{\lambda_I}{\lambda_S} F(j_S,i)  -\gamma_I j_I -\frac{j_I}{\tau_I},\\
	\frac{\partial j_R}{\partial t} + \lambda_R^2 \frac{\partial r}{\partial x} &= \frac{\lambda_R}{\lambda_I}\gamma_I j_I -\frac{j_R}{\tau_R}.
\end{split}
\ee
To analyze the behavior of this model in the zero-relaxation limit \cite{LT}, we introduce the space-dependent diffusion coefficients $D_J=\lambda_J^2 \tau_J$, $J\in\{S,I,R\}$,
which characterize the diffusive transport mechanism of susceptible, infectious and removed. Keeping the above quantities fixed while letting the relaxation times $\tau_{J}\to 0$ (and so the characteristic velocities $\lambda_{J} \to \infty$), from the last three equations in \eqref{eq.SIR_macro} we obtain
\[
j_S = -D_S \frac{\partial s}{\partial x},\quad  j_I = -D_I\frac{\partial i}{\partial x},\quad j_R = -D_R\frac{\partial r}{\partial x},
\]
which inserted into the first three equations in \eqref{eq.SIR_macro} permit to recover the following parabolic reaction-diffusion model, widely used in literature to study the spread of infectious diseases\cite{Berestycki,Magal,Sun,Vig1}:
\be
\begin{split}\label{eq:SIR_diff}
\frac{\partial s}{\partial t} &=  -F(s,i) +\frac{\partial}{\partial x} \left({D_S}\frac{\partial s}{\partial x}\right), \\
\frac{\partial i}{\partial t} &=  F(s,i)-\gamma_I i+\frac{\partial}{\partial x} \left({D_I}\frac{\partial i}{\partial x}\right), \\
\frac{\partial r}{\partial t} &=  \gamma_I i+\frac{\partial}{\partial x}\left({D_R}\frac{\partial r}{\partial x}\right).
\end{split}
\ee

%\begin{remark}
The model's flexibility in describing different regimes, hyperbolic or parabolic, depending on spatially dependent relaxation times $\tau$, makes it appropriate for characterizing human dynamics. It is clear that daily routines are the result of a complicated mix of people typically moving only within the urban areas of a city and people moving to the suburbs or neighboring cities. In this case, it makes sense to simulate the specifics of movements within an urban region using a diffusion operator rather than describing them in detail because of the high complexity of the dynamics and the lack of microscopic information. On the other hand, commuters moving from one city to another follow well-established paths, for which a description provided by transport operators is more suitable, avoiding an unrealistic diffusion of them in suburban space at infinite speeds typical of parabolic models\cite{ABBDPTZ21,Bert3,Bert,[BDP21]}.
%\end{remark}

%%%%%%%%%%%%%%%%%
\begin{figure}[t!]
\centering
\includegraphics[width=0.48\textwidth]{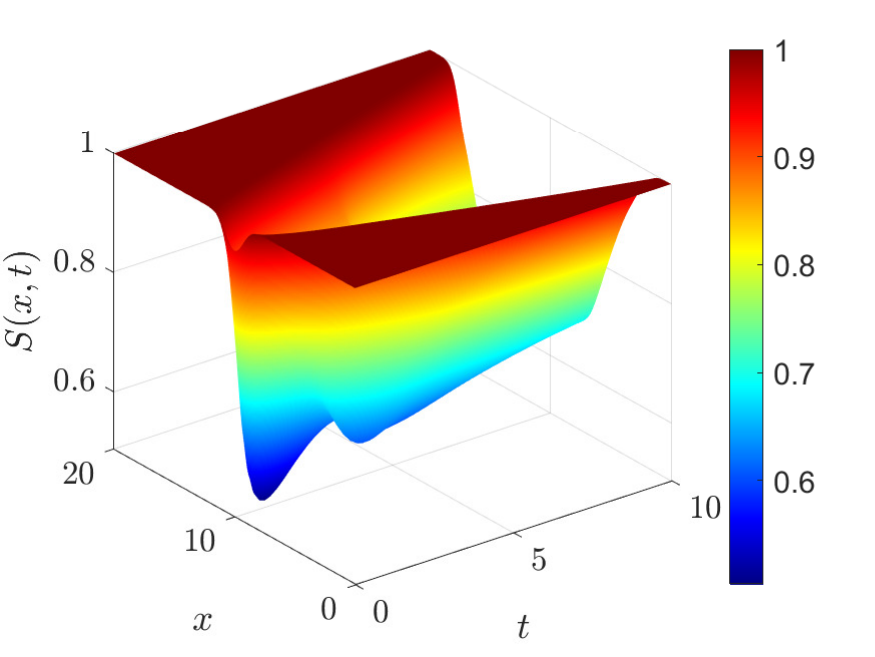}
\includegraphics[width=0.48\textwidth]{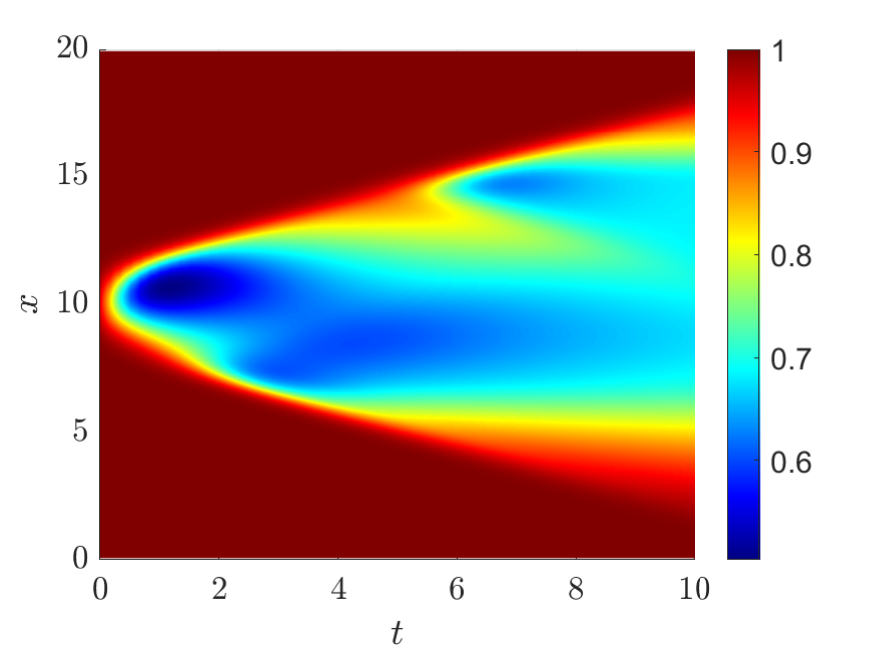}
\includegraphics[width=0.48\textwidth]{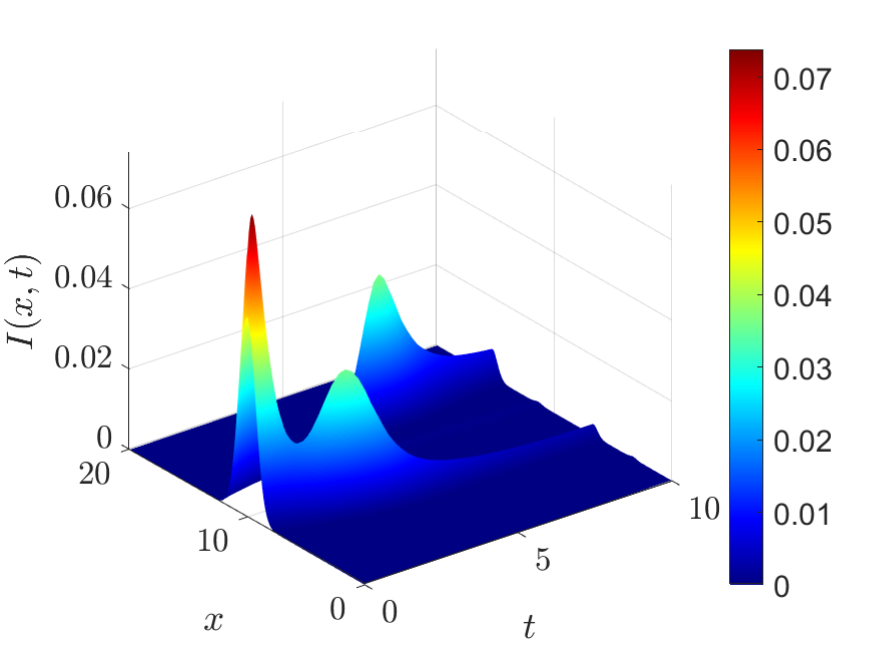}
\includegraphics[width=0.48\textwidth]{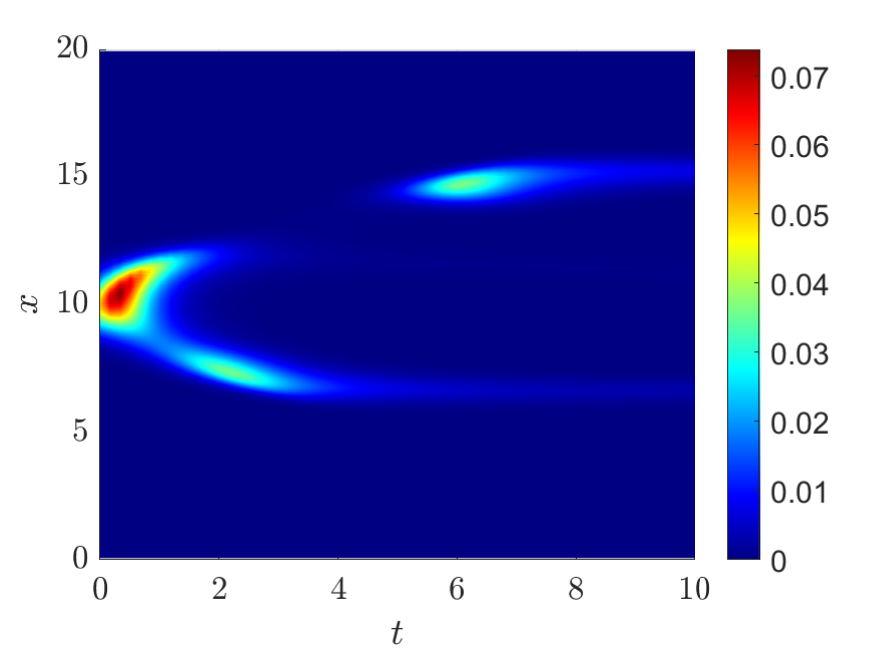}
\caption{Spread of an epidemic with spatially-heterogeneous transmission rate with model \eqref{eq.SIR_kinetic} in a hyperbolic configuration. Space-time evolution of susceptible population (top row) and infectious individuals (bottom row).}
\label{fig.SIhyp}
\end{figure}

\begin{figure}[t!]
\centering
\includegraphics[width=0.48\textwidth]{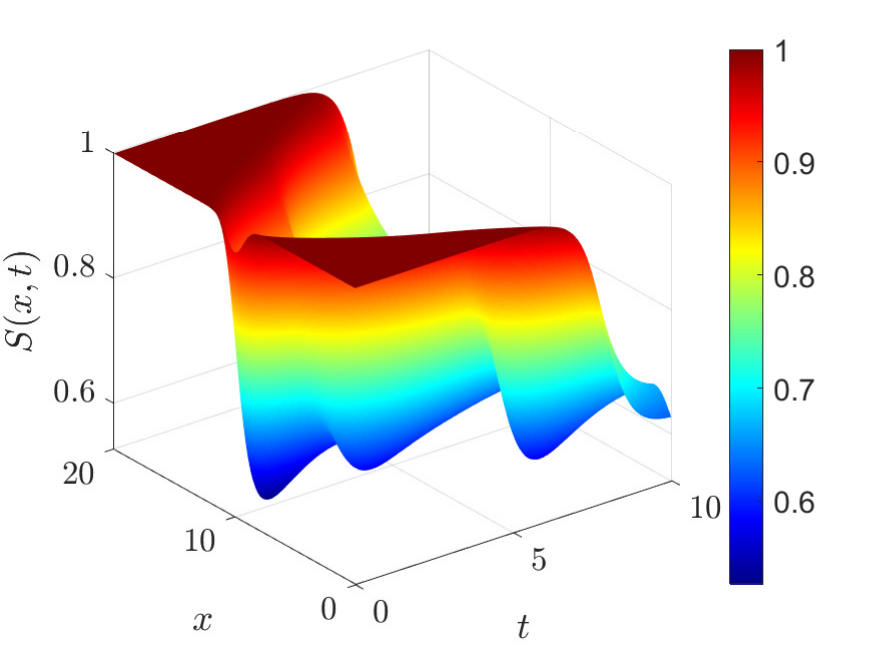}
\includegraphics[width=0.48\textwidth]{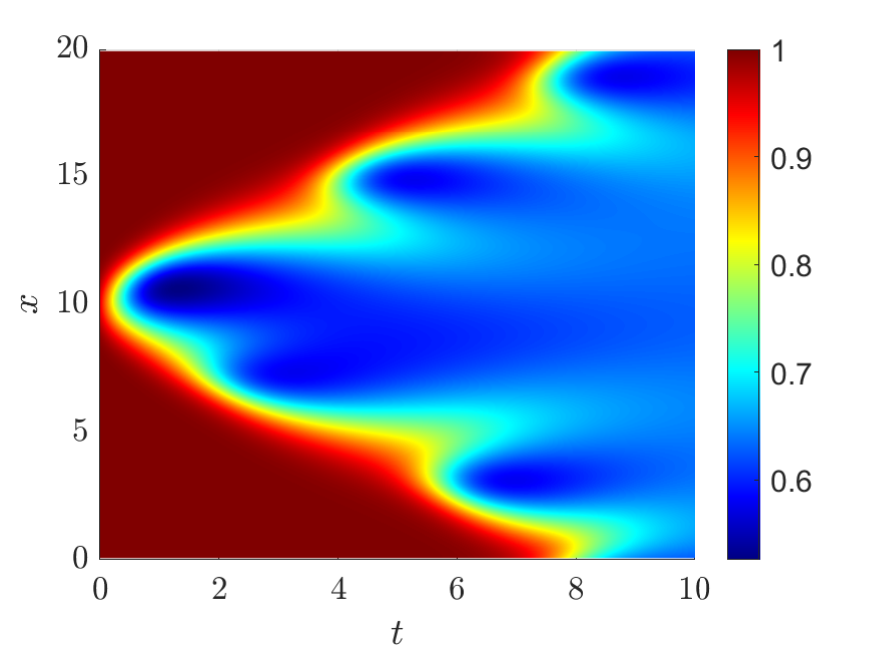}
\includegraphics[width=0.48\textwidth]{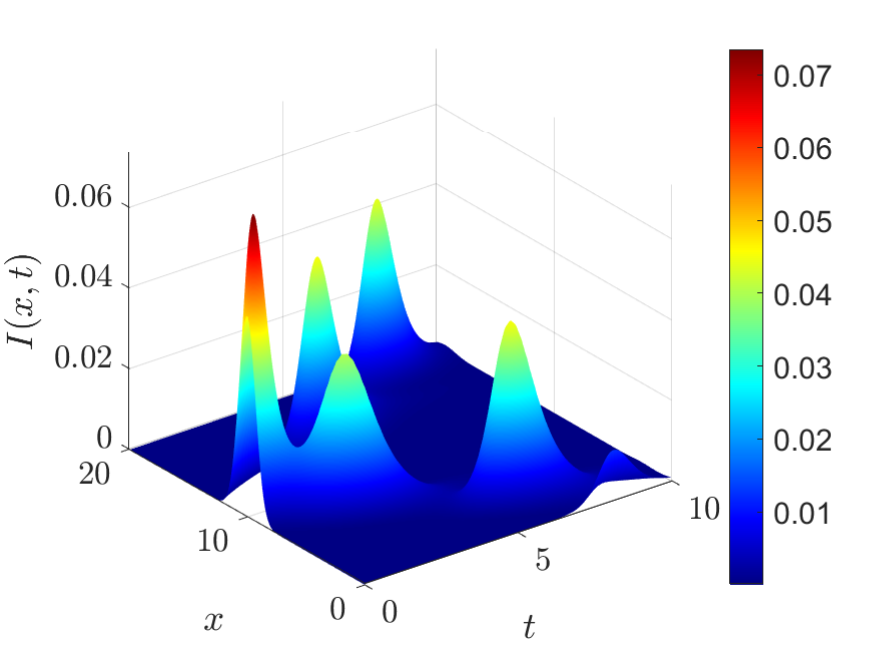}
\includegraphics[width=0.48\textwidth]{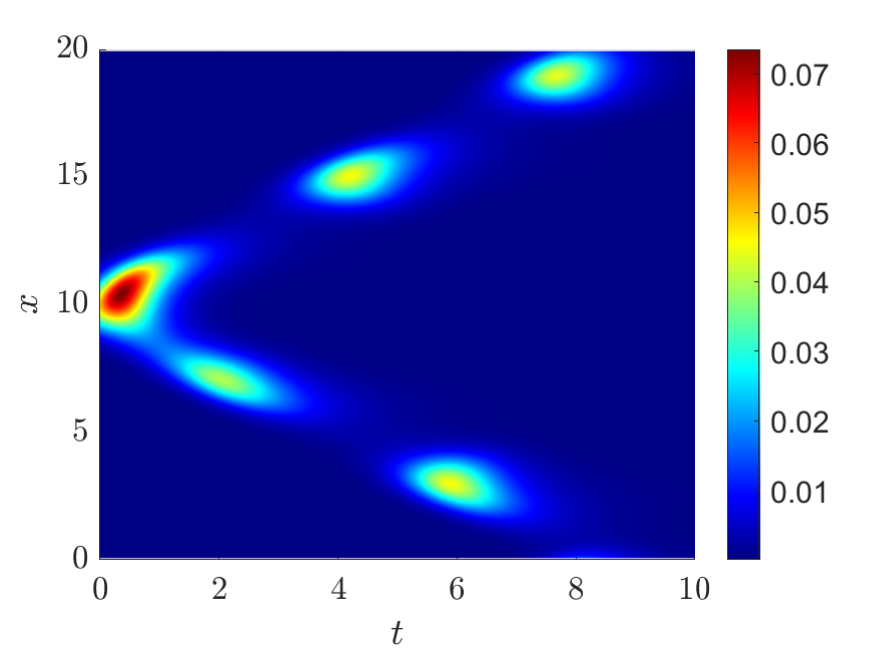}
\caption{Spread of an epidemic with spatially-heterogeneous transmission rate with model \eqref{eq.SIR_kinetic} in a parabolic configuration. Space-time evolution of susceptible population (top row) and infectious individuals (bottom row).}
\label{fig.SIpar}
\end{figure}
%%%%%%%%%%%%%%%%%

To numerically solve the multiscale system of equations \eqref{eq.SIR_kinetic}, we employ an Asymptotic-Preserving (AP) Implicit-Explicit (IMEX) Runge Kutta scheme in time coupled with a Finite Volume method for the space discretization\cite{[BP21],Bert}.
In the context of this paper, we simply mention the importance of the AP property of the chosen numerical method, which is fundamental for solving multiscale systems in general. Indeed, this property makes the numerical scheme consistent with the asymptotic limits of the system of equations of interest. In addition, the usage of an IMEX scheme prevents the choice of computational time step from being limited by the magnitude of the scaling parameter\cite{BPR,GT2002,JPT,NP}.

In Figures \ref{fig.SIhyp} and \ref{fig.SIpar} we present an example of numerical results obtained simulating the spread of an epidemic considering either a hyperbolic configuration of relaxation parameters ($\tau_J=1$, $J \in \{S,I,R\}$) and characteristic velocities ($\lambda_S^2 =\lambda_I^2 = 0.5$, $\lambda_I^2 = 0.1$) or a fully parabolic configuration ($\tau_J = 0$, $\lambda_J = \infty$, $J \in \{S,I,R\}$). In these tests, we use a bi-linear incidence function, fixing $p=1$ and $\kappa=0$, accounting for a spatially variable transmission rate that follows a sinusoidal distribution in space, being $\beta_T(x) = \beta_0\left(1-0.1\sin\left(0.5\pi x\right)\right)$ with $\beta_0 =13$. Finally, we set $\gamma_I = 10$ in the whole domain. From these plots, one can appreciate the different propagation dynamics determined by the choice of scaling parameters, observing how much more rapidly and aggressively the epidemic spreads through the territory in a configuration of parabolic regime (Fig. \ref{fig.SIpar}). In contrast, in the hyperbolic case the spread of the virus remains slightly confined to a central area of the domain (Fig. \ref{fig.SIhyp}).

%%%
\subsection{Characterization of the non-commuting population}
The modeling presented in the previous section takes into account only those individuals who move over long distances around the domain. However, the daily routine of a group of individuals that makes up an entire society is the result of a mix of very different habits: while a part of the population turns out to be commuting every day, for example for work reasons, moving from the city of domicile to the city of work, a conspicuous part of the population is not in the habit of frequently leaving their residence, staying mostly at home or moving only within their neighborhood (think, for example, of elderly individuals).
Therefore, in order to make the modeling more realistic, it is evidently necessary to include a proportion of individuals with zero mobility, i.e., ``non-commuting" individuals (which we will distinguish from those who instead commute in the spatial domain), who do not act on the epidemic transport mechanism but nevertheless play a role in the spread of the virus.
%The presence of a group of non-commuting agents in the modeling also prevents the entire population from moving indiscriminately throughout the spatial domain, resulting in a decidedly implausible mass migration effect \cite{Bert3,[BP21]}.

To account for stationary agents, starting from the framework of system \eqref{eq.SIR_kinetic}, we introduce an additional subgroup of individuals for each epidemic compartment, re-defining
\be\label{eq.defSIR}
s = f_S^++f_S^-+f_S^0, \quad i = f_I^++f_I^-+f_I^0, \quad r=f_R^++f_R^-+f_R^0,
\ee
where $f_S^0$, $f_I^0$, and $f_R^0$ identify the non-commuting susceptible, infectious and removed individuals, respectively. This population is characterized by a velocity $\lambda_J=0$, $J \in \{S,I,R\}$, thus its dynamics is actually governed by a system of ODEs, which recalls the standard SIR model\cite{BertProc,[BP21]}:
\be
\begin{split}\label{eq.SIR0}
	\frac{d f_S^0}{d t} &= -F(f_S^0,i) ,\\
	\frac{d f_I^0}{d t} &= F(f_S^0,i) -\gamma_I f_I^0,\\
	\frac{d f_R^0}{d t} &= \gamma_I f_I^0 .
\end{split}
\ee
The system of equations described above, combined with system \eqref{eq.SIR_kinetic} and  definitions \eqref{eq.defSIR}, provides a new mathematical model capable of describing the space-time epidemic evolution of a society composed of both commuting and non-commuting individuals. From this model, in analogy to what presented in the previous section, it is possible to derive the macroscopic formulation and evaluate its asymptotic behavior, recovering again a parabolic diffusion limit. For the detailed analysis the reader is directed to Refs. \cite{BertProc,[BP21]}.

%%%
\subsection{Extension of the modeling to spatial network structures}
\begin{figure}[t!]
\centering
\includegraphics[width=0.40\textwidth]{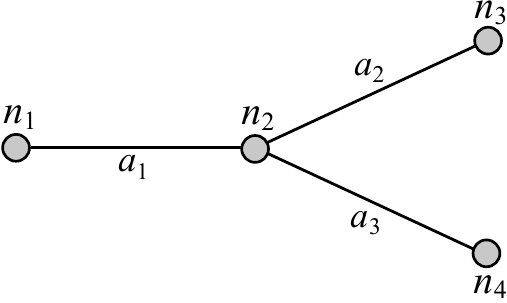}
\caption{Schematic representation of a network consisting of $N=4$ nodes ($n_1$, $n_2$, $n_3$, $n_4$) and $A=3$ arcs ($a_1$, $a_2$, $a_3$) in a Y configuration.}
\label{fig.network_scheme}
\end{figure}

The modeling presented in the previous sections can be extended to spatial networks similar to those discussed in Refs. \cite{Borsche2016,Borsche2018,Bretti2006,Bretti2014,Piccioli2021,Piccoli}.
% for different applications concerning the study of traffic flows, chemotaxis processes or blood blow dynamics.
The network structure that we consider in this context is that of a connected graph $\mathcal{G = (N,A)}$ formed by a finite set $\mathcal{N}=\{n_1,\ldots,n_N\}$ of $N$ nodes (or vertices) and a finite set $\mathcal{A}=\{a_1,\ldots,a_A\}$ of $A$ bi-directional arcs (or edges) having length $L_1,\ldots,L_A$ \cite{Piccoli}. A schematic representation of such a network structure is shown in Figure \ref{fig.network_scheme}. Each arc-node interface constitutes a junction of the network, which must be characterized by appropriate conditions able to guarantee the satisfaction of specific conservation properties of the system. These junction conditions depend on the modeling framework; still, they always arise from the solution of a Riemann problem at the interface, resorting to the Riemann invariants of the system, and applying the necessary conservation requirements.
For example, a different characterization of the type of node considered in the network results in the imposition of different transmission conditions at the junctions. Indeed, each node can be viewed as either an \emph{active} node, with its own localized dynamics, or a \emph{passive} node, which plays only an information hub role for related arcs.

%%%%%%%%%%%%%%%%%
%%%%%%%%%%%%%%%%%

According to Refs. \cite{[BP21]} and \cite{Bert}, a 1D epidemic network can be constructed by considering that network's (active) nodes represent localities of interest, such as cities, provinces, or, on a larger scale, regions, countries; while the network's arcs (which enclose the 1D spatial dynamics) represent the ensemble of paths connecting each locality to the others. As a result, the epidemic state of every node changes over time influenced by the movement of people who commute from other locations in the network.

At each junction, we impose conservation of total mass (population) and fluxes. In particular, the distribution of the flux at each branching between the various outgoing arcs is managed by ad-hoc $\alpha$ coefficients, always set respecting the conservation of the total flux at the junction \cite{[BP21],Bert}.
Then, the solution of the Riemann problem arising from the discontinuity of variables at the interface is performed using the Riemann invariants of system \eqref{eq.SIR_kinetic}, which correspond to the kinetic variables
\[
f_S^{\pm} = \frac1{2} \left(s_c \pm \frac{j_S}{\lambda_S}\right), \quad
f_I^{\pm} = \frac1{2} \left(i_c \pm \frac{j_I}{\lambda_I}\right), \quad
f_R^{\pm} = \frac1{2} \left(r_c \pm \frac{j_R}{\lambda_R}\right),
\]
where $s_c = f_S^++f_S^-$, $i_c = f_I^++f_I^-$, and $r_c = f_R^++f_R^-$ are the macroscopic densities of commuters, expressed for each epidemic compartment.

In addition, it must be taken into account that nodes located at the inlet/outlet of the domain do not present ingoing/outgoing arcs. At these nodes, to ensure that there are no individuals entering or leaving the network (thus to ensure the preservation of the total population), the classical no-flux boundary condition is applied.
For more details on the derivation and imposition of junction conditions in network epidemic modeling the reader can directly refer to Ref. \cite{Bert}.

\begin{remark}
Network structures modeling the spatial spread of an epidemic are quite common in the literature (see Refs. \cite{[GBM20],Malizia} and references therein). These models are based on systems of ODEs, which describe the epidemic evolution of each node (region). These nodes are influenced by mobility mechanisms, with individuals traveling between them via connecting arcs. However, the arcs of the network do not possess their own evolutionary dynamics. Essentially, the transport mechanism of individuals is not described but is instead replaced by appropriate node-to-node transmission conditions (fluxes) of individuals based on commuting matrices.
\end{remark}

%%%%%%%%%%%%%%%%%
\begin{figure}[t!]
\centering
\includegraphics[width=0.48\textwidth]{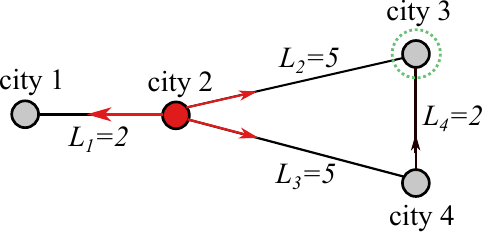}
\includegraphics[width=0.48\textwidth]{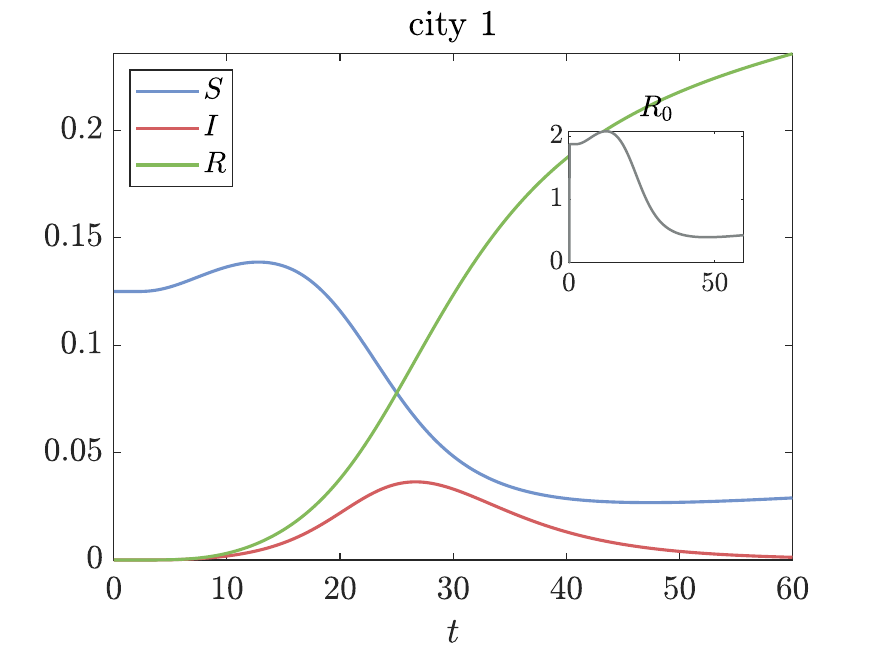}
\includegraphics[width=0.48\textwidth]{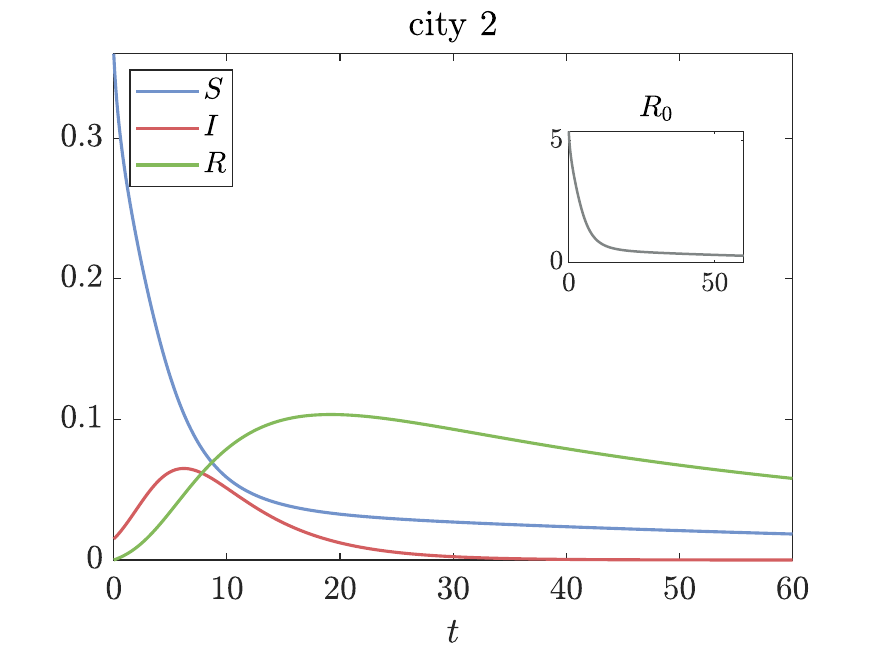}
\includegraphics[width=0.48\textwidth]{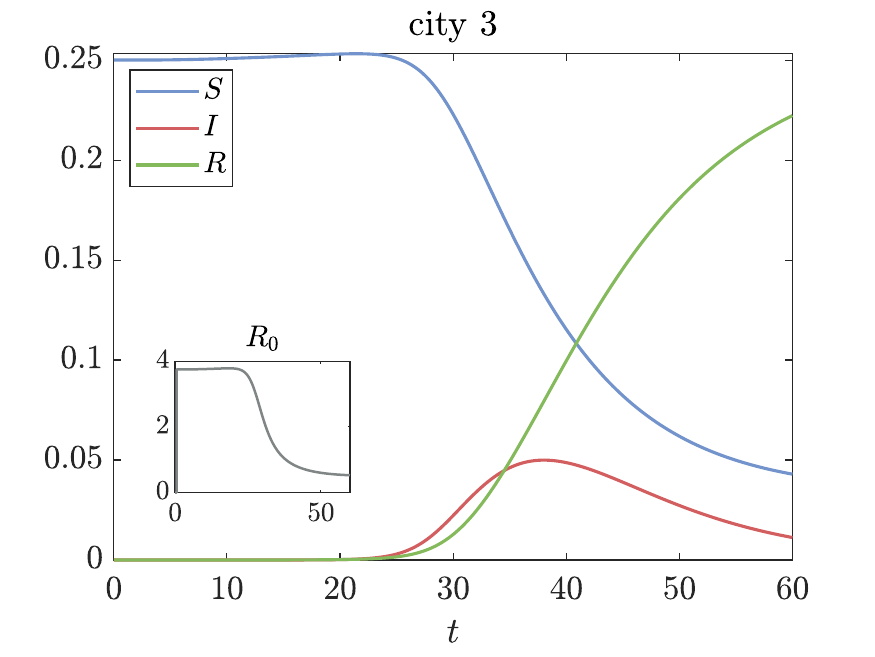}
\includegraphics[width=0.48\textwidth]{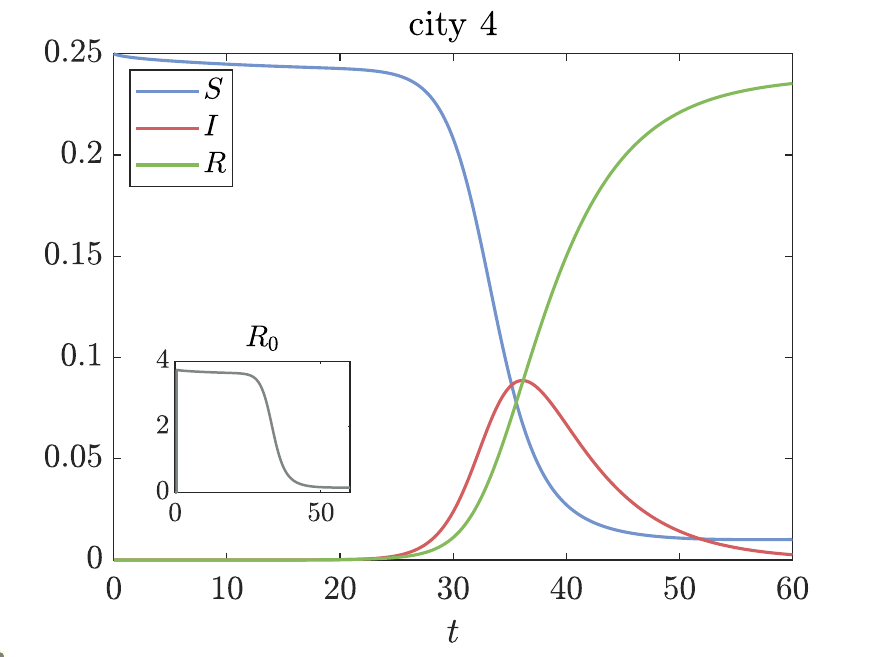}
\includegraphics[width=0.48\textwidth]{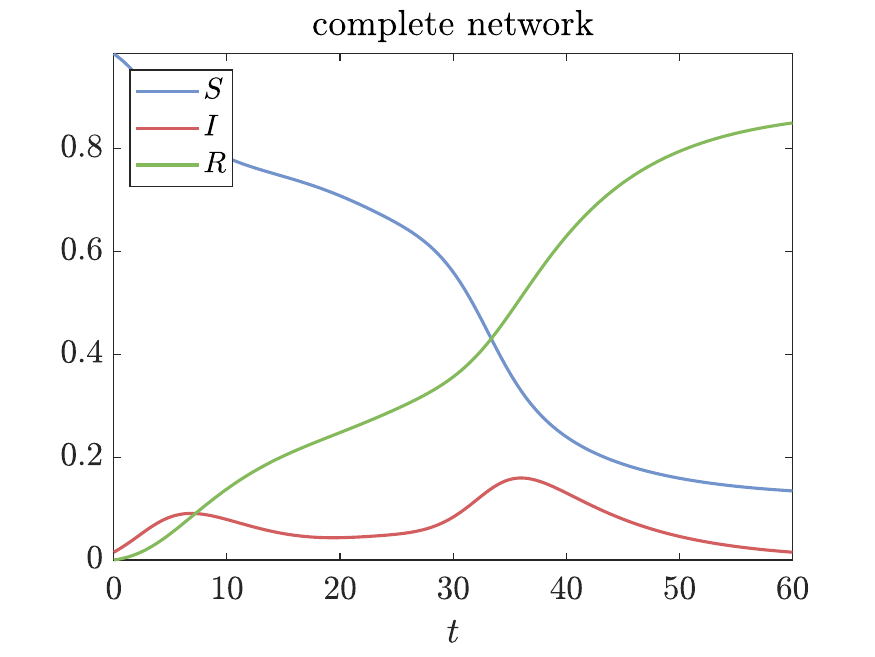}
\caption{Numerical test simulating the spread of an epidemic accounting for a spatial network structure and model \eqref{eq.SIR_kinetic}. The virus starts spreading from city 2 to the rest of the areas. The temporal evolution of the epidemic is presented for each city (node) of the network together with that of the reproduction number $R_0(t)$, as well as in cumulative sense for the whole domain.}
\label{fig.testY}
\end{figure}
%%%%%%%%%%%%%%%%%
To make the idea of the network structure clearer, we present an illustrative numerical test in which we consider a population consisting only of commuting individuals who move between four cities connected to each other according to the network shown in the top-right plot of Figure \ref{fig.testY}. At the beginning of the simulation, only in city 2 there are infected individuals, while in the rest there are only susceptible people. City 2 is the most populated in the network, having 37.5\% of the entire population (among which 4\% is infected at the initial time), while the lowest density is in city 1, with 12.5\% of population. Cities 3 and 4 present the same density of individuals (25\% of population), but in city 3 we consider a parameter of social distancing $\kappa = 1$, while in the rest of the network $\kappa=0$. Furthermore, we fix $\beta_T = 0.15$ and $\gamma_I= 0.2$ everywhere. Finally, we impose $\lambda_J = 1$, $J=\{S,I,R\}$ in the whole network, while we set in all the nodes $\tau_J = 10^{-3}$ and in the arcs $\tau_J = 10^{-1}$.

From Figure \ref{fig.testY}, it can be seen that, due to the movement of the agents, the epidemic begins to spread from city 2 to city 1 very rapidly, and then also reaches cities 3 and 4 shortly thereafter. However, although in cities 3 and 4 the virus begins to spread at the same time, it can be seen that the propagation remains much more contained in city 3 due to the $\kappa>0$ parameter.

%%%
\subsection{Accounting for parameter sensitivity and data uncertainty}
It is evident that the enrichment of dynamics given by moving from purely temporal modeling through ODEs to spatio-temporal modeling described by PDEs results in greater model complexity associated with an increase in the number of parameters involved. Similarly, extending the modeling to additional epidemic compartments, moving from a classical SIR partitioning of the population to more complex ones, such as those proposed in Refs. \cite{[GBM20],Giordano,Eli,PZ20,Parolini21} to study the first waves of COVID-19 by exploiting as much as possible the available data (taking into account exposed, pre-symptomatic, asymptomatic, hospitalized, quarantined, deceased, and so on), also leads to greater complexity and additional parameters that require a delicate pre-processing calibration phase.

In parallel to this, one can hardly overlook the inevitable uncertainty in the data and measurements used to calibrate and validate mathematical models and numerical simulations. In the specific case of epidemiological modeling, it is well known, for instance, that the actual number of infected individuals is always underestimated by official resources\cite{APZ,APZ2,DellaRossa,[GBM20]}. This is due in part to the lack of a sufficiently structured screening process for managing pandemics, but also to the difficulties in identifying and tracking asymptomatic or weakly infected individuals\cite{Buonomo,Giordano,Peirlinck}. Similar considerations concern the difficulties in estimating the contact rate, the incubation rate and the same infectious period.

Therefore, the application of effective computational methods to quantify the impact of parametric fluctuations on numerical solutions appears necessary for proper interpretation of the results. From a mathematical viewpoint, one introduces a vector of uncertain input parameters $\mathbf{z} = (z_1, \ldots, z_{d})^T \in \Omega_z \subseteq \mathbb{R}^{d}$, where $z_k$, $k = 1, \ldots, d$ is a random variable with some known or unknown distribution. As a consequence, the solution $u$ of the model under investigation depends not only on the physical variables but also on the random vector, i.e., $u = u(x,t;\mathbf{z})$.

There is a vast literature on uncertainty quantification (UQ) methods for systems of ODEs and PDEs, with recent years seeing a notable increase in works on the subject\cite{BLMS,HI,Pettersson,Poette,Xiu}. See also Ref. \cite{PareschiUQ} for a recent introduction in the case of kinetic equations.
Roughly speaking, UQ methods for PDEs fall into two categories: \emph{intrusive} and \emph{non-intrusive}. Intrusive methods involve modifying models to incorporate uncertain inputs, which can be complex and challenging, particularly for systems with multidimensional uncertainty\cite{JP,Pettersson,Xiu}. Furthermore, intrusive approaches for hyperbolic systems can also cause the loss of important structural properties of the original problem, like hyperbolicity, well-balancing, positivity preservation, and consistency of the behavior at large times\cite{GHI,Iacomini,Poette}. On the other hand, non-intrusive methods, like Monte Carlo (MC) and stochastic Collocation (sC), do not require altering deterministic codes, making them more suitable for certain systems\cite{JP,Xiu}. One of the main problems with MC methods is that the solution statistics have a slow rate of convergence. Different techniques have been developed to accelerate their convergence, like Multi-Level Monte Carlo\cite{DP,Mishra,PTZ} and Multi-Fidelity methods\cite{BertBiFi,DLPZ,Liu,LPZ,Zhu}.
In contrast, sC offers high accuracy with exponential convergence rates, but requires the knowledge of the probability density function (PDF) of the uncertain inputs and faces computational challenges with increasing dimensionality (known as the \emph{curse of dimensionality}). Sparse grid techniques alleviate this issue, enabling efficient handling of numerous random inputs\cite{Mishra1,XH}.

An example of the development and application of a Bi-Fidelity method to account for uncertain inputs in kinetic epidemic models in line with the framework discussed in Section \ref{Sec:5.1} can be found in Ref. \cite{BertBiFi}. The proposed method gain efficiency by making use of the correlations in the random space of a surrogate low-fidelity model (cheap to evaluate), to effectively inform the selection of representative points in the parameter space and then uses this information to construct accurate approximations of the (expensive) high-fidelity solutions\cite{Liu,LPZ}. Therefore, the approach is based on the evaluation of a high-fidelity model on a small number of samples, appropriately selected from many evaluations of a low-fidelity model. For the details of the method we direct the reader to Ref. \cite{BertBiFi}.

%%%%%%%%%%%%%%%%%
\begin{figure}[t!]
\centering
\includegraphics[width=0.55\textwidth]{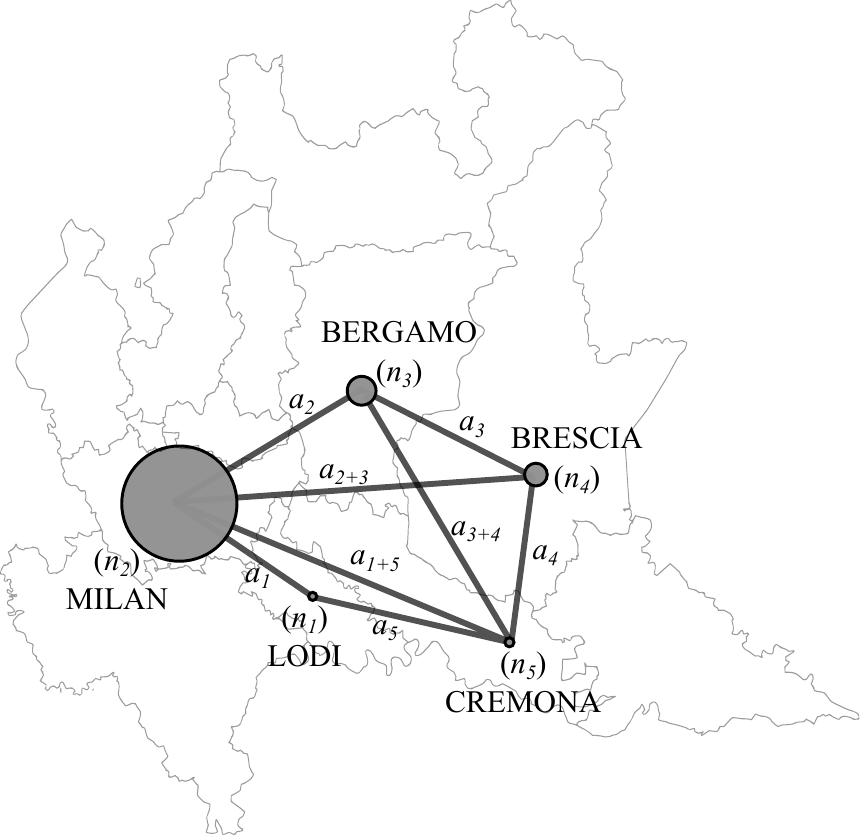}
\caption{Network structure adopted for the simulation of the outbreak of COVID-19 pandemic in the Italian Lombardy Region in 2020 (see Ref. \protect\cite{[BP21]}).}
%Extract from Ref. \protect\cite{[BP21]}.}
\label{fig.networkLombardia}
\end{figure}
%%%%%%%%%%%%%%%%%

In the following, we present an example of numerical results obtained using an sC technique when simulating the outbreak of COVID-19 in the Lombardy Region of Italy\cite{BertProc,[BP21]}. The methodology used couples the sC approach to an AP-IMEX Runge Kutta Finite Volume scheme. We leave the discussion of the details of the method to Refs. \cite{Bert4,[BP21]}. We underline that, given the non-intrusive nature of the sC method, the AP property of the deterministic IMEX scheme is preserved also in the stochastic framework.

%%%%%%%%%%%%%%%%%
\begin{figure}[tp!]
\centering
\begin{subfigure}[t]{0.48\textwidth}
\caption{Milan (network node $n_2$)}
\includegraphics[width=\textwidth]{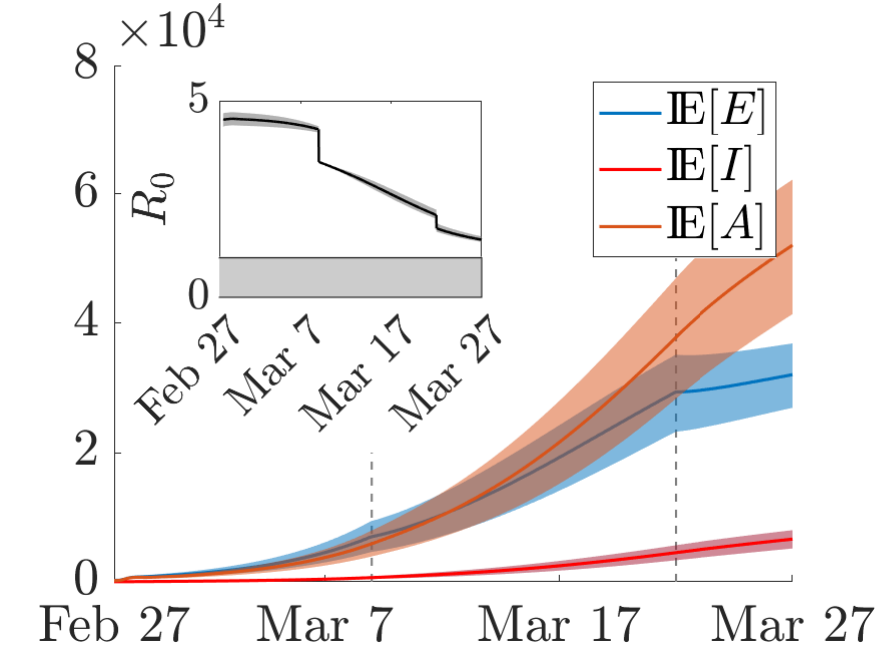}
\end{subfigure}
\begin{subfigure}[t]{0.48\textwidth}
\caption{Lombardy (entire network)}
\includegraphics[width=\textwidth]{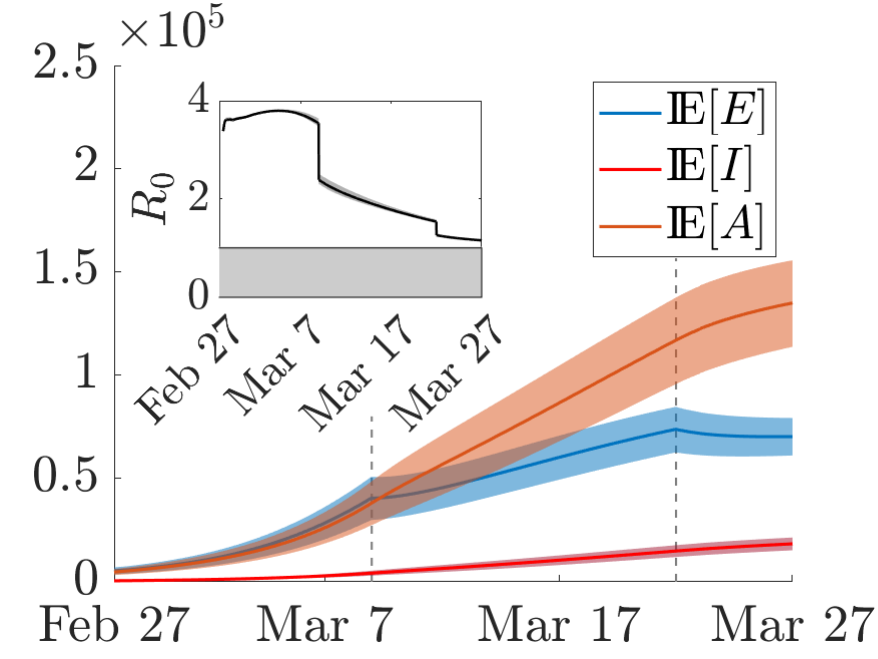}
\end{subfigure}
\includegraphics[width=0.48\textwidth]{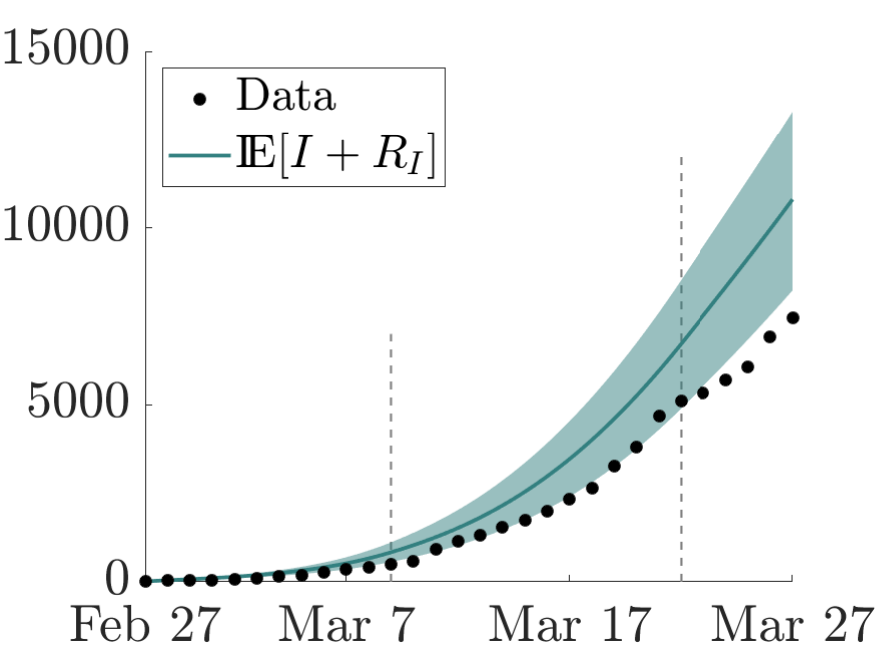}
\includegraphics[width=0.48\textwidth]{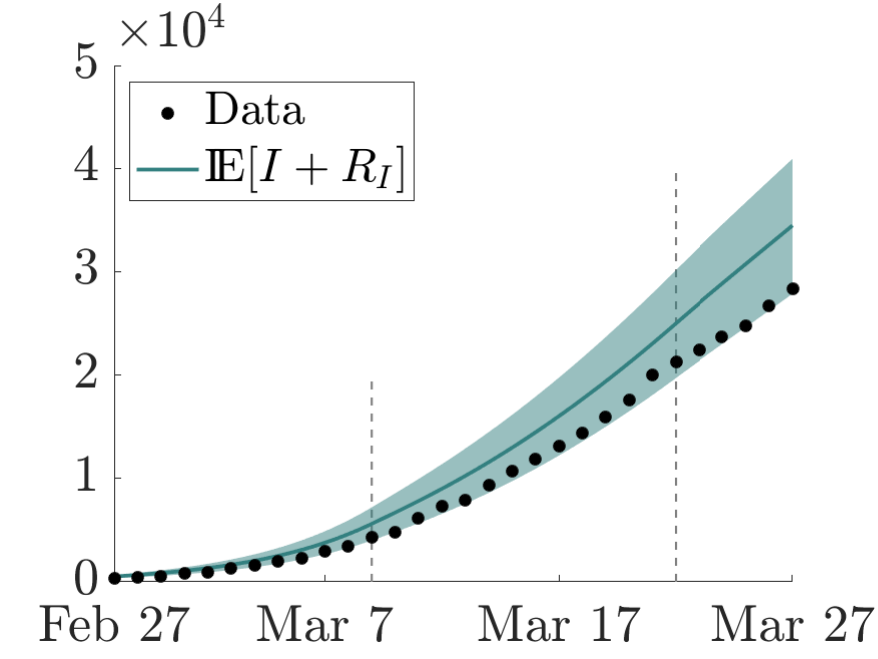}
\includegraphics[width=0.48\textwidth]{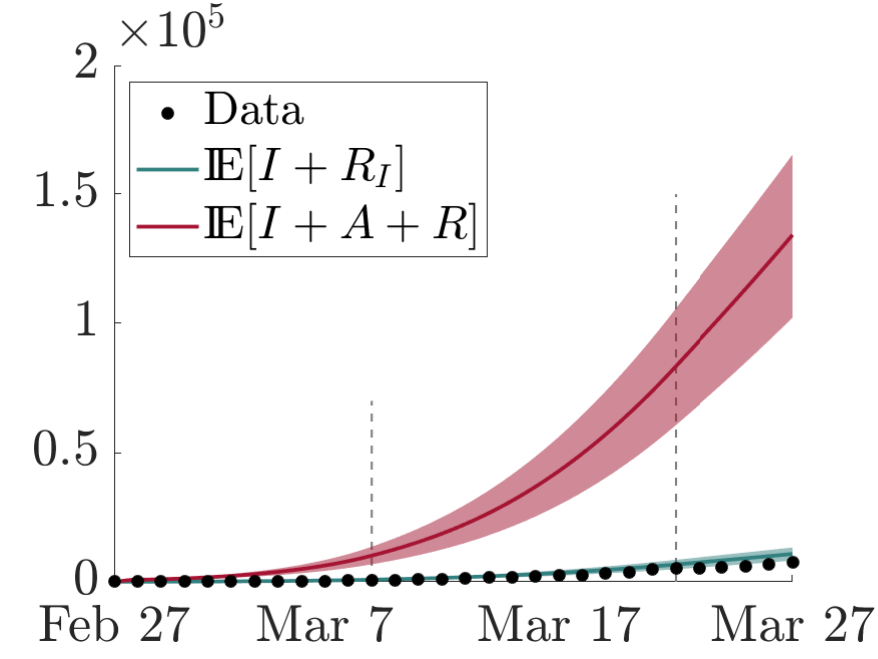}
\includegraphics[width=0.48\textwidth]{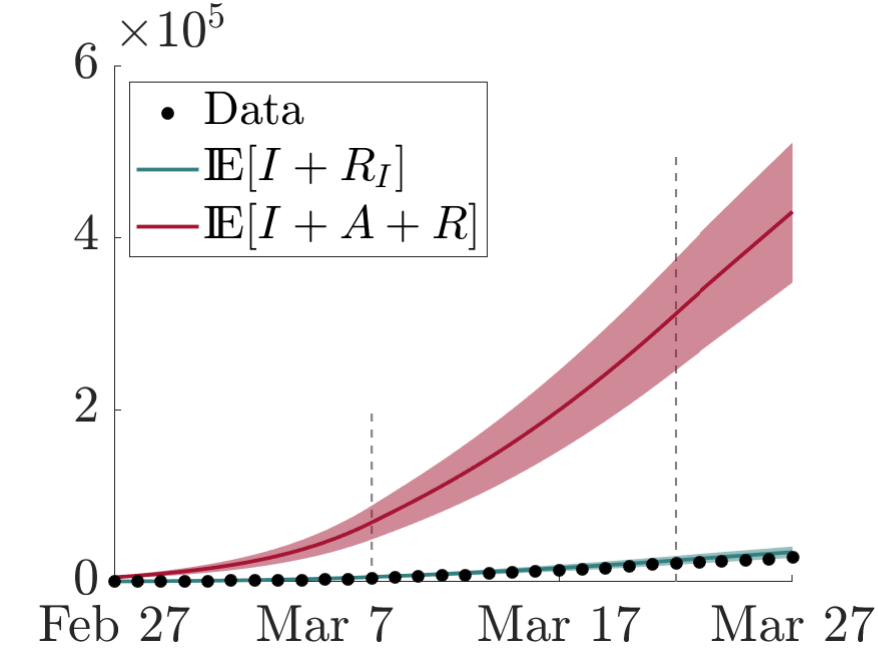}
\caption{Numerical simulation of the outbreak of COVID-19 pandemic in the Italian Lombardy Region in 2020 accounting for uncertain initial conditions and epidemic parameters. Evolution of the pandemic in the city of Milan (left column) and in the entire network (right column). We report the expected values together with 95\% confidence intervals. First row: evolution of exposed $E$, symptomatic $I$ and asymptomatic or weakly symptomatic $A$, associated with the temporal evolution of the reproduction number; second row: evolution of the cumulative number of symptomatic $I+R_I$; third row: evolution of the cumulative number of symptomatic $I+R_I$ versus that of all infected in the system $I+A+R$ (see Ref. \protect\cite{[BP21]}).
}
\label{fig.testLombardia}
\end{figure}
%%%%%%%%%%%%%%%%%

For this test, a SEIAR compartmental partitioning of the population is considered. To this aim we include in the modeling the compartment of exposed $E$, which are infected but not infectious, and we distinguish between infectious individuals $A$, who are infected but asymptomatic or mildly symptomatic, and highly symptomatic individuals $I$. Then, a network with five nodes is examined, representing the five main provinces that were affected by the pandemic spread in the first months of 2020: Lodi, Milan, Bergamo, Brescia, and Cremona (see Figure \ref{fig.networkLombardia}). Each arc's characteristic speed is set to allow a complete round journey in each origin-destination portion in a single day, and the relaxation time is chosen so that the model works in hyperbolic regime. Conversely, we assign to metropolitan areas, coinciding with the nodes of the network, a parabolic setting to accurately represent the diffusive nature of infectious propagation usually occurring in densely populated areas. Then, we set $\lambda_I = 0$ across all network nodes, based on the assumption that all subjects with severe symptoms should be identified and placed under quarantine.

The amount of total population in each province is given by data from the National Institute of Statistics of Italy. Furthermore, the transmission coefficients at the arc-node interfaces and the percentages of commuters belonging to each province are imposed by resorting to the official national assessment of mobility flows.
For the model calibration we use data reported by the Italian Civil Protection Department and an iterative process over the numerical method.
Due to the screening policy adopted in Italy during the first wave of COVID-19, we chose to
associate all detected infected individuals to the $I$ compartment, still considering that the tracking of positive individuals cannot be considered exempt from uncertainty. Thus, we introduce a random input with uniform distribution, $z \sim \mathcal{U}(0,1)$, and define initial conditions for compartment $I$, at each node, accounting for the underestimation of these values:
$ i(x,0,z) = i_0(1+z) $,
with $i_0$ density of infectious people at the beginning of the simulation as given by data recorded by the Civil Protection Department. In addition, also the transmission rate of the virus is considered affected by the uncertain input.
For further details on the setup of the test, see Ref. \cite{[BP21]}.

In Figure \ref{fig.testLombardia} we report the temporal evolution of the epidemic compartments, in terms of expected values with 95\% confidence intervals, in the sole Milan city (left) and in the whole network (right).
As can be seen from the second row of the Figure, the lower bound of the confidence band of the cumulative amount over time of severely symptomatic individuals is comparable with the data recorded by the Italian Civil Protection Department. However, it appears clearly that the value of infected individuals is consistently underestimated by the data. This can be observed from the comparison of the expected evolution over time of the cumulative amount of infected $I$ (severely symptomatic) versus the actual cumulative amount of total infectious persons, including also the $A$ compartment, shown in the third row of the Figure. These plots confirm the impact that the presence of asymptomatic or undetected individuals had on the first wave of the COVID-19 pandemic.

%%%
\subsection{Research perspectives about spatial spread of epidemics}
Research on the spatial spread of epidemics has evolved significantly in recent years due to advances in data collection, computational methods, and interdisciplinary collaboration. Understanding how diseases spread across geographic regions is critical for effective public health interventions, resource allocation, and policy formulation.
Emerging research leverages several approaches, including mathematical modeling using PDEs, network analysis, and also machine learning, to capture the complex interplay between population dynamics, mobility patterns, and environmental factors. Among the many possible research directions we would like to mention two main lines of development in this concluding section.

\vskip.2cm \noindent $\bullet$ \textit{Coupling spatial and social structures.}
A first natural extension of the modeling discussed in the last sections is to couple spatial dynamics with the heterogeneous social structure of the population discussed in Section \ref{Sec:4}. For the simple SIR partitioning this corresponds to blending \eqref{eq.SIR_kinetic} with \eqref{eq:kin-main}, by defining new statistical density variables $f_S^\pm=f_S^\pm(x,w,t)$, $f_I^\pm=f_I^\pm(x,w,t)$, and $f_R^\pm=f_R^\pm(x,w,t)$, which characterize the individuals moving in the different directions in the one-dimensional domain $\Omega$ and possessing the social feature $w \in \mathcal{I}$. Clearly, this extension would allow the two starting models to be recovered through the definition of the marginal distributions. In particular, definitions used in \eqref{eq:kin-main} can be obtained by integrating in space, being
\[f_J(w,t) = \int_\Omega (f_J^+(x,w,t)+f_J^-(x,w,t))\, dx, \qquad J \in \{ S,I,R\};\]
while densities \eqref{eq.macro-densities} would result from integration in $\mathcal{I}$:
\begin{equation*}
\begin{split}
s(x,t)=\int_\mathcal{I}(f_S^+(x,w,t)+f_S^-(x,w,t))\,dw,\\
i(x,t)=\int_\mathcal{I}(f_I^+(x,w,t)+f_I^-(x,w,t))\,dw,\\
r(x,t)=\int_\mathcal{I}(f_R^+(x,w,t)+f_R^-(x,w,t))\,dw.
\end{split}
\end{equation*}

This coupling would provide a more holistic understanding of epidemic spread and would greatly enhance our ability to develop effective strategies for disease control and prevention. In fact, from one side, incorporating social features allows for a more realistic representation of mobility patterns, better capturing the complexities of human interactions and their impact on the  propagation of viruses. On the other hand, coupled spatial-social models can help identify potential disease hot-spots by considering not only geographical locations but also other individual features like the presence of the so-called super spreaders or social vulnerabilities, permitting to prioritize surveillance and prevention efforts accordingly.

%to get
%\be
%\begin{split}
%	\frac{\partial f_S^{\pm}}{\partial t} \pm \lambda_S \frac{\partial f_S^{\pm}}{\partial x} &= - -K_T(f_S^\pm,i) \mp \frac{1}{2\tau_S}\left(f_S^+ - f_S^-\right)+\frac 1{\tau_S} Q_S(f_S^\pm)(\cc,t),\\
%	\frac{\partial f_I^{\pm}}{\partial t} \pm \lambda_I \frac{\partial f_I^{\pm}}{\partial x} &=  K_T(f_S^\pm,i) -\gamma_I f_I^{\pm} \mp \frac{1}{2\tau_I}\left(f_I^+ - f_I^-\right)+\frac 1{\tau_I} Q_I(f_I^\pm)(\cc,t),\\
%	\frac{\partial f_R^{\pm}}{\partial t} \pm \lambda_R \frac{\partial f_R^{\pm}}{\partial x} &= \gamma_I f_I^{\pm} \mp \frac{1}{2\tau_R}\left(f_R^+ - f_R^-\right)+\frac 1{\tau_R} Q_R(f_R^\pm)(\cc,t).									
%\end{split}	
%\ee
%where now $f_S^\pm=f_S^\pm(x,w,t)$, $f_I^\pm=f_I^\pm(x,w,t)$, and $f_R^\pm=f_R^\pm(x,w,t)$ characterize the individuals moving in the different directions and possessing the social feature $w \in I$.

\vskip.2cm \noindent $\bullet$ \textit{Advancing data-driven modeling.}
%As emphasized throughout the paper, computational simulations of epidemic dynamics play a key role in informing public health strategies and policy decisions. However,
Another line of development is that related to the need to align models with observations taking into account the limited quantity and quality of data provided by official sources\cite{Eli1,Piazzola,Ziarelli}.
A data-driven perspective is a crucial step to understand the emerging features of the dynamic and to match models parameters  with real-world observations\cite{WE}.
%The inherent complexities of this process highlight the need for innovative, data-driven approaches \cite{WE}.
In this context, physics-informed neural networks (PINNs) has emerged as a promising solution to bridge the gap between differential models and collected data\cite{Raissi}. Indeed, PINNs aim at merging physics-based knowledge of the phenomenon under study and machine learning techniques, in which known governing principles of the differential model are integrated into the learning process\cite{Karniadakis,Raissi}.

More specifically, let us define a neural network representation of a spatio-temporal dynamics as $u_{NN}(x,t;\theta)$. To find the optimal values for the neural network's weights and bias and the model parameters that need to be inferred, whose set is uniquely identified by $\theta$, the neural network is trained by minimizing the following type of loss function:
\be
\label{eq:general-pinn-loss}
\mathcal{L}(\theta) = w_d^T \mathcal{L}_{d}(\theta) +  w_r^T \mathcal{L}_{r}(\theta) + w_b^T \mathcal{L}_{b}(\theta).
\ee
Here, $\mathcal{L}_{r}$ and $\mathcal{L}_{b}$ quantify the discrepancy of the neural network surrogate with the underlying differential system and its initial and/or boundary conditions, respectively. Furthermore, $\mathcal{L}_{d}$ measures the mismatch with respect to data, when observations are available. Finally, $w_{r}$, $w_{b}$, $w_{d}$ are the weight vectors corresponding to each loss term.
After finding the optimal set of parameters $\theta^*$ by minimizing the PINN loss \eqref{eq:general-pinn-loss}, i.e. solving the problem $\theta^* = \argmin \mathcal{L}(\theta)$ (normally recurring to stochastic gradient based techniques), the neural network surrogate $u_{NN}( x,t; \theta^*)$ can be evaluated at any given spatio-temporal point to get the complete approximate solution\cite{BertAPNNChap}.
The structure of the loss term \eqref{eq:general-pinn-loss}
 allows the neural network to assimilate both explicit physical laws and available observational data, ensuring a more informed learning process and enabling it to effectively address not only the inverse problem of parameter estimation, but also the forward problem of predicting future developments in the dynamics being investigated\cite{Biazzo,Kharazmi,Ning}.

%%%%%%%%%%%%%%%%%
\begin{table}[b!]
    \centering
    \begin{tabular}{@{}ccccc@{}}\toprule
    Parameter & Ground Truth & Initial Guess & Estimation & Relative Error \\ \hline
    $\beta_0$ & 9.0 & 5.0 & 9.0170 & $1.89\times 10^{-3}$ \\
    $\beta_1$ & 2.5 & 1.5 & 2.4512 & $1.95\times 10^{-2}$ \\
    $\zeta$ & 0.55 & 0.5 & 0.5508 & $1.45\times 10^{-3}$ \\
    \bottomrule
    \end{tabular}
    \caption{Spatial spread of an epidemic with heterogeneous transmission rate in the territory in diffusive regime ($\lambda^2_J =10^5$, $\tau_J=10^{-5}$, $J \in \{S,I,R\}$). Inferred results for the three different unknown coefficients in the incidence function and relative error with respect to the correct solution. }
    \label{tab.testPINN}
\end{table}
%%%%%%%%%%%%%%%%%
\begin{figure}[tp!]
%\begin{subfigure}[t]{0.48\textwidth}
%\caption{Ground truth}
\includegraphics[width=0.48\textwidth]{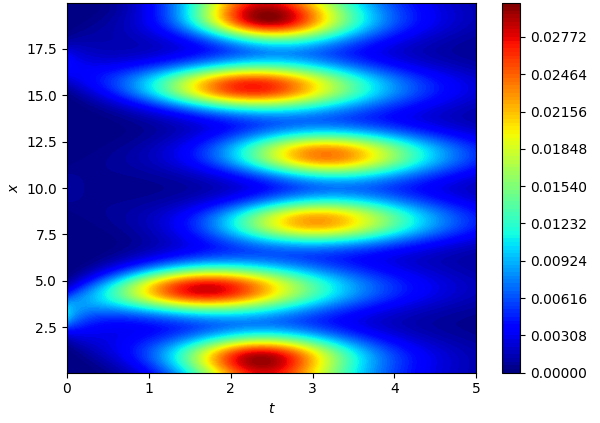}
%\end{subfigure}
%\begin{subfigure}[t]{0.48\textwidth}
%\caption{APNN forecast}
\includegraphics[width=0.48\textwidth]{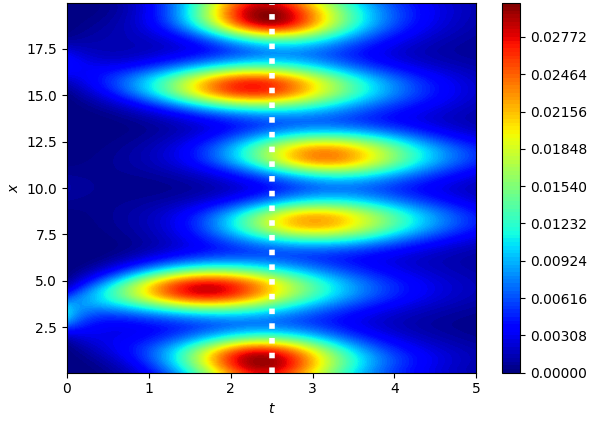}
%\end{subfigure}
%\centering
%\begin{subfigure}[t]{0.48\textwidth}
%\caption{$s(x,t)$ - ground truth}
%\includegraphics[width=\textwidth]{figure/GB/test3_parabolic_scattered_S_true}
%\end{subfigure}
%\begin{subfigure}[t]{0.48\textwidth}
%\caption{$i(x,t)$ - ground truth}
%\includegraphics[width=\textwidth]{figure/GB/test3_parabolic_scattered_I_true}
%\end{subfigure}
%\begin{subfigure}[t]{0.48\textwidth}
%\caption{$s(x,t)$ - APNN reconstruction}
%\includegraphics[width=\textwidth]{figure/GB/test3_parabolic_scattered_S_pred}
%\end{subfigure}
%\begin{subfigure}[t]{0.48\textwidth}
%\caption{$i(x,t)$ - APNN reconstruction}
%\includegraphics[width=\textwidth]{figure/GB/test3_parabolic_scattered_I_pred}
%\end{subfigure}
%\begin{subfigure}[t]{0.48\textwidth}
%\caption{$s(x,t)$ - APNN forecast}
%\includegraphics[width=\textwidth]{figure/GB/test3_parabolic_extrapolation_S_pred}
%\end{subfigure}
%\begin{subfigure}[t]{0.48\textwidth}
%\caption{$i(x,t)$ - APNN forecast}
%\includegraphics[width=\textwidth]{figure/GB/test3_parabolic_extrapolation_I_pred}
%\end{subfigure}
\caption{Spatial spread of an epidemic with heterogeneous transmission rate in the territory in diffusive regime ($\lambda^2_J =10^5$, $\tau_J=10^{-5}$, $J \in \{S,I,R\}$). Spatio-temporal evolution of infectious individuals. Left: reference solution (ground truth); right: APNN forecast for $t\in[2.5,5]$ after sparse data training on $t\in [0,2.5]$.}
\label{fig.testPINN}
\end{figure}
%%%%%%%%%%%%%%%%%

However, the use of standard PINNs in the context of multiscale problems, as in the case of the epidemic system \eqref{eq.SIR_kinetic}, can still lead to incorrect inferences and predictions. This occurs because the equations defines small spatio-temporal scales (governed by the smallness of the scaling parameter), which result in reduced or simplified models that must be applied consistently throughout the learning process of the neural network. In these situations, only the dynamics at the leading order can be accurately described, because the PINN loses accuracy in the asymptotic limit regimes\cite{BertAPNNChap,BertAPNN,Shi,JMW}.
A possible remedy to this issue, as recently proposed in Refs. \cite{BertAPNN,JMW}, is to modify the loss function \eqref{eq:general-pinn-loss} to include AP properties during the training phase, thereby developing what has been called an asymptotic-preserving neural network (APNN).
%The actualization of this kind of AP loss function will therefore rely on the specific problem being addressed and be supported by a suitable asymptotic analysis of the model.
In essence, the residual term $\mathcal{L}_{r}(\theta)$ must permit the recovery of the asymptotic limits that the model presents in the continuum even at the level of the neural network\cite{BertAPNNChap,BertAPNN}. For the specific case of the spatial epidemic system discussed in Section \ref{Sec:5.1}, this coincides with enforcing the learning of the model written in macroscopic form, i.e. equations \eqref{eq.SIR_macro}.

To show the capabilities of the APNN proposed in Ref. \cite{BertAPNN} to analyze the spatio-temporal evolution of infectious diseases modeled by \eqref{eq.SIR_kinetic}, we report numerical results obtained considering a heterogeneous epidemic environment, characterized by the transmission rate
$\beta(x) = \beta_0 + \beta_1\sin\left(\zeta\pi x\right)$, in a context of scaling parameters that reproduce a diffusive regime ($\lambda^2_J =10^5$, $\tau_J=10^{-5}$, $J \in \{S,I,R\}$).
In this test, first, a relatively sparse availability of $1000$ sample measurements from synthetic data over the complete spatio-temporal domain is considered, aiming to infer unknown physical parameters $\beta_0, \beta_1$ and $\zeta$ with the APNN.
Secondly, the forecasting performance of the neural network in predicting the spread of the infectious disease from $t=2.5$ to $t=5$ is investigated.

The results of parameters inference based on the sparse measurements are shown in Table \ref{tab.testPINN}, where it can be noticed that the APNN accurately recovers the correct values of the unknown parameters characterizing the epidemic incidence function with satisfactory errors. Also the space-time APNN reconstruction of the densities of infectious people is in very good agreement with the reference solution (ground truth) even in predicting the epidemic spread outside the training region, as it can be observed from Fig. \ref{fig.testPINN}.

%\begin{remark}
%It is important to underline that these results could not have been obtained with a vanilla PINN. In fact, the asymptotic diffusive behavior is captured by the neural network only because an it has been used an APNN, trained considering an appropriate AP loss function, which allows it to remain consistent with the asymptotic behaviors of the model (see Refs. \cite{BertAPNNChap,BertAPNN}).
%\end{remark}

Further steps forward can be achieved by integrating data uncertainty into the context of APNNs, considering for instance Ref. \cite{Zhang2019}. Of particular interest is also the transition to modeling Asymptotic-Preserving DeepONets proposed in Ref. \cite{JinDeepONet}. In this paper, the concept of learning operators (particularly differential operators) from the available data through a parametric approach, introduced in Ref. \cite{DeepONet} with the definition of ``Deep Operator Networks'' (DeepONet), is extended to the case of multiscale systems. Indeed, the DeepONet architecture appears to be particularly useful for studying scenarios in which the underlying dynamics are described by complex differential operators that cannot be easily postulated and when data is limited or noisy, as is often the case with epidemic dynamics.

%%%%%%%%%%%%%%%%%%%%%%%%%%%%%%%%%%%%%%%%%%%%
\section{Conclusion}\label{Sec:6}
%%%%%%%%%%%%%%%%%%%%%%%%%%%%%%%%%%%%%%%%%%%%

This paper proposed a critical overview on multiscale methods to describe an epidemic by SARS-CoV-2. The presentation starts from the three key questions posed in Section \ref{Sec:1} and has been developed consistently with the philosophy proposed in Section \ref{Sec:2}, which motivates how the study should take into account the needs of our society and the specific features of the territory where the epidemic develops.

The answers to the above questions are provided, sequentially, in Section \ref{Sec:3}, focusing on in-host dynamics, Section \ref{Sec:4}, focusing on the interactions between social dynamics and epidemics, and Section \ref{Sec:5} regarding the spread across the territory at the regional level. In fact, the study of Section \ref{Sec:3} can contribute to Section \ref{Sec:4}, while Section \ref{Sec:5} can also integrate with what discussed in the previous sections.

At the end of each section, some considerations on research perspectives were discussed. We trust that the set of contents presented can lead to a new view of epidemics that goes beyond deterministic population dynamics to improve our understanding and prediction of viral spread dynamics, as well as renew their control and prevention.

%%%%%%%%%%%%%%%%%%%%%%%%%%%%%%%%%%%%%%%%%%%%
\section*{Acknowledgment}
%%%%%%%%%%%%%%%%%%%%%%%%%%%%%%%%%%%%%%%%%%%%
This work has been written within the activities of GNCS and GNFM groups of INdAM (Italian National Institute of High Mathematics).
GB has been partially funded by the call ``Bando Giovani anno 2023 per progetti di ricerca finanziati con il contributo 5x1000 anno 2021'' of the University of Ferrara,
and by the European Union -- NextGenerationEU, MUR--PRIN 2022 PNRR Project No. P2022JC95T ``Data-driven discovery and control of multi-scale interacting artificial agent systems''.
AB acknowledges the support from the MUR--PRIN2020 Project No. 2020JLWP23 ``Integrated Mathematical Approaches to Socio-Epidemiological Dynamics'' and from the European Union's Horizon Europe research and innovation programme, under the Marie Skłodowska-Curie grant agreement No. 101110920 ``MesoCroMo – A Mesoscopic approach to Cross-diffusion Modelling in population dynamics''.
RE was partially funded by the MODCOV19 platform of the French National Institute of Mathematical Sciences and their Interactions (CNRS).
LP has been supported by the Royal Society under the Wolfson Fellowship ``Uncertainty quantification, data-driven simulations and learning of multiscale complex systems governed by PDEs". The partial support from ICSC -- Centro Nazionale di Ricerca in High Performance Computing, Big Data and Quantum Computing, funded by European Union -- NextGenerationEU, and MUR--PRIN 2022 Project No. 2022KKJP4X ``Advanced numerical methods for time dependent parametric partial differential equations with applications" is also acknowledged. 

%%%%%%%%%%%%%%%%%%%%%%%%%

\vspace*{7mm}
\includegraphics[width=0.5\textwidth]{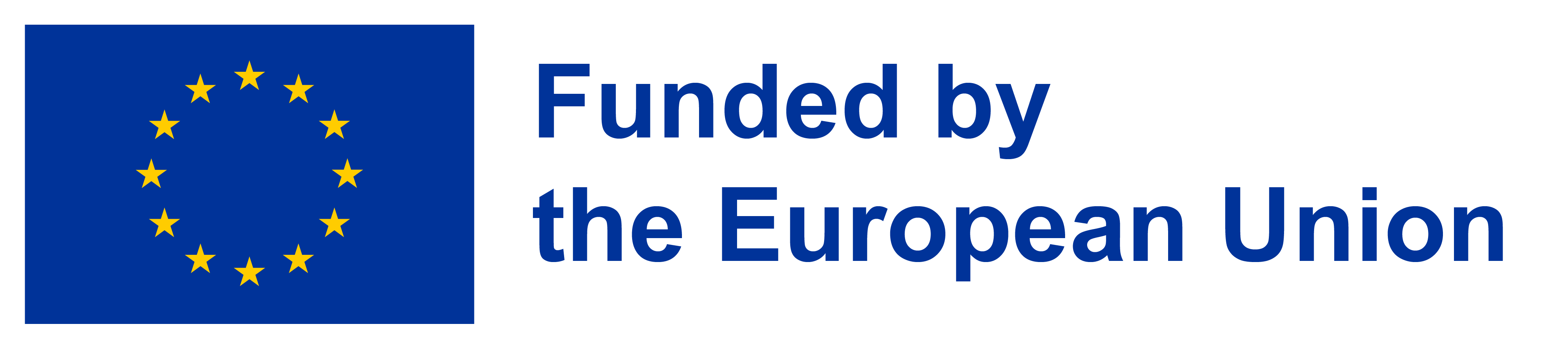}

%%%%%%%%%%%%%%%%%%%%%%%%%%%%%%%%%%%%%%%%%%%%
\section*{Disclaimer}
%%%%%%%%%%%%%%%%%%%%%%%%%%%%%%%%%%%%%%%%%%%%
Funded by the European Union. Views and opinions expressed are however those of the author(s) only and do not necessarily reflect those of the European Union or of the European Research Executive Agency (REA). Neither the European Union nor the granting authority can be held responsible for them.

%%%%%%%%%%%%%%%%%%%%%%%%%%%%%%%%%%%%%%%%%%%%

%%%%%%%%%%%%%%%%%%%%%%%%%%%%%%%%%%%%%%%%%%%%
\end{document}